\documentclass[extra]{gji}
\pdfoutput=1
\renewcommand{\arraystretch}{1.2}
\usepackage{longtable}
\usepackage{amsmath}
\usepackage{amscd}
\usepackage{longtable}
\usepackage{color}
\usepackage{graphicx}
\usepackage{float}
\usepackage{subfigure}
\newcommand{\di}{\boldsymbol{\delta}}

\def\numpoints{2217}
\def\optimJo{412}
\def\rmseo{0.142} 
\def\optimJt{472}
\def\rmset{0.058} 

\newcommand{\Hpsymbol}{H}

\newcommand{\Yfun}{Y} 

\newcommand{\Gfun}{G} 
\newcommand{\Hfun}{\boldsymbol{H}} 
\newcommand{\Pfun}{\boldsymbol{P}}
\newcommand{\Bfun}{\boldsymbol{B}}
\newcommand{\Cfun}{\boldsymbol{C}}
\newcommand{\Efun}{\boldsymbol{E}}
\newcommand{\Ffun}{\boldsymbol{F}}
\newcommand{\Hpfun}{\Hpsymbol}

\newcommand{\srs}{\estSlepcoef^\satalt}
\newcommand{\srsj}{\estSlepcoef^\satalt_J}
\newcommand{\trs}{\estvSlepcoef^\satalt}
\newcommand{\trsj}{\estvSlepcoef^\satalt_J}
\newcommand{\datarad}{d}
\newcommand{\datavec}{\boldsymbol{d}}
\newcommand{\signal}{V}
\newcommand{\signall}{V}
\newcommand{\vecsignal}{\bnabla \signal}
\newcommand{\vecsignall}{\bnabla \signall}
\newcommand{\scalsignal}{\partial_r \signal}
\newcommand{\scalsignalt}{\partial_\theta \signal}
\newcommand{\scalsignalp}{\partial_\phi \signal}
\newcommand{\noise}{n}
\newcommand{\vecnoise}{\boldsymbol{n}}
\newcommand{\estsignal}{\tilde \signal}
\newcommand{\variance}{\nu}
\newcommand{\bias}{\beta}
\newcommand{\esterr}{\epsilon}

\newcommand{\sigpower}{S}
\newcommand{\noisepower}{N}
\newcommand{\DmatL}{\hat\Dmat_{L}}
\newcommand{\DmatLL}{\hat\Dmat_{>L,L}}
\newcommand{\KmatLL}{\hat\Kmat_{>L,L}}
\newcommand{\GmatLJ}{\hat\Gmat_{>L,J}}
\newcommand{\HmatLJ}{\hat\Hmat_{\hat\Efunvec,>L,J}}
\newcommand{\GmatJ}{\Gmat_J}
\newcommand{\HmatJ}{\Hmat_J}
\newcommand{\GmatuJ}{\Gmat_{>J}}
\newcommand{\GLa}{\hat\Gfun_{>L,\alpha}}
\newcommand{\GLJ}{\hat\Gfunvec_{>L,J}}
\newcommand{\HLJ}{\hat\Hfunvec_{\hat\Efunvec,>L,J}}

\newcommand{\intrnorm}[1]{\int_R{\left(#1\right)^2}\dOmega}

\newcommand{\sigpoints}{\mathsf{\signal}}
\newcommand{\gradsigpoints}{\mathsf{\signal}'}
\newcommand{\noisepoints}{\mathsf{n}}
\newcommand{\Ypoints}{\boldsymbol{\mathsf{Y}}}
\newcommand{\Gpoints}{\boldsymbol{\mathsf{G}}}
\newcommand{\Epoints}{\boldsymbol{\mathsf{E}}}
\newcommand{\Hpoints}{\boldsymbol{\mathsf{H}}}

\newcommand{\Yfunvec}{\mathcal{Y}}
\newcommand{\Gfunvec}{\mathcal{G}}
\newcommand{\Hfunvec}{\boldsymbol{\mathcal{H}}}

\newcommand{\Efunvec}{\boldsymbol{\mathcal{E}}}

\newcommand{\Hpfunvec}{\mathcal{\Hpsymbol}}
\newcommand{\Gmat}{\mathbf{G}} 
\newcommand{\Hmat}{\mathbf{H}} 
\newcommand{\Amat}{\mathbf{A}} 
\newcommand{\Bmat}{\mathbf{B}} 
\newcommand{\Aelm}{A}
\newcommand{\Belm}{B}
\newcommand{\Dmat}{\mathbf{D}} 
\newcommand{\Delm}{D} 
\newcommand{\Kmat}{\mathbf{K}} 
\newcommand{\Kelm}{K} 
\newcommand{\sphcoef}{\mathrm{u}}
\newcommand{\estsphcoef}{\mathrm{\tilde u}}
\newcommand{\vsphcoef}{\mathrm{v}}
\newcommand{\estvsphcoef}{\mathrm{\tilde v}}
\newcommand{\Slepcoef}{\mathrm{s}}
\newcommand{\estSlepcoef}{\tilde{\mathrm{s}}}
\newcommand{\vSlepcoef}{\mathrm{t}}
\newcommand{\estvSlepcoef}{{\tilde{\mathrm{t}}}}
\newcommand{\slepfuncoef}{\mathrm{g}}
\newcommand{\vslepfuncoef}{\mathrm{h}}

\newcommand{\ssphcoef}{u}
\newcommand{\sestsphcoef}{\tilde{u}}
\newcommand{\svsphcoef}{v}
\newcommand{\sestvsphcoef}{\tilde v}
\newcommand{\sSlepcoef}{s}

\newcommand{\svSlepcoef}{t}
\newcommand{\sslepfuncoef}{g}
\newcommand{\svslepfuncoef}{h}
\newcommand{\matT}{\mathrm{T}} 
\newcommand{\pointT}{{\sf T}} 
\newcommand{\funT}{\mathcal{T}} 

\newcommand{\Earthrad}{{r_e}}
\newcommand{\satalt}{{r_s}}
\newcommand{\randrada}{{r_a}}

\newcommand{\region}{R}
\newcommand{\npoints}{k}

\newcommand{\diag}{\text{diag}} 
\newcommand{\mse}{\text{mse}} 
\newcommand{\mss}{\text{mss}}

\newcommand{\Imat}{\mathbf{I}}
\newcommand{\Lamat}{\mathbf{\Lambda}}
\newcommand{\Sigmat}{\mathbf{\Sigma}}
\newcommand{\dOmega}{\,d\Omega}

\newcommand{\dpoints}{\mathsf{d}}

\newcommand{\rvec}{\boldsymbol{\hat{r}}}
\newcommand{\xvec}{\boldsymbol{x}}
\newcommand{\zerovec}{\boldsymbol{0}}

\newcommand{\bnabla}{\mbox{\boldmath$\nabla$}}
\newcommand{\thvec}{\boldsymbol{\hat \theta}}
\newcommand{\phvec}{\boldsymbol{\hat \phi}}
\newcommand{\hsom}{\hspace{0.1em}}
\newcommand{\hsomm}{\hspace{-0.1em}}
\renewcommand{\mid}{:}

\newcommand{\afact}{-(l+1)\,\Earthrad^{-1}}
\newcommand{\bfact}{-\sqrt{(l+1)(2l+1)}\,\Earthrad^{-1}}

\newcommand{\Pone}{\mathbf{P 1}}
\newcommand{\Ptwo}{\mathbf{P 2}}
\newcommand{\Pthree}{\mathbf{P 3}}
\newcommand{\Pfour}{\mathbf{P 4}}

\begin{document}
\onecolumn

\title[Potential-field Estimation with Slepian
  Functions]{Potential-field estimation using scalar and vector
  Slepian functions at satellite altitude}
\author{Alain Plattner and Frederik J.~Simons}

\maketitle
\tableofcontents


\begin{abstract}
In the last few decades a series of increasingly sophisticated
satellite missions has brought us gravity and magnetometry data of
ever improving quality. To make optimal use of this rich source of
information on the structure of Earth and other celestial bodies, our
computational algorithms should be well matched to the specific
properties of the data. In particular, inversion methods require
specialized adaptation if the data are only locally available, their
quality varies spatially, or if we are interested in model recovery
only for a specific spatial region. Here, we present two approaches to
estimate potential fields on a spherical Earth, from gradient data
collected at satellite altitude. Our context is that of the estimation
of the gravitational or magnetic potential from vector-valued
measurements. Both of our approaches utilize spherical Slepian
functions to produce an approximation of local data at satellite
altitude, which is subsequently transformed to the Earth's spherical
reference surface. The first approach is designed for radial-component
data only, and uses scalar Slepian functions. The second approach uses
all three components of the gradient data and incorporates a new type
of vectorial spherical Slepian functions which we introduce in this
chapter.
\end{abstract}

\section{Introduction}

The estimation of the gravity potential \cite[e.g.][]{Moritz2010,Nutz2002} or
that of the magnetic potential on a spherical Earth \cite[e.g.][]{Sabaka+2010} 
from gradient data at satellite altitude
can be stated as a ``reevaluation'', of a three-dimensional function
that is harmonic in a spherical shell, given values of its gradient
within the harmonic shell \citep{Freeden+2009}.  The reevaluation on
the surface of a spherical Earth or planet is to be interpreted as a
transformation, between the gradient at satellite altitude on the one
hand, and the potential function on the surface on the other
hand. Such an operation is entwined with the notion of a (global)
basis of functions in which to carry it out. When expressed in
spherical harmonics, its numerical conditioning depends exponentially
on the spherical-harmonic bandwidth of the data
\citep{Freeden+2009}. The better the data quality, the higher the
spherical-harmonic degrees that can be resolved
\citep[e.g.][]{Maus+2006b}, but also, the poorer the conditioning of
the transformation.  Scalar and vector spherical harmonics
\citep[e.g.][]{Arkani-Hamed2001,Maus+2006c,Arkani-Hamed2004,
  Olsen+2009,Gubbins+2011b} are only a few among the many basis
functions that can be used for magnetic-field estimation. Alternatives
include ellipsoidal harmonics
\citep[e.g.][]{Boelling+2005,Maus2010,Lowes+2012}, monopoles
\citep[e.g.][]{OBrien+94}, spherical wavelets
\citep[e.g.][]{Mayer+2006, Chambodut+2005}, spherical-cap harmonics
\citep[e.g.][]{Haines85a,Hwang+97,Korte+2003}, and their relatives
\citep[e.g.][]{DeSantis91,Thebault+2006}. Specifically for
gravity-field estimation, besides the spherical harmonics
\citep[e.g.][]{Freeden+2009,Eshagh2009b}, we can also list spherical
wavelets \citep[e.g.][]{Chambodut+2005,Fengler+2006a}, ellipsoidal
harmonics \citep[e.g.][]{Lowes+2012}, and mascons
\citep[e.g.][]{Rowlands+2005}.

Data quality might not be evenly distributed over the entire sphere or
may even only be locally available
\cite[]{Arkani-Hamed+86,Arkani-Hamed2002,Maus+2006a}. For this reason,
methods that take the locality of the data into account are of great
value. Unfortunately, a function, and hence a method of analysis, can
not be bandlimited and spacelimited at the same time. Every localized
method that transforms data at satellite altitude into a potential
field on Earth's surface needs to circumvent or embrace this
fact. \citet{Schachtschneider+2010, Schachtschneider+2012} analyze the
errors introduced by local approximation in a general framework.

The method that we present here builds on the localized function bases
first described by \citet{Slepian+61} for problems in time-series
analysis. They constructed one-dimensional functions that are
bandlimited but optimally concentrated within a target interval, and
later extended the concept of what became known as the \emph{Slepian
  functions} to multidimensional Cartesian cases
\citep{Slepian64}. \cite{Albertella+99} and then \citet{Simons+2006a}
ushered in the realm of scalar spherical Slepian functions, and
\citet{Jahn+2012} and \citet{Plattner+2013} first described vectorial
spherical Slepian functions --- all of these ideally suited for
applications in geomathematics, and fitting neatly with the general
notions of signal concentration and the uncertainty principle espoused
by \cite{Freeden+2004b} and \cite{Kennedy+2013}, among others. A more
detailed introduction to scalar and vectorial Slepian functions can be
found in the chapter ``Scalar and Vector Slepian Functions, Spherical
Signal Estimation and Spectral Analysis'' by Simons and Plattner in
this book. Theoretical considerations on the application of scalar
Slepian functions to potential-field estimation from scalar potential
data at satellite altitude was presented by \citet{Simons+2006b}, and
some very practical cases in oceanography, terrestrial geodesy, and
planetary science, can be found elsewhere
\citep{Slobbe+2012,Harig+2012,Lewis+2012}.

In this chapter, after emphasizing some preliminaries in
Section~\ref{section preliminaries}, stating the problems to be solved
in Section~\ref{problo}, and introducing the scalar and a special type of
vector Slepian functions in Section~\ref{section Slepian functions},
we extend the approach presented by \citet{Simons+2006b} to the
potential estimation from radial-derivative data, in
Section~\ref{section radial data inversion}. Subsequently, we present
a method to estimate the potential field from local three-component
gradient data using vector Slepian functions in Section~\ref{section
  vectorial data inversion}. Finally, in Section~\ref{section
  examples} we present numerical examples for both, the
radial-component method and the fully vectorial gradient-data method.


\section{Scalar and Vector Spherical Harmonics and Harmonic Continuation}\label{section preliminaries}

In this chapter we employ a notation that is similar to the one used
in the chapter ``Scalar and Vector Slepian Functions, Spherical Signal
Estimation and Spectral Analysis'', by Simons and Plattner in this
book. We adapted the notation to transparently account for scalar and
vector-valued functions.  Scalar-valued functions are italicized, with
capital letters such as for example $\Yfun_{lm}$ for the classical
spherical-harmonic functions. Vector-valued functions are italic but
boldfaced, with capital letters, such as $\Efun_{lm}$ for the
gradient-vector harmonics that we define. Column vectors containing
scalar functions are in a calligraphic font, for example~$\Yfunvec$,
whereas column vectors that contain vector functions are calligraphic
but bold, as in~$\Efunvec$. Column vectors of expansion coefficients
are roman and lower-case, such as~$\sphcoef$, and their scalar
entries are in lowercase italics, such as~$\ssphcoef_{lm}$. If
functions or coefficients are estimated from the data, they receive a
tilde, such as $\estsignal$ or $\estsphcoef$.  Matrices containing
coefficients or multiplicative factors are roman and bold, such
as~$\Amat$. Matrices containing functions evaluated at specific points
are sans-serif bold, such as $\Ypoints$. 

\subsection{Scalar Spherical Harmonics}

As customary we define, for a point $\rvec$ on the surface of the unit
sphere $\Omega=\{\xvec \mid \lVert \xvec \rVert = 1\}$ with
colatitudinal value $0\leq \theta\leq \pi$ and longitudinal value
$0\leq \phi < 2\pi$, the real spherical-harmonic functions
\begin{align}
\Yfun_{l m}(\rvec)=\Yfun_{l m}(\theta,\phi)&=
\begin{cases}  
\sqrt{2}X_{l \lvert m \rvert}(\theta) \cos m\phi &\text{if } -l\leq m<0,\\
X_{l 0}(\theta)&\text{if } m=0,\\
\sqrt{2}X_{l m}(\theta)\sin m\phi &\text{if } 0<m\leq l,
\end{cases}\label{Y definition}
\end{align}
\begin{align}
X_{l m}(\theta)&=(-1)^m\left(\frac{2l+1}{4\pi}\right)^{1/2}
\left[\frac{(l-m)!}{(l+m)!}\right]^{1/2} P_{lm}(\cos \theta),\label{X
definition}\\ 
P_{lm}(\mu)&=\frac{1}{2^ll!}(1-\mu^2)^{m/2}\left(\frac{d}{d\mu}\right)^{l+m}(\mu^2-1)^l.
\label{P definition}
\end{align}
With this definition of the surface spherical harmonics~$\Yfun_{lm}$
we may learn from \cite{Backus+96}, \cite{Dahlen+98} or
\citet{Freeden+2009} that they are the orthonormal eigenfunctions of
the scalar Laplace-Beltrami operator
\begin{equation}\label{definition Laplace-Beltrami}
\nabla_1^2  = \partial_\theta^2 +\cot{\theta}\,\partial_\theta + 
(\sin{\theta})^{-2}\partial_\phi^2,
\end{equation}
with eigenvalues $-l(l+1)$, thus $\nabla_1^2 \Yfun_{l m} = -l(l+1)\Yfun_{lm}$. 
In spherical coordinates we can define the three-dimensional Laplace operator
\begin{equation}
\nabla^2 = \partial^2_r + 2r^{-1}\partial_r + r^{-2} \nabla_1^2,
\end{equation}
and the Laplace  equation by which we define a three-dimensional
function $\signal(r\rvec)$ to be harmonic, 
\begin{equation}\label{scalar Laplace equation}
\nabla^2 \signal(r\rvec)=0.
\end{equation}
The general solution of Eq.~(\ref{scalar Laplace equation}) comprises
one component that vanishes at the origin $r=0$ and another that is
regular by going to zero at infinity. The inner, $r^l \Yfun_{lm}$,
and outer, $r^{-l-1}\Yfun_{lm}$, solid spherical harmonics form a
basis for all solutions of Laplace's equation and serve to approximate
external-field and internal-field scalar potentials \cite[]{Olsen+2010}, respectively
\cite[]{Blakely95,Langel+98}.

The spherical harmonics~$\Yfun_{lm}$ defined in (\ref{Y definition})
form an orthonormal basis for square-integrable real-valued functions on the unit
sphere~$\Omega$. We can describe any such function~$\signal(\rvec)$  as a
unique linear combination of spherical harmonics via the expansion
\begin{equation}\label{spherical harmonics unit sphere}
\signal(\rvec)=\sum_{l=0}^\infty \sum_{m=-l}^l \ssphcoef_{lm} \Yfun_{lm}(\rvec),\quad \text{ where } \quad
\ssphcoef_{lm}=
\int_{\Omega} \signal(\rvec)\Yfun_{lm}(\rvec)\dOmega.
\end{equation}
Now let $\signal(r\rvec)$ be a three-dimensional function that
satisfies the Laplace equation (\ref{scalar Laplace equation}) outside
of the unit sphere, and which is regular at infinity.  If we know the
spherical-harmonic coefficients of $\signal(r\rvec)$ on the unit
sphere ($r=1$), from Eq.~(\ref{spherical harmonics unit sphere}), then
we can describe the function at any point~$r\ge 1$ outside of the unit
sphere using the outer harmonics by writing
\begin{equation}\label{scalar reevaluation from unit sphere}
\signal(r\rvec) = \sum_{l=0}^\infty\sum_{m=-l}^l  r^{-l-1} \ssphcoef_{lm} \Yfun _{lm}(\rvec).
\end{equation}
More generally, for a function $\signal(r\rvec)$ that satisfies 
 Eq.~(\ref{scalar Laplace equation}) 
outside a ball of radius~$\Earthrad$, and which is regular at
infinity, its evaluation on a sphere~$\Omega_\randrada$ of radius
$\randrada \ge \Earthrad$
is an expansion of spherical harmonics in the following way
\begin{equation}\label{scalar decomposition at radius randrada}
\signal(\randrada\rvec)=\sum_{l=0}^\infty\sum_{m=-l}^l \ssphcoef_{lm}^\randrada
\Yfun_{lm}(\rvec), 
\quad \text{ where } \quad \ssphcoef_{lm}^\randrada=\int_{\Omega} \signal(\randrada\rvec) \Yfun_{lm}
(\rvec) \dOmega,
\end{equation} 
In order to evaluate $\signal(r\rvec)$ at any other radius $r\geq
\Earthrad$ given the spherical-harmonic coefficient values
$\ssphcoef_{lm}^{\randrada}$ at radius $\randrada\ge \Earthrad$, we
can use Eq.~(\ref{scalar reevaluation from unit sphere}) twice, to
first evaluate $\signal(r\rvec)$ on the unit sphere and then, at
radius $r$, to obtain
\begin{equation}\label{potential field at other altitude}
\signal(r\rvec)=\sum_{l=0}^\infty\sum_{m=-l}^l 
\left( \frac{r}{\randrada}\right)
^{-l-1}\ssphcoef_{lm}^\randrada \Yfun_{lm}(\rvec).
\end{equation}

\subsection{Gradient-Vector Spherical Harmonics}\label{section vector spherical harmonics}

\begin{figure}
\centering
\subfigure{\includegraphics[width=0.4\textwidth,
trim = 13mm 7mm 59.5mm 0mm, clip]{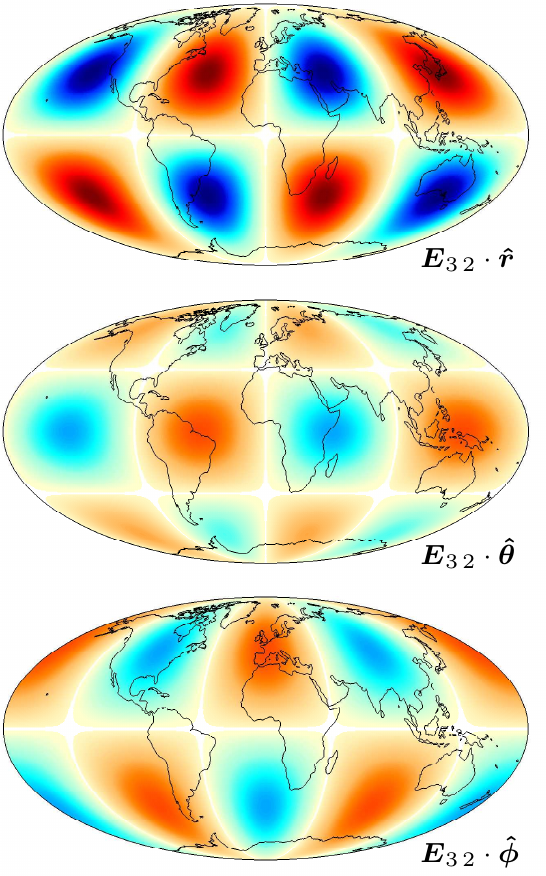}} \hspace{1cm}
\subfigure{\includegraphics[width=0.4\textwidth,
trim = 13mm 7mm 59.5mm 0mm, clip]{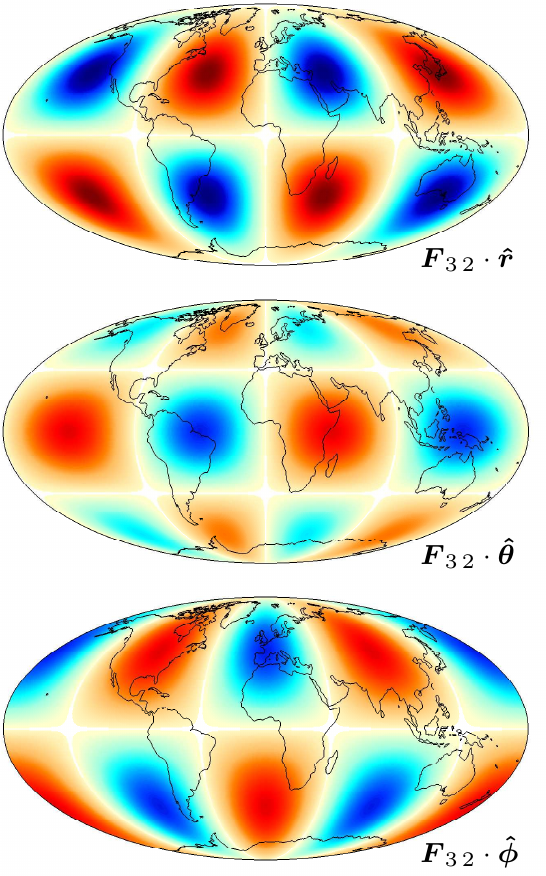}}
 \caption{\label{figure Elm} The gradient-vector spherical harmonics
   of Eq.~(\ref{E definition})--(\ref{F definition}), more specifically
   $\Efun_{3\, 2}$ and $\Ffun_{3\,2}$. Shown are the radial components
   $\Efun_{3\,2}\cdot \rvec$ and $\Ffun_{3\, 2}\cdot
   \rvec$, the tangential (colatitudinal) components
   $\Efun_{3\,2}\cdot \thvec$ and $
   \Ffun_{3\,2}\cdot \thvec$, and the tangential (longitudinal) components
   $\Efun_{3\,2}\cdot \phvec$ and
   $\Ffun_{3\,2}\cdot \phvec$.  }
\end{figure}

From the scalar spherical harmonics $\Yfun_{lm}(\rvec)$ we may define
vector spherical-harmonic functions on the unit sphere using the
Helmholtz decomposition in the usual way
\citep{Backus+96,Dahlen+98,Freeden+2009} as the fully normalized $\Pfun_{00}(\rvec)=
\rvec\hsom \Yfun_{00}(\rvec)$ and, for $l\geq 1$ and $-l \leq m\leq
l$,
\begin{align}
\Pfun_{lm}(\rvec)& = \rvec\hsom \Yfun_{lm}(\rvec),\label{Plm} \\
\Bfun_{lm}(\rvec)& = \frac{ \bnabla_1 \Yfun_{lm}(\rvec)}{\sqrt{l(l+1)}} = \frac{[
  \thvec\hsom\partial_\theta + \phvec\hsom (\sin \theta)^{-1} \partial_\phi]
  \Yfun_{lm}(\rvec)}{\sqrt{l(l+1)}},\label{Blm}\\ 
\Cfun_{lm}(\rvec)& = \frac{-\rvec\times \bnabla_1
  \Yfun_{lm}(\rvec)}{\sqrt{l(l+1)}}=\frac{[\thvec\hsom (\sin \theta)^{-1} \partial_\phi-\phvec
\hsom\partial_\theta]
  \Yfun_{lm}(\rvec)}{\sqrt{l(l+1)}}, \label{Clm}
\end{align}
where the relevant surface and the three-dimensional gradient operators are
\begin{align}
\bnabla_1&= \thvec\hsom\partial_\theta + \phvec \hsom(\sin \theta)^{-1} \partial_\phi,\\
\bnabla &= \rvec\hsom 
\partial_r + r^{-1}\bnabla_1
.
\end{align}
For our purposes, we use an alternative basis of normalized vector
spherical harmonics \citep{Nutz2002, Mayer+2006, Freeden+2009}.  We define
$\Efun_{00}=\Pfun_{00}$, and, for $l\geq 1$ and $-l\leq m 
\leq l$, 
\begin{align}
\Efun_{lm}&=\sqrt{\frac{l+1}{2l+1}}\Pfun_{lm} - \sqrt{\frac{l}{2l+1}}\Bfun_{lm},\label{E 
definition}\\
\Ffun_{lm}&=\sqrt{\frac{l}{2l+1}}\Pfun_{lm} + \sqrt{\frac{l+1}{2l+1}}\Bfun_{lm}.\label{F 
definition}
\end{align}
This alternative orthonormal basis of vector spherical harmonics
$\Efun_{lm},\Ffun_{lm}$, and $\Cfun_{lm}$, is identical to the $\tilde
y^{(1)}_{n,m}$, $\tilde y^{(2)}_{n,m}$, $-\tilde y^{(3)}_{n,m}$ in the
notation of \citet{Freeden+2009} and to the $\mathbf{u}^{(1)}_{n,k}$,
$\mathbf{u}^{(2)}_{n,k}$, $-\mathbf{u}^{(3)}_{n,k}$ of
\citet{Mayer+2006}. The functions $\mathbf{\Pi}_{ni}^{m,(c,s)}$ by 
\cite{Sabaka+2010} are scaled variants of the functions
$\Efun_{lm}$. Fig.~\ref{figure Elm} shows three-component spatial
renditions of two of the basis elements, $\Efun_{3\,2}$ and $\Ffun_{3\,2}$.

\subsection{Harmonic Continuation of Scalar and Vector Fields}\label{section potential field from 
satellite data}  

From now on we will always assume that Earth's surface is a
sphere~$\Omega_\Earthrad$ of fixed radius~$\Earthrad$, and that the
satellite altitude is a sphere~$\Omega_{\satalt}$ of radius~$\satalt
\geq \Earthrad$.  Using Eqs~(\ref{scalar decomposition at radius
  randrada})--(\ref{potential field at other altitude}) we can express
the potential field $\signal(\satalt\rvec)$ at the satellite
altitude $\satalt$ via the spherical-harmonic coefficients
$\ssphcoef^\Earthrad_{lm}$ on Earth's surface~$\Earthrad$ by
\begin{equation}\label{continued potential}
\signal(\satalt\rvec)=\sum_{l=0}^\infty\sum_{m=-l}^l
\left( \frac{\satalt}{\Earthrad}\right)
^{-l-1}\ssphcoef_{lm}^\Earthrad \Yfun_{lm}(\rvec),
\end{equation} 
where the coefficients are the entries of a vector~$\sphcoef^\Earthrad$, given by
\begin{equation}\label{definition uEarthrad}
 \ssphcoef_{lm}^\Earthrad=\int_{\Omega} \signal(\Earthrad \rvec) \Yfun_{lm}(\rvec) \dOmega.
 \end{equation}
The gradient of the potential at satellite altitude will then be given
by the expression
\begin{align}\label{potential gradient}
\vecsignal(\satalt\rvec) = 
\sum_{l=0}^{\infty}\sum_{m=-l}^{l}&
\afact \left(\frac{\satalt}{\Earthrad}\right)^{-l-2}
\ssphcoef_{lm}^\Earthrad  \,\rvec \hsom\Yfun_{lm}(\rvec) \\\nonumber
& +  \Earthrad^{-1} \left( \frac{\satalt}{\Earthrad} \right)^{-l-2} \ssphcoef_{lm}^\Earthrad 
\,\bnabla_1\Yfun_{lm}(\rvec).
\end{align}
Eq.~(\ref{potential gradient}) reveals that the potential coefficients $\ssphcoef_{lm}^
\Earthrad$ are uniquely determined from the radial component of its
gradient, as is well known \citep{Lowes+95},
\begin{equation}\label{radial derivative}
\vecsignal(\satalt\rvec)\cdot \rvec= \partial_r
 \signal(\satalt\rvec)  = \sum_{l=0}^\infty\sum_{m=-l}^l  \afact
 \left(\frac{\satalt} {\Earthrad}\right)^{-
l-2} \ssphcoef_{lm}^\Earthrad \Yfun_{lm}(\rvec).
\end{equation}
If we had perfect knowledge of the radial component
of the field~$\vecsignal$, 
the potential~$\signal$ 
would be uniquely
determined. When the data are contaminated by noise, we might gain by
taking the radial and both tangential components into account.

As shown, for example, by \citet{Freeden+2009}, we can reformulate
Eq.~(\ref{potential gradient}) by inserting the definitions
(\ref{Plm})--(\ref{Blm}) of the vector spherical harmonics
$\Pfun_{lm}$ and $\Bfun_{lm} $ and then using the definition~(\ref{E
 definition}) of the vector spherical harmonic~$\Efun_{lm}$ to write  
 \begin{align}\nonumber
 \vecsignal(\satalt\rvec) =& \sum_{l=0}^\infty\sum_{m=-l}^l
 \Earthrad^{-1} \left(\frac{\satalt}{\Earthrad}
\right)^{-l-2} 
\ssphcoef_{lm}^\Earthrad \big[(-l-1)\Pfun_{lm}(\rvec)+  \bnabla_1Y_{lm}(\rvec)\big]\\
\label{gradient potential satellite altitude}
 =& \sum_{l=0}^\infty\sum_{m=-l}^l \bfact \left(\frac{\satalt}{\Earthrad}\right)^{-l-2}\ssphcoef_{lm}
^\Earthrad 
\Efun_{lm}(\rvec).
 \end{align}
Eq.~(\ref{gradient potential satellite altitude}) thus shows that the
gradient~$\vecsignal(r\rvec)$ of a potential~$\signal(r\rvec)$ that
satisfies the Laplace equation $\nabla^2 \signal(r\rvec)=0$ outside
the sphere $r > \Earthrad$ and which vanishes at infinity, can be
expressed as a linear combination of the vector spherical
harmonics~$\Efun_{lm}(\rvec)$ of Eq.~(\ref{E definition}). For this
reason we will dub those \emph{gradient vector spherical harmonics} in
this paper. We can expand $\vecsignal(\satalt\rvec)$ as
\begin{equation}\label{vector decomposition at radius randrada pt1}
\vecsignal(\satalt\rvec) = \sum_{l=0}^\infty \sum_{m=-l}^l  
\svsphcoef_{lm}^\satalt
\Efun_{lm}(\rvec), 
\end{equation}
where the entries of the vector $\vsphcoef^\satalt$ are given by
\begin{equation}\label{vector decomposition at radius randrada pt2}
 \svsphcoef_{lm}^\satalt = \int_{\Omega} \vecsignal (\satalt\rvec) \cdot \Efun_{lm}
(\rvec) \dOmega.
\end{equation}
 
The relationships between the spherical-harmonic expansion
coefficients of the scalar potential~$\signal(r\rvec)$ and the radial
component of the gradient~$\scalsignal(r\rvec)$, and the
gradient-vector expansion coefficients of the
gradient~$\vecsignal(r\rvec)$, on Earth's surface~$r=\Earthrad$, and
at satellite altitude~$r=\satalt$, can be described in the following
(extended) ``Meissl'' scheme \citep{Rummel+95,Nutz2002, Freeden+2009}
which identifies the basis transformations and the multiplicative
factors for the expansion coefficients needed to interrelate them:
\begin{equation}\label{Meissl scheme}
\begin{CD}
\scalsignal(\satalt\rvec) 
@<\times (-l-1)/\satalt<<
\signal(\satalt\rvec) 
@>\times\left(-\sqrt{(l+1)(2l+1)}/\satalt\right)>\Yfun_{lm}\to\Efun_{lm}>
\vecsignal(\satalt\rvec)\\ 
@AA\times\left(\frac{\satalt}{\Earthrad}\right)^{-l-2}A
@AA\times\left(\frac{\satalt}{\Earthrad}\right)^{- l-1}A 
@AA\times\left(\frac{\satalt}{\Earthrad}\right)^{-l-2}A\\
\scalsignal(\Earthrad\rvec)
@<<\times(-l-1)/\Earthrad<\signal(\Earthrad\rvec) 
@>\Yfun_{lm}\to\Efun_{lm}>\times\left(-\sqrt{(l+1)(2l+1)}/\Earthrad\right)>
\vecsignal(\Earthrad\rvec)
\end{CD}
\end{equation}
From the spherical-harmonic coefficients of $\signal(\Earthrad\rvec)$
we can obtain the spherical-harmonic coefficients of
$\signal(\satalt\rvec)$ as $\ssphcoef_{lm}^\satalt =
(\satalt/\Earthrad)^{-l-1}\ssphcoef_{lm}^\Earthrad$. In order to
obtain the spherical-harmonic coefficients of
$\scalsignal(\satalt\rvec)$ from those of $\signal(\Earthrad\rvec)$ we
can either first follow $\signal(\Earthrad\rvec) \rightarrow
\scalsignal(\Earthrad\rvec)$ and then $\scalsignal(\Earthrad\rvec)
\rightarrow\scalsignal(\satalt\rvec)$, or first
$\signal(\Earthrad\rvec) \rightarrow \signal(\satalt\rvec)$ and then
$\signal(\satalt\rvec)\rightarrow\scalsignal(\satalt\rvec)$. Either
way we obtain the spherical-harmonic coefficients of
$\scalsignal(\satalt\rvec)$ as
$\afact(\satalt/\Earthrad)^{-l-2}\ssphcoef_{lm}^\Earthrad$.  To obtain
$\vecsignal(\satalt\rvec)$ from $\signal(\Earthrad\rvec)$ we replace
the spherical-harmonic functions~$\Yfun_{lm}$ by the gradient-vector
spherical harmonics~$\Efun_{lm}$ and multiply their coefficients with
$\bfact (\satalt/\Earthrad)^{-l-1}$. Similarly, we can obtain the
coefficients for any function in this scheme from the coefficients of
any other function by following the arrows: replacing, if necessary,
basis functions and multiplying the coefficients with the
corresponding factors, as shown.

\section{Potential-Field Estimation Using Spherical Harmonics}
\label{problo}

With the preliminaries out of the way we now turn our attention to
problems of geomathematical and geophysical interest. We distinguish
and treat the following four problems in potential-field estimation:
\begin{enumerate}
\item[$\Pone$] Estimating the spherical-harmonic potential-field
  coefficients from scalar data collected at the same altitude.
\item[$\Ptwo$] Estimating spherical-harmonic potential-field
  coefficients at source level from radial data collected at satellite 
  altitude.
\item[$\Pthree$] Estimating the gradient-vector spherical-harmonic
  coefficients from vector data collected at the same altitude. 
\item[$\Pfour$] Estimating spherical-harmonic potential-field
  coefficients at source level from vector data 
  at satellite altitude.
\end{enumerate}

Problems $\Pone$ and~$\Pthree$ will serve as problems introductory to
the more involved but practically more relevant $\Ptwo$
and~$\Pfour$. We will provide numerical solutions as estimations based
on data point values for all four problems. For problems $\Ptwo$ and
$\Pfour$ we will also provide analytical solutions which will then
enable us to calculate the effects of localization and bandlimitation
on the estimation process. When discussing, in Sections~\ref{section
  radial data inversion} and~\ref{section vectorial data inversion},
the use of localized basis functions as a means of regularizing
problems $\Ptwo$ and $\Pfour$, we will provide an analysis of the
effect of making bandlimited reconstructions of non-bandlimited
functions explicitly, in Sections~\ref{section analytical formulation}
and~\ref{section vector analytical formulation}.

\subsection{Discrete Formulation and Unregularized Solutions}\label{section point wise}

In this section we describe classical least-squares approaches to
estimating the spherical-harmonic (problems $\Pone$, $\Ptwo$, and~$\Pfour$)
or gradient-vector spherical-harmonic (problem $\Pthree$) coefficients
of potential fields and their gradients from discretely available,
noiseless data. 

\subsubsection{Problem $\Pone$: Scalar potential data, scalar-harmonic
  potential coefficients, equal
altitude}

Let there be $\npoints$ scalar function values
\begin{equation}\label{definition sigpoints}
  \sigpoints = 
  \Big(\signal(\satalt\rvec_1)\quad\cdots\quad\signal(\satalt\rvec_\npoints)\Big)^\pointT,
\end{equation}
evaluated at positions $\satalt\rvec_1,\ldots,
\satalt\rvec_\npoints$ on a sphere $\Omega_\satalt$. These are the samples 
\begin{equation}\label{bloL}
\signal(\satalt\rvec_i)=\sum_{l=0}^\infty\sum_{m=-l}^l
\ssphcoef^\satalt_{lm}\Yfun_{lm}(\rvec_i)
.
\end{equation}
Our objective is to estimate the spherical-harmonic
coefficients~$\ssphcoef^\satalt_{lm}$ within a certain bandwidth~$L$,
i.e. for $0\leq l\leq L$ and $-l\leq m\leq l$.  This can be performed
using least-squares, assuming that the number of data exceeds the
number of degrees of freedom in the system, $(L+1)^2\leq \npoints$.
Defining the matrix of point evaluations on the unit sphere
\begin{equation}\label{Y point matrix}
\Ypoints
=\begin{pmatrix}
\Yfun_{00}(\rvec_1)&\cdots&\Yfun_{00}(\rvec_\npoints)\\
\vdots&&\vdots\\
\Yfun_{L L}(\rvec_1)&\cdots&\Yfun_{L L}(\rvec_\npoints)			
,
\end{pmatrix}
\end{equation}
and the bandlimited vector of estimated coefficients
\begin{equation}\label{urs}
\estsphcoef^\satalt=
\begin{pmatrix}
\sestsphcoef^\satalt_{00}&\cdots&
\sestsphcoef^\satalt_{LL} 
\end{pmatrix}^T
,
\end{equation}
the statement of our first problem is to solve
\begin{equation}\label{scalar to scalar equal alt equation}
\arg\min_{\estsphcoef^\satalt}\lVert \Ypoints^\pointT \estsphcoef^\satalt -\sigpoints  \rVert^2
,
\end{equation} 
and the solution is given by
\begin{equation}\label{scalar to scalar equal alt solution} 
 \estsphcoef^\satalt = (\Ypoints\Ypoints^\pointT)^{-1} \Ypoints\,\sigpoints
 \qquad  \text{(solution to problem $\Pone$).} 
\end{equation}

\subsubsection{Problem $\Ptwo$: Scalar derivative data,
  scalar-harmonic potential coefficients,
  different altitudes}

Next, we wish to turn the equal-altitude problem~$\Pone$ described in
Eq.~(\ref{scalar to scalar equal alt equation}) and solved in
Eq.~(\ref{scalar to scalar equal alt solution}) into a
$\satalt$-to-$\Earthrad$ downward-continuation, radial-derivative
component-to-potential problem $\Ptwo$.  We define a diagonal upward
transformation matrix~$\Amat$, which includes the effects of harmonic
continuation and radial differentiation (see Eqs~\ref{radial
  derivative} and~\ref{Meissl scheme}), by its elements 
\begin{equation}\label{definition Amat}
\Aelm_{lm,l'm'} = \afact\left(\frac{\satalt}{\Earthrad}\right)^{-l-2}\delta_{ll'}\delta_{mm'}.
\end{equation}
The discrete set of point values from which we desire to recover the
spherical-harmonic potential coefficients on the surface of the Earth,
$\ssphcoef^\Earthrad_{lm}$, are the sampled radial components of the gradient of the
potential~(\ref{bloL}) evaluated at satellite altitude~$\satalt$,
\begin{equation}\label{radial data}
\gradsigpoints_r = 
\Big(\vecsignal(\satalt\rvec_1)\cdot \rvec \quad \cdots \quad \vecsignal(\satalt\rvec_
\npoints) \cdot \rvec 
\Big)^\pointT,
\end{equation}
Problem $\Ptwo$, estimating the spherical-harmonic coefficients
$\ssphcoef^\Earthrad_{lm}$ of the potential on Earth's
surface~$\Omega_\Earthrad$, collected in the vector
\begin{equation}\label{ure}
\estsphcoef^\Earthrad=
\begin{pmatrix}
\sestsphcoef^\Earthrad_{00}&\cdots&
\sestsphcoef^\Earthrad_{LL} 
\end{pmatrix}^T
,
\end{equation}
from potential-field data collected at satellite altitude on
$\Omega_\satalt$, is then formulated as 
\begin{equation}
\label{Equation problem 2}
\arg\min_{\estsphcoef^\Earthrad}\lVert
\Ypoints^\pointT\Amat\estsphcoef^\Earthrad - \gradsigpoints_r \rVert^2
,
\end{equation}
and is found to be
\begin{equation}\label{solution 1 to problem 2}
\estsphcoef^\Earthrad =
\Amat^{-1}(\Ypoints\Ypoints^\pointT)^{-1}\Ypoints\,\gradsigpoints_r
\qquad \text{(solution 1 to problem $\Ptwo$).} 
\end{equation}

\subsubsection{Problem $\Pthree$: Vector data, vector-harmonic
  coefficients, equal altitude}

In a third problem we seek to estimate the coefficients of the
gradient function $\vecsignal(\satalt \rvec)$ at satellite altitude,
all together
\begin{equation}\label{definition vsphcoef}
\estvsphcoef^\satalt=
\begin{pmatrix}
\sestvsphcoef^\satalt_{00}&\cdots&\sestvsphcoef^\satalt_{LL}
\end{pmatrix}^T
,
\end{equation}
in the basis of the gradient-vector spherical harmonics~$\Efun_{lm}$,
from discrete function values of 
$\vecsignal(\satalt\rvec)$ given at the points
$\satalt\rvec_1,\ldots,\satalt\rvec_\npoints$. Introducing
\begin{equation}\label{full vectorial data}
\gradsigpoints = \begin{pmatrix}{\gradsigpoints_r}^\pointT & {\gradsigpoints_\theta}^\pointT & {\gradsigpoints_\phi}^\pointT
\end{pmatrix}^
\pointT,
\end{equation}
with $\gradsigpoints_r$ as defined previously in Eq.~(\ref{radial
  data}), and, analogously, 
\begin{align}
\gradsigpoints_\theta =& 
\Big( \vecsignal(\satalt\rvec_1) \cdot \thvec \quad \cdots \quad
\vecsignal(\satalt\rvec_\npoints)\cdot \thvec\Big)^\pointT,\label{vecbla1} \\
 \gradsigpoints_\phi =& 
\Big( \vecsignal(\satalt\rvec_1)\cdot \phvec \quad \cdots \quad
\vecsignal(\satalt\rvec_
\npoints)\cdot \phvec \Big)^\pointT.
\label{vecbla2}
\end{align}
To formulate problem $\Pthree$ for the pointwise evaluated functions
given in Eqs~(\ref{radial data}) and~(\ref{vecbla1})--(\ref{vecbla2}),
namely the samples
\begin{equation}\label{blaL}
\vecsignal(\satalt\rvec_i) = \sum_{l=0}^\infty \sum_{m=-l}^l \svsphcoef_{lm}^\satalt\Efun_{lm}(\rvec_i), 
\end{equation}
we also define the matrix of point-evaluations of the gradient-vector
spherical harmonics
\begin{equation}\label{E point matrix}
\Epoints = \begin{pmatrix} \Epoints_r&\Epoints_\theta & \Epoints_\phi \end{pmatrix},
\end{equation}
where the constituent matrices are given by
\begin{align}
  \Epoints_r=&
  \begin{pmatrix}
    \Efun_{00}(\rvec_1)\cdot \rvec &\cdots&\Efun_{00}(\rvec_\npoints)\cdot\rvec\\
    \vdots&&\vdots\\
    \Efun_{L L}(\rvec_1)\cdot \rvec&\cdots&\Efun_{L L}(\rvec_\npoints)\cdot 
    \rvec
  \end{pmatrix},\\ \nonumber \\ 
  \Epoints_\theta=&
  \begin{pmatrix}
    \Efun_{00}(\rvec_1)\cdot \thvec &\cdots&\Efun_{00}(\rvec_\npoints)\cdot \thvec\\
    \vdots&&\vdots\\
    \Efun_{L L}(\rvec_1)\cdot \thvec&\cdots&\Efun_{L L}(\rvec_\npoints)\cdot \thvec
  \end{pmatrix},\\ \nonumber \\ 
  \Epoints_\phi=&
  \begin{pmatrix}
    \Efun_{00}(\rvec_1)\cdot \phvec&\cdots&\Efun_{00}(\rvec_\npoints)\cdot \phvec\\
    \vdots&&\vdots\\
    \Efun_{L L}(\rvec_1)\cdot \phvec&\cdots&\Efun_{L L}(\rvec_\npoints)\cdot 
    \phvec
  \end{pmatrix}.
\end{align}
Using the definitions in Eqs~(\ref{definition vsphcoef}), (\ref{full
  vectorial data}), and~(\ref{E point matrix}), problem $\Pthree$ is
stated as
\begin{equation}
\label{problem 3}
\arg\min_{\estvsphcoef^\satalt}\lVert \Epoints^\pointT\estvsphcoef^\satalt -
\gradsigpoints \rVert^2
,
\end{equation}
and easily seen to be solved by 
\begin{equation}\label{solution to problem 3}
\estvsphcoef^\satalt = (\Epoints\Epoints^\pointT)^{-1}
\Epoints\gradsigpoints \qquad \text{(solution to problem $\Pthree$)}. 
\end{equation}

\subsubsection{Problem $\Pfour$: Vector data, scalar-harmonic potential coefficients, different altitudes}

Finally, in order to transform the equal-altitude gradient-vector
problem $\Pthree$ into a downward-continuation, gradient-data to
scalar-potential problem $\Pfour$ we introduce the upward
transformation matrix~$\Bmat$. This diagonal matrix contains the
effect of harmonic continuation and differentiation (see
Eqs~\ref{gradient potential satellite altitude} and~\ref{Meissl scheme}), and has the elements 
\begin{equation}\label{definition B}
\Belm_{lm,l'm'} =  \bfact \left(\frac{\satalt}{\Earthrad}\right)^{-l-2}\delta_{ll'}\delta_{mm'}.
\end{equation}
Problem $\Pfour$, estimating the spherical-harmonic coefficients
$\ssphcoef^\Earthrad_{lm}$ of the potential on Earth's surface~$\Omega_\Earthrad$, from
gradient data collected at satellite altitude on $\Omega_\satalt$, can
hence be formulated as
\begin{equation}
\label{Equation problem 4}
\arg\min_{\estsphcoef^\Earthrad} \lVert
\Epoints^\pointT\Bmat\estsphcoef^\Earthrad - \gradsigpoints \rVert^2
,
\end{equation} 
with the solution
\begin{equation}\label{solution 1 to problem 4}
\estsphcoef^\Earthrad=  \Bmat^{-1}
(\Epoints\Epoints^\pointT)^{-1}\Epoints\gradsigpoints \qquad
 \text{(solution 1 to problem $\Pfour$).}
\end{equation}

For every one of the solutions listed thus far in Eqs~(\ref{scalar to
  scalar equal alt solution}), (\ref{solution 1 to problem 2}),
(\ref{solution to problem 3}), and~(\ref{solution 1 to problem 4}), we
require at least as many data points as there are coefficients to
estimate, $\npoints\geq(L+1)^2$, or $3\npoints\geq(L+1)^2$ for the 
vectorial case, otherwise the matrices
$(\Ypoints\Ypoints^\pointT)$ and $(\Epoints\Epoints^\pointT)$ will not
be invertible. If we have data distributed only over a certain
concentration region~$\region$, the matrices
$(\Ypoints\Ypoints^\pointT)$ or $(\Epoints\Epoints^\pointT)$ will
usually be badly conditioned and require regularization
\citep{Simons+2006b}. Furthermore, we have sidestepped issues of bias
due to making bandlimited estimates (Eqs~\ref{urs}, \ref{ure}
and~\ref{definition vsphcoef}) from intrinsically wideband field
observations (\ref{bloL}) and~(\ref{blaL}). Lastly, we have so far
blithely ignored any observational noise. For the more realistic
practical cases of the problems $\Ptwo$ and $\Pfour$ we will develop
regularization methods, in Sections~\ref{section radial data
  inversion} and~\ref{section vectorial data inversion}, that take the
target region $\region$ explicitly into account, and whose performance
we assess using detailed statistical considerations. Before doing so,
however, we first establish some more notation.

\subsection{Continuous Formulation and  Bandwidth Considerations}\label{preliminaries functional
  formulation}

Let us define the $(L+1)$-dimensional vector $\Yfunvec$ to contain the
spherical-harmonic functions~$\Yfun_{lm}$ up to a bandlimit~$L$,
\begin{equation}\label{Yfunvec definition}
\Yfunvec=
\begin{pmatrix} 
\Yfun_{00} &\cdots&\Yfun_{LL}
\end{pmatrix}^T.
\end{equation} 
In the same manner we shall define the vector of all
spherical-harmonic functions up to infinite bandwidth as, simply,
$\hat \Yfunvec$. The symbol $\hat\Yfunvec_{>L}$ will denote the vector
of spherical harmonics with degrees higher than~$L$. Using this
notation we write the column vector with the complete basis
\begin{equation}\label{Yfunvec all degrees}
\hat\Yfunvec=\begin{pmatrix}
\Yfunvec\\\hat\Yfunvec_{>L}
\end{pmatrix}.
\end{equation}
Up to a certain bandlimit~$L$, we can describe the spherical-harmonic
coefficients of a potential field~$\signal(\satalt\rvec)$ on the
sphere~$\Omega_\satalt$, whose estimates we encountered previously in
Eq.~(\ref{urs}), as 
\begin{equation}\label{functional sphcoef calculation}
  \sphcoef^\satalt=\int_\Omega\Yfunvec\, \signal(\satalt\rvec)
\dOmega,
\end{equation}
and their infinite-dimensional counterparts will be
\begin{align}\label{bluinf}
  \hat\sphcoef^\satalt&=\int_\Omega\hat\Yfunvec\,\signal(\satalt\rvec)\dOmega,\\
  \hat\sphcoef^\satalt_{>L}&=\int_\Omega\hat\Yfunvec_{>L}\signal(\satalt\rvec)\dOmega.
\end{align}
With these definitions we rewrite a representation similar to
Eq.~(\ref{bloL}), for a potential field that is not
bandlimited, as
\begin{equation}\label{non-bandlimited scalar}
\signal(\satalt\rvec)=\hat\Yfunvec^\funT\hat\sphcoef^\satalt = \begin{pmatrix} \Yfunvec^\funT & \hat
\Yfunvec_{>L}^\funT 
\end{pmatrix} \begin{pmatrix} \sphcoef^\satalt\\ \hat\sphcoef^\satalt_{>L} \end{pmatrix}=  \Yfunvec^
\funT\sphcoef^\satalt + \hat
\Yfunvec_{>L}^\funT\hat\sphcoef^\satalt_{>L},
\end{equation}
and for future reference we also write the equivalent of
Eq.~(\ref{radial derivative}), using Eq.~(\ref{definition Amat}), 
in broadband and bandlimited form as
\begin{equation}\label{popo}
\partial_r\signal(\satalt\rvec)=
\hat\Yfunvec^\funT\hsomm\Amat\hat\sphcoef^\Earthrad
=\Yfunvec^\funT\hsomm\Amat\sphcoef^\Earthrad+
\hat\Yfunvec_{>L}^\funT\Amat\hat\sphcoef^\Earthrad_{>L}.
\end{equation}
The double duty performed of $\Amat$ is not likely to be confusing: its
dimensions simply adapt to those of the vectors that it multiplies.
Eq.~(\ref{non-bandlimited scalar}) contains an estimation problem
that, assuming continuity of global data coverage, is solved by
Eq.~(\ref{bluinf}), owing to the orthonormality of the $\Yfun_{lm}$
over the entire sphere, $\int_\Omega
\Yfun_{lm}\Yfun_{l'm'}\dOmega= \delta_{ll'}\delta_{mm'}$. For complete data coverage,
Eq.~(\ref{functional sphcoef calculation}) solves the bandlimited
portion of the estimation problem, and we can see that in that case
Eq.~(\ref{functional sphcoef calculation}) is indeed the continuous
equivalent of Eq.~(\ref{scalar to scalar equal alt solution}), as
pointed out also in the chapter ``Scalar and Vector Slepian Functions,
Spherical Signal Estimation and Spectral Analysis'' by Simons and
Plattner elsewhere in this book.

For the gradient-vector spherical harmonics we define the
$(L+1)^2$-dimensional vector of functions containing the $\Efun_{lm}$
up to a certain bandlimit~$L$ as
\begin{equation}\label{Efunvec definition} 
\Efunvec =  \begin{pmatrix}
  \Efun_{00}&\cdots&\Efun_{LL}\end{pmatrix}^T
.
\end{equation}
Using a similar notation as for the scalar harmonics, the
infinite-dimensional vector containing all gradient-vector spherical
harmonics to infinite bandlimit will be $\hat\Efunvec$, and the
infinite-dimensional vector  with all gradient
vector spherical harmonics for degrees $l>L$ will be
$\hat\Efunvec_{>L}$. The column vector with the complete vector basis
is thus
\begin{equation}\label{Efunvec all degrees}
\hat\Efunvec=\begin{pmatrix}
\Efunvec\\\hat\Efunvec_{>L}
\end{pmatrix}.
\end{equation}
Up to a given bandwidth~$L$ we can calculate the gradient-vector
spherical-harmonic coefficients of a gradient field
$\vecsignal(\satalt\rvec)$ at satellite altitude, previously known in
the form of Eq.~(\ref{vector decomposition at radius randrada pt2}), via  the expression
\begin{equation}\label{functional vsphcoef calculation}
\vsphcoef^\satalt=\int_\Omega\Efunvec\cdot\vecsignal(\satalt\rvec)\dOmega.
\end{equation} 
The corresponding infinite-dimensional vectors of gradient-vector spherical-harmonic
coefficients are
\begin{align}\label{buli}
\hat\vsphcoef^\satalt&=\int_\Omega\hat\Efunvec\cdot\vecsignal(\satalt\rvec)\dOmega,\\
\hat\vsphcoef^\satalt_{>L}&=\int_\Omega\hat\Efunvec_{>L}\cdot\vecsignal(\satalt\rvec)\dOmega
.
\end{align}
Our definition of the inner product between a vector of vector-valued
functions and a vector-valued function is 
\begin{equation}\label{definition gradient times Efunvec}
\Efunvec\cdot\vecsignal = \begin{pmatrix}
\Efun_{00}\cdot \vecsignal\\
\vdots\\
\Efun_{LL}\cdot\vecsignal
\end{pmatrix}.
\end{equation}
In the same way we define the outer product between two vectors of
vector-valued functions as
\begin{equation}\label{definition Efunvec times Efunvec}
\Efunvec\cdot\Efunvec^\funT=\begin{pmatrix} \Efun_{00}\cdot\Efun_{00}&\cdots&
\Efun_{00}\cdot\Efun_{LL}\\
\vdots&&\vdots\\
\Efun_{LL}\cdot\Efun_{00}&\cdots&\Efun_{LL}\cdot\Efun_{LL}
 \end{pmatrix}.
\end{equation}
We can represent the non-bandlimited gradient function~$\vecsignal(\satalt\rvec)$ via its 
gradient-vector spherical-harmonic coefficients
\begin{equation}\label{bliblu}
\vecsignal(\satalt\rvec) = \hat\Efunvec^\funT\hat\vsphcoef^\satalt =\begin{pmatrix} \Efunvec^\funT& \hat
\Efunvec_{>L}^\funT 
\end{pmatrix} \begin{pmatrix} \vsphcoef^\satalt \\ \hat\vsphcoef^\satalt_{>L} \end{pmatrix} = \Efunvec^
\funT\vsphcoef^\satalt + \hat
\Efunvec_{>L}^\funT \hat\vsphcoef^\satalt_{>L},
\end{equation}
and, via eq.~(\ref{definition B}) with the dimensions of $\Bmat$
stretched appropriately as in Eq.~(\ref{popo}), the equivalent of
Eq.~(\ref{gradient potential satellite altitude}),  
\begin{equation}
\vecsignal(\satalt\rvec)=
\hat\Efunvec^\funT\hsomm\Bmat\hat\sphcoef^\Earthrad\label{popa}=
\Efunvec^\funT\hsomm\Bmat\sphcoef^\Earthrad+
\hat\Efunvec_{>L}^\funT\Bmat\hat\sphcoef^\Earthrad_{>L}
.
\end{equation}
Eq.~(\ref{bliblu}) again contains an estimation problem solved by
Eq.~(\ref{buli}) in the scenario of noiseless, continuous, and
complete data-coverage, as can be seen from the orthonormality
orthonormality relation $\int_\Omega
\Efun_{lm}\cdot\Efun_{l'm'}\dOmega= \delta_{ll'}\delta_{mm'}$. As with
the scalar problem described above, the bandlimited coefficient
set~(\ref{functional vsphcoef calculation}) is approximated by the
discrete solution~(\ref{solution to problem 3}) in the case of
complete data coverage.

\section{Scalar and Vector Spherical Slepian Functions}\label{section Slepian functions}

In this section we summarize the derivation and properties of scalar
spherical Slepian functions developed by \citet{Simons+2006a} and
further discussed in the chapter
``Scalar and Vector Slepian Functions, Spherical Signal Estimation and
Spectral Analysis'' by Simons and Plattner in this book. The scalar
Slepian functions will play a key role in the solution to
problem~$\Ptwo$, the estimation of scalar spherical-harmonic
coefficients of the potential on Earth's surface from radial-component
data at satellite altitude, in a spatially localized setting. To be
able to consider spatial localization in the context of problem
$\Pfour$, the estimation of the scalar potential on Earth's surface
from vectorial gradient data at altitude, we introduce a special case
of the vectorial Slepian functions constructed by
\citet{Plattner+2013} and further discussed in the chapter ``Scalar
and Vector Slepian Functions, Spherical Signal Estimation and Spectral
Analysis'' by Simons and Plattner in this book. 

\subsection{Scalar Slepian Functions}\label{section scalar Slepian functions}

We design functions that are bandlimited to a maximum spherical
harmonic degree~$L$ but at the same time spatially concentrated inside
a target region~$\region$. Via optimization of a local energy
criterion we obtain a new basis of functions in the sense of
\citet{Slepian83}, as a particular linear combination of spherical
harmonics. Unlike the latter, which are global functions
indexed by their degree and order, the ``Slepian'' functions can be sorted
according to their energy concentration inside of the target
region. Local approximations to scalar functions can be made from the
first few well-concentrated Slepian functions, as we will be needing
for the solution to problem~$\Ptwo$, where the spherical-harmonic
coefficients of a potential field are determined from radial data only. 

Scalar spherical Slepian functions~$\Gfun$ are bandlimited
spherical-harmonic expansions
\begin{equation}\label{definition single Gfun}
\Gfun(\rvec)=\sum_{l=0}^L
\sum_{l=-m}^m 
\sslepfuncoef_{lm}\Yfun_{lm}(\rvec)
=\Yfunvec^\funT\slepfuncoef
\end{equation}
that are constructed by solving the quadratic optimization problem
\begin{equation}\label{scalar concentration problem}
\lambda=\max_{\Gfun} \frac{\displaystyle\int_{\region} \Gfun^2(\rvec)\dOmega}
{\displaystyle \int_\Omega  
\Gfun^2(\rvec)\dOmega}=
\max_{\slepfuncoef} \frac{\slepfuncoef^\matT \Dmat\, \slepfuncoef}
{\slepfuncoef^\matT \slepfuncoef},
\end{equation}
for the expansion coefficients $\sslepfuncoef_{lm}$  in the $(L+1)^2$-dimensional
column vectors
\begin{equation}
\slepfuncoef=
\begin{pmatrix}
\sslepfuncoef_{00}&\cdots&\sslepfuncoef_{LL} 
\end{pmatrix}^T,
\end{equation}
with $\Yfunvec$ as in Eq.~(\ref{Yfunvec definition}). The symmetric
positive-definite kernel matrix $\Dmat$ is defined by its elements
\begin{equation}\label{definition scalar kernel}
\Delm_{lm,l'm'}=\int_\region\Yfun_{lm}(\rvec)\Yfun_{l'm'}(\rvec)\dOmega,\qquad
\Dmat=\int_{\region}\Yfunvec\Yfunvec^\funT\dOmega.
\end{equation}
The stationary solutions of Eq.~(\ref{scalar concentration problem}) are the 
eigenvectors $\slepfuncoef_1,\ldots,\slepfuncoef_\alpha,\ldots,\slepfuncoef_{(L+1)^2}$ that
constitute an orthogonal coefficient matrix
\begin{equation}\label{definition Gmat}
\Gmat = 
\Big(
\slepfuncoef_1\quad \cdots\quad \slepfuncoef_\alpha\quad \cdots\quad \slepfuncoef_{(L+1)^2} 
\Big)
,\qquad
\Gmat\Gmat^\matT=\Gmat^\matT\Gmat=\Imat
=\int_{\Omega}\Yfunvec\Yfunvec^\funT\dOmega,
\end{equation} 
defined by the eigenvalue problem
\begin{equation}\label{scalar eigenvalue problem}
\Dmat\hsom\Gmat = \Gmat\Lamat,
\qquad
\Dmat=\Gmat\Lamat\Gmat^\matT
,
\end{equation}
with the eigenvalues
$\Lamat=\diag(\lambda_1,\ldots,\lambda_{(L+1)^2})$ the concentration
values of Eq.~(\ref{scalar concentration problem}), many of which are
near one, and many near zero. We index the 
individual elements $\slepfuncoef_{lm,\alpha}\in\Gmat$ by
$\alpha=1,\ldots,(L+1)^2$ and order them according to their
eigenvalues in decreasing order $1> \lambda_1\geq \cdots \geq
\lambda_{(L+1)^2}>0$, to obtain a global basis for the space of
spherical functions with bandlimit~$L$, given by
\begin{equation}\label{definition Gfun}
\Gfun_\alpha(\rvec)=\sum_{l=0}^L\sum_{m=-l}^l
\sslepfuncoef_{lm,\alpha}\Yfun_{lm}(\rvec)=\Yfunvec^\funT\slepfuncoef_\alpha
.
\end{equation}
We normalize the different eigenvectors $\slepfuncoef_\alpha$ so that
the newly constructed basis $\Gfun_1,\ldots,\Gfun_{(L+1)^2}$ remains
orthonormal over the entire sphere $\Omega$, but it is now also
orthogonal over the region $\region$,
\begin{equation}\label{slepor}
  \int_\Omega \Gfun_\alpha\Gfun_\beta\dOmega=\delta_{\alpha\beta}, \qquad 
  \int_\region \Gfun_\alpha\Gfun_\beta\dOmega=\lambda_\alpha\delta_{\alpha\beta} 
.
\end{equation} 
To further the notation introduced in and after (\ref{Yfunvec
  definition}) we now define the $(L+1)^2$-dimensional function vector
containing all Slepian functions, for a bandlimit~$L$ and a
region~$\region$, to be
\begin{equation}\label{definition Gfunvec}
\Gfunvec=
\Big(
\Gfun_1\quad \cdots\quad\Gfun_{(L+1)^2} 
\Big)^T=\Gmat^\matT\Yfunvec.
\end{equation}
Identifying the Slepian transformation matrix~$\Gmat$ in this way, we can then
write the representation of a bandlimited function~$\signall(\rvec)$ by
involving the spherical-harmonic expansion coefficients~$\sphcoef$, or
the Slepian-function expansion coefficients
$\Slepcoef=\Gmat^\matT\sphcoef$, in the equivalent forms
\begin{equation}\label{scalar slepian transformation}
\signall(\rvec)=
\sum_{l=0}^L\sum_{m=-l}^l
\ssphcoef_{lm}\Yfun_{lm}(\rvec)
=\Yfunvec^\funT\sphcoef
=\Yfunvec^\funT\Gmat\Gmat^\matT\sphcoef=\Gfunvec^\funT\Slepcoef=\sum_{\alpha=1}^{(L+1)^2}\sSlepcoef_\alpha
\Gfun_\alpha(\rvec).
\end{equation}

Writing the $(L+1)^2\times J$-dimensional matrix containing the
$(L+1)^2$ spherical-harmonic coefficients of the $J$
best-concentrated Slepian functions~$\GmatJ$ and its
$(L+1)^2\times(L+1)^2-J$-dimensional complement~$\GmatuJ$ as 
\begin{equation}\label{definition GmatJ}
  \GmatJ=
  \begin{pmatrix} 
    \slepfuncoef_1&\cdots&
    \slepfuncoef_J
  \end{pmatrix},\qquad
  \GmatuJ=
  \begin{pmatrix} 
    \slepfuncoef_{J+1}&\cdots&
    \slepfuncoef_{(L+1)^2}
  \end{pmatrix},
\end{equation}
the $J$-dimensional vector of functions containing the $J$
best-concentrated bandlimited Slepian functions~$\Gfunvec_J$, and its
complement~$\Gfunvec_{>J}$ as
\begin{equation}\label{definition GfunvecJ}
  \Gfunvec_J = \GmatJ^\matT\Yfunvec = \begin{pmatrix}
    \Gfun_1&\cdots&\Gfun_J  \end{pmatrix}^T,
\qquad
\Gfunvec_{>J} = \GmatuJ^\matT\Yfunvec,
\end{equation}
and denoting the $J\times J$-dimensional diagonal matrix containing
the $J$ largest concentration ratios by $\Lamat_J$, 
Eqs~(\ref{definition scalar kernel}), (\ref{scalar eigenvalue
  problem}) and~(\ref{definition  GfunvecJ}) together imply that 
\begin{equation}\label{scalar integration Slepian function}
  \Lamat_J= \diag(\lambda_1,\ldots,\lambda_J)= 
\int_\region  \Gfunvec_J^{}\hsom\Gfunvec_J^\funT\dOmega. 
\end{equation}
The orthonormality of the eigenvectors
$\slepfuncoef_1,\ldots,\slepfuncoef_{(L+1)^2}$ in Eqs~(\ref{definition
  Gmat})--(\ref{scalar eigenvalue problem}) guarantees that
$\GmatJ^\matT\GmatJ^{}=\Imat_{J \times J}$. In contrast, the
 matrix $ \GmatJ^{}\GmatJ^\matT$ is a
 $(L+1)^2\times(L+1)^2$-dimensional noninvertible projection,
$(\GmatJ^{}\GmatJ^\matT)^2=\GmatJ^{}\GmatJ^\matT\GmatJ^{}\GmatJ^\matT=\GmatJ^{}\GmatJ^\matT$.
The Slepian functions allow for a constructive approximation of
bandlimited functions of the kind $\signall(\rvec)$, locally within the
target region~$\region$, by restricting the expansion~(\ref{scalar
  slepian transformation}) to the~$J$ best-concentrated Slepian
functions \citep{Simons+2009b,Beggan+2013},
\begin{equation}\label{bloppie}
\signall(\rvec)\approx\sum_{\alpha=1}^J\sSlepcoef_\alpha
\Gfun_\alpha(\rvec)=\Gfunvec_J^\funT\Slepcoef_J=
\Yfunvec^\funT\GmatJ^{}\GmatJ^\matT\sphcoef
,\qquad \rvec\in R
.
\end{equation}
The greater the number of terms~$J$, the less well localized the
approximation, but the smaller the approximation error.


Instead of spatially concentrating spectrally limited functions, we can
also spectrally concentrate spatially limited functions. The
spacelimited Slepian functions can be obtained by restricting the
bandlimited Slepian functions to the space domain of interest:
\begin{equation}\label{scalar spatial truncation}
\hat \Gfun_\alpha(\rvec) = \begin{cases}
\Gfun_\alpha(\rvec)&\text{if } \rvec \in \region,\\
0&\text{if }\rvec \in \Omega\setminus \region.
\end{cases}
\end{equation}
The spherical-harmonic coefficients of the Slepian functions $\hat
\Gfun_\alpha=\hat\Yfunvec^\funT\hat\slepfuncoef_\alpha$, using the
notation of Eq.~(\ref{Yfunvec all degrees}), form the
infinite-dimensional vector
\begin{equation}\label{definition hat slepfuncoef}
\hat\slepfuncoef_\alpha=
\begin{pmatrix}
\hat\sslepfuncoef_{00,\alpha}&\cdots&
\hat\sslepfuncoef_{LL,\alpha}&\cdots \quad
\end{pmatrix}^T
,
\end{equation}
and thus, using the orthonormality of the spherical harmonics and
Eqs~(\ref{scalar spatial truncation}) and~(\ref{definition 
  Gfun}), they are given by
\begin{equation}\label{scalar spacelimited relation}
\hat\slepfuncoef_\alpha=\int_\Omega
\hat\Yfunvec\hat\Gfun_\alpha\dOmega= \int_\region
\hat\Yfunvec\hat\Gfun_\alpha\dOmega= \int_\region
\hat\Yfunvec\Gfun_\alpha\dOmega=\left(\int_\region
\hat\Yfunvec\Yfunvec^\funT\dOmega\right)\slepfuncoef_\alpha=\DmatL\slepfuncoef_\alpha
\end{equation}
where we have defined the $\infty\times(L+1)^2$-dimensional
rectangular counterpart of the localization kernel~(\ref{definition
  scalar kernel}), namely 
\begin{equation}\label{definition DmatL}
\DmatL=\int_\region
\hat\Yfunvec\Yfunvec^\funT\dOmega.
\end{equation}

To prepare for what is yet to come, in Section~\ref{section analytical
 formulation}, we now also introduce another rectangular kernel, 
\begin{equation}\label{definition hatDmat L+1 L}
\DmatLL=\int_\region \hat\Yfunvec_{>L}\Yfunvec^\funT\dOmega,
\end{equation}
an infinite-dimensional vector containing the spherical-harmonic
coefficients of $\hat\slepfuncoef_{\alpha}$ for degrees higher
than~$L$,
\begin{equation}\label{definition single g >L}
  \hat\slepfuncoef_{>L,\alpha}=
  \begin{pmatrix}
    \hat\sslepfuncoef_{L+1\, -L,\alpha} &\hat\sslepfuncoef_{L+1\,  -L+1,\alpha}&\cdots \quad
  \end{pmatrix}^T
  ,
\end{equation}
and the $\infty\times J$-dimensional matrix containing the expansion
coefficients $\hat\slepfuncoef_{>L,\alpha}$, for $\alpha=1\rightarrow J$, as
\begin{equation}\label{definition hatGmat J L}
\GmatLJ=
\begin{pmatrix} 
\hat\slepfuncoef_{>L,1}&\cdots&\hat\slepfuncoef_{>L,J} 
\end{pmatrix}=\DmatLL\GmatJ.
\end{equation}
The vector of coefficients $\hat\slepfuncoef_{>L,\alpha}$ defined in
Eq.~(\ref{definition single g >L}) spectrally truncates the
spacelimited Slepian function $\hat\Gfun_{\alpha}$ to a function
\begin{equation}\label{definition Gfun >L}
\GLa=\sum_{l=L+1}^\infty\sum_{m=-l}^l\hat\sslepfuncoef_{lm,\alpha}\Yfun_{lm}
= \hat\Yfunvec_{>L}^\funT\,\hat\slepfuncoef_{>L,\alpha},
\end{equation}
the $\alpha$th element of the set $\hat\Gfunvec_{>L}$, and
finally, we also define the $J$-dimensional vector of functions with
contributions confined to the degrees higher than~$L$, using
Eqs~(\ref{definition hatGmat J L}), (\ref{definition hatDmat L+1 L})
and~(\ref{definition GfunvecJ}) again, in the equivalent formulations
\begin{equation}\label{definition hatGfun J L}
\GLJ=
\begin{pmatrix}
\hat\Gfun_{>L,1}&\cdots&\hat\Gfun_{>L,J}
\end{pmatrix}^T 
=\GmatLJ^\matT \hat\Yfunvec_{>L}
=\GmatJ^\matT \DmatLL^\matT\hat\Yfunvec_{>L}
=\left(\int_\region \Gfunvec_J^{}\hat\Yfunvec_{>L}^\funT\dOmega\right)\hat\Yfunvec_{>L}.
\end{equation}


\subsection{Gradient-Vector Slepian Functions}\label{section gradient vector
  Slepian functions}

Similarly to the scalar Slepian functions in Section~\ref{section
  scalar Slepian functions} we can construct Slepian functions from
vector spherical harmonics, as described by \citet{Plattner+2013} and
in the chapter ``Scalar and Vector Slepian Functions, Spherical Signal
Estimation and Spectral Analysis'' by Simons and Plattner in this
book. However, in Section~\ref{section potential field from satellite
  data} we showed that the estimation of a scalar potential field from
vectorial data only depends on the gradient-vector spherical
harmonics~$\Efun_{lm}$ defined in Section~\ref{section vector
  spherical harmonics}. In the following we will therefore construct
vector Slepian functions from gradient-vector spherical
harmonics~$\Efun_{lm}$ only. These new so-called \emph{gradient-vector
  Slepian functions} will be useful for problem~$\Pfour$, the
estimation of a scalar potential from vectorial data.

\begin{figure}
\begin{centering}
 \includegraphics[width=0.47\textwidth,
 trim = 13mm 7mm 59.5mm 0mm, clip]{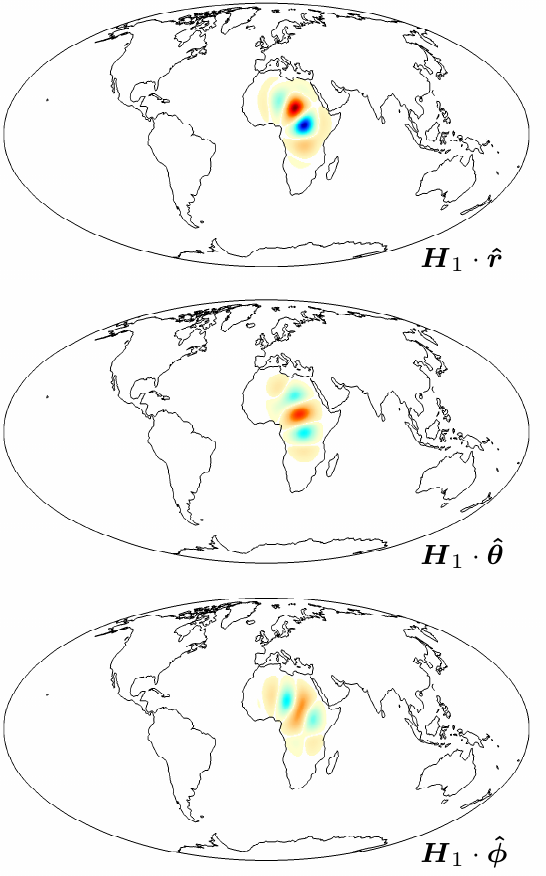}
 \caption{\label{figure GVS} The three vectorial components of the
   gradient-vector Slepian function~$\Hfun_1$ best concentrated to
   Africa at a maximum spherical-harmonic degree $L=30$. Top panel
   shows the radial component $\Hfun_1 \cdot \rvec$, center panel the
   tangential (colatitudinal) component $\Hfun_1\cdot \thvec$, and
   bottom panel the tangential (longitudinal) component~$\Hfun_1 \cdot
   \phvec$. The concentration coefficient is $\sigma=0.999892$.}
\end{centering}
\end{figure}

We construct the gradient-vector Slepian functions
\begin{equation}\label{definition Hfun}
\Hfun(\rvec)=\sum_{l=0}^L\sum_{m=-l}^l \svslepfuncoef_{lm}\Efun_{lm}(\rvec) = \Efunvec^
\funT\vslepfuncoef,
\end{equation}
as the stationary solutions of the maximization problem
\begin{equation}\label{vector Slepian optimization problem}
\sigma=\max_{\Hfun}\frac{\displaystyle \int_\region \Hfun(\rvec)\cdot\Hfun(\rvec)\dOmega }
{\displaystyle \int_
\Omega \Hfun(\rvec)\cdot\Hfun(\rvec)\dOmega}
=\max_{\vslepfuncoef}\frac{\vslepfuncoef^\matT \Kmat\, \vslepfuncoef}
{\vslepfuncoef^\matT
\vslepfuncoef},
\end{equation}
for the expansion coefficients $\svslepfuncoef_{lm}$ in the
$(L+1)^2$-dimensional vector
\begin{equation}\label{definition hvec}
\vslepfuncoef = \begin{pmatrix}
\svslepfuncoef_{00} & \cdots &
\svslepfuncoef_{LL}\end{pmatrix}^T
,
\end{equation}
and $\Efunvec$ was defined in Eq.~(\ref{Efunvec definition}). The
symmetric positive-definite matrix $\Kmat$ is given by its elements
\begin{equation}\label{definition vector Kernel}
\Kelm_{lm,l'm'} = \int_\region
\Efun_{lm}(\rvec)\cdot\Efun_{l'm'}(\rvec) \dOmega,
\qquad
\Kmat=\int_\region \Efunvec\cdot\Efunvec^\funT\dOmega.
\end{equation}
using Eq.~(\ref{definition Efunvec times Efunvec}). The stationary
solutions of Eq.~(\ref{vector Slepian optimization problem}) are the
eigenvectors
$\vslepfuncoef_1,\ldots,\vslepfuncoef_{\alpha},\ldots,\vslepfuncoef_{(L+1)^2}$
in the matrix
 \begin{equation}\label{definition Hmat}
 \Hmat=
\Big(
\vslepfuncoef_1\quad\cdots\quad\vslepfuncoef_\alpha\quad\cdots\quad\vslepfuncoef_{(L+1)^2}\Big),
\qquad
\Hmat\Hmat^\matT=\Hmat^\matT\Hmat=\Imat=\int_\Omega \Efunvec\cdot\Efunvec^\funT\dOmega,
\end{equation}
defined by the eigenvalue problem 
\begin{equation}\label{definition gradient vector Slepian function coefficients}
\Kmat\Hmat=\Hmat\Sigmat
,\qquad
 \Kmat=\Hmat\Sigmat\Hmat^\matT
,
\end{equation}  
with the eigenvalues $\Sigmat=\diag(\sigma_1,\ldots,\sigma_{(L+1)^2})$
the concentration values of Eq.~(\ref{vector Slepian optimization
  problem}), of which most are near unity or near zero. We index and order the
$\svslepfuncoef_{lm,\alpha}\in\Hmat$ according to their eigenvalues in
decreasing order such that $1>\sigma_1\geq\ldots\geq
\sigma_{(L+1)^2}>0$ to obtain a concentration-ordered basis of
gradient-vector functions bandlimited to~$L$ given by
\begin{equation}\label{vslepor}
  \Hfun_\alpha(\rvec)=
\sum_{l=0}^L\sum_{m=-l}^l\svslepfuncoef_{lm,\alpha}\Efunvec_{lm}(\rvec)=\Efunvec^\funT\vslepfuncoef_\alpha.
\end{equation}
See Fig.~\ref{figure GVS} for a three-component space-domain example. 
We normalize the eigenvectors $\vslepfuncoef_\alpha$ of
Eq.~(\ref{definition gradient vector Slepian function coefficients})
so that the new basis $\Hfun_1,\ldots,\Hfun_{(L+1)^2}$ is orthonormal
over the entire sphere $\Omega$ and orthogonal over the region
$\region$,
\begin{equation}\label{foxot}
  \int_\Omega \Hfun_\alpha\cdot\Hfun_\beta\dOmega=\delta_{\alpha\beta}, \qquad 
  \int_\region \Hfun_\alpha\cdot\Hfun_\beta\dOmega=\sigma_\alpha\delta_{\alpha\beta}.
\end{equation} 
In the notation of Eq.~(\ref{Efunvec definition}) and beyond, the
vector containing all gradient-vector Slepian functions for
bandlimit~$L$ and region~$\region$ is given by
\begin{equation}\label{definition Hfunvec}
\Hfunvec=
\Big(\Hfun_1\quad\cdots\quad\Hfun_{(L+1)^2}\Big)^T=\Hmat^\matT\Efunvec.
\end{equation}
The transformation of a bandlimited gradient-vector function into its
equivalent gradient-vector Slepian-function expansion happens via the
gradient-vector Slepian transformation matrix $\Hmat$  as
$\vSlepcoef=\Hmat^\matT\vsphcoef$ and
\begin{equation}\label{vector Slepian expansion}
\vecsignall(\rvec)=\sum_{l=0}^L\sum_{m=-l}^l \svsphcoef_{lm}\Efun_{lm}(\rvec) =
\Efunvec^\funT\vsphcoef
=\Efunvec^\funT\Hmat\Hmat^\matT\vsphcoef
=\Hfunvec^\funT\vSlepcoef
=\sum_{\alpha=1}^{(L+1)^2}\vSlepcoef_\alpha\Hfun_{\alpha}(\rvec).
\end{equation}
We introduce the $(L+1)^2\times J$ dimensional matrix containing the
$(L+1)^2$ gradient-vector spherical-harmonic coefficients for each of
the $J$ best-concentrated gradient-vector Slepian functions
\begin{equation}\label{definition HmatJ}
\HmatJ=\begin{pmatrix} \vslepfuncoef_1 & \cdots&\vslepfuncoef_J \end{pmatrix},
\end{equation}
the $J$-dimensional vector of vector-valued functions containing the
$J$ best-concentrated gradient-vector Slepian functions  
\begin{equation}\label{definition HfunvecJ}
\Hfunvec_J=\HmatJ^\matT\Efunvec=\begin{pmatrix} \Hfun_1&\cdots&\Hfun_J \end{pmatrix}^T,
\end{equation}
and the $J\times J$ dimensional diagonal matrix containing the $J$
largest concentration ratios 
\begin{equation}\label{definition SigmatJ}
\Sigmat_J=\diag(\sigma_1,\ldots,\sigma_J)
= \int_\region \Hfunvec_J^{}\cdot\Hfunvec_J^\funT\dOmega.
\end{equation}
where the  last equality is a consequence of Eqs~(\ref{definition vector
  Kernel}), (\ref{definition gradient vector Slepian function
  coefficients}) and (\ref{definition   HfunvecJ}). 

The orthonormality of the $\vslepfuncoef_1,\ldots,\vslepfuncoef_{(L+1)^2}$ in
Eqs~(\ref{definition Hmat})--(\ref{definition gradient vector Slepian function coefficients})
ensures that $\HmatJ^\matT\HmatJ^{}=\Imat_{J\times  J}$, but the
$(L+1)^2\times (L+1)^2$-dimensional projection matrix
$\HmatJ^{}\HmatJ^\matT$ is not invertible. A local approximation of the
gradient function can be obtained from 
\begin{equation}
\vecsignall(\rvec)\approx\sum_{\alpha=1}^J
\svSlepcoef_\alpha\Hfun_\alpha(\rvec)=\Hfunvec_J^\funT\vSlepcoef_J
=\Efunvec^\funT\HmatJ\HmatJ^\matT\vsphcoef 
,\qquad \rvec\in R.
\end{equation} 

For use in Section~\ref{section vector analytical formulation} we
finally define the $\infty\times(L+1)^2$-dimensional matrix
\begin{equation}\label{definition hat Kmat L+1 >L}
\KmatLL=\int_\region \hat\Efunvec_{>L} \cdot \Efunvec^\funT\dOmega
,
\end{equation}
using the notation in Eqs~(\ref{Efunvec definition})--(\ref{Efunvec
  all degrees}), and from this, we derive an expression for the $J$
components of the spacelimited gradient-vector Slepian functions in
the gradient-vector spherical-harmonic basis at degrees larger than~$L$,  
\begin{equation}\label{definition hatHfun J L}
\HLJ=
\HmatLJ^\matT\hat\Efunvec_{>L}
=\HmatJ^\matT \KmatLL^\matT\hat\Efunvec_{>L}
=\left(\int_\region \Hfunvec_J^{}\cdot
\hat\Efunvec_{>L}^\funT\dOmega\right)\hat\Efunvec_{>L}.
\end{equation}
The analogy with the scalar Eq.~(\ref{definition hatGfun J L}) is only
partial since the spacelimited versions of~$\Hfunvec$ also have
non-vanishing components in the span of $\Ffun$ of Eq.~(\ref{F
  definition}) and $\Cfun$ of Eq.~(\ref{Clm}) --- not just $\Efun$.

\section{Potential-Field Estimation from Radial Data Using Slepian
  Functions }\label{section radial data inversion} 

With the scalar Slepian functions defined in Section~\ref{section
  scalar Slepian functions} we can now formulate the solution to
problem~$\Ptwo$ as a localized bandlimited potential-field estimation
problem, from noisy radial-derivative data at satellite altitude. More
precisely we will use the Slepian functions to localize the
radial-field analysis at satellite altitude and then, in a second
step, downward transform the resulting spherical-harmonic coefficients
using the notions developed in Section~\ref{section potential field
  from satellite data}.

As in the exposition of the classical spherical-harmonics based
solutions described in Sections~\ref{section point wise}
and~\ref{preliminaries functional formulation}, we start with a
description of the numerical estimation procedure based on pointwise
data in Section~\ref{section scalar numerical} before proceeding to a
functional formulation that will facilitate the statistical analysis
of the performance of the methods, in Section~\ref{section analytical
  formulation}. Throughout this section we do not assume that the
target signal~$\signal(\rvec)$ is bandlimited, but a bandwidth~$L$
does need to be chosen to form the
approximation~$\estsignal(\rvec)$. The bias that arises from this
choice of bandlimitation will be discussed in Section~\ref{section
  analytical formulation}.

\subsection{Discrete Formulation and Truncated Solutions}\label{section scalar numerical} 

From pointwise data values of the radial derivative of the potential
at satellite altitude, given at the points
$\satalt\rvec_1,\ldots,\satalt\rvec_\npoints$, and polluted by noise, 
\begin{equation}\label{scalar noisy data}
\dpoints_r=\gradsigpoints_r + \noisepoints_r,
\end{equation} 
we seek to estimate the bandlimited partial set of corresponding
spherical-harmonic coefficients $\sphcoef^\Earthrad= (
\ssphcoef_{00}^\Earthrad\,\cdots\,\ssphcoef_{LL}^\Earthrad )^T$ of the
scalar potential~$\signal$ on Earth's surface $\Omega_\Earthrad$, as in the
original statement~(\ref{Equation problem 2}) of Problem~$\Ptwo$.  In
Eq.~(\ref{scalar noisy data}), $\gradsigpoints_r$ is defined as in
Eq.~(\ref{radial data}), and $\noisepoints_r$ is a vector of noise
values at the evaluation points.

As seen in Eq.~(\ref{solution 1 to problem 2}), the solution to
problem~$\Ptwo$ involves the inversion of a ``normal'' matrix,
$(\Ypoints\Ypoints^\pointT)^{-1}$, that is reminiscent of the
localization kernel in Eq.~(\ref{definition scalar kernel}), and
therefore has many near-zero eigenvalues, and the additional
accounting for the effects of altitude via the term $\Amat^{-1}$,
which will potentially unstably inflate the smallest-scale noise terms
\cite[][]{Maus+2006a}. Instead of regularization by damping (in the
spherical-harmonic basis), the approach we propose is based on
truncation (in the Slepian basis). We focus on the estimation of the
radial field at satellite altitude in a chosen target
region~$\region$, by estimating only its $J$ best-concentrated Slepian
coefficients. The hard truncation level~$J$ is a regularization
parameter whose value needs to be chosen based on signal-to-noise
considerations and an optimality criterion, much as a proper damping
parameter would
\citep{Mallat2008,Kaula67a,Simons+2006b,Wieczorek+2007}.

Define the $(L+1)^2\times\npoints$-dimensional matrix containing the
Slepian functions $\Gfun_1,\ldots,\Gfun_{(L+1)^2}$ evaluated at the
latitudinal and longitudinal locations of the data (on the unit
sphere),
\begin{equation}\label{definition Gpoints}
\Gpoints=\Gmat^\matT\Ypoints,
\end{equation}
where the scalar Slepian transformation matrix~$\Gmat$ is defined in
Eq.~(\ref{definition Gmat}). Note the change in (serif vs sans)
type. The matrix~$\Ypoints$ contains the spherical harmonics evaluated
at the data locations on the unit sphere, as in Eq.~(\ref{Y point
  matrix}). Problem~$\Ptwo$ is restated from its original formulation
in Eq.~(\ref{Equation problem 2}) via a bandlimited Slepian
transformation at altitude, to
\begin{equation}\label{point data}
\arg\min_{\estsphcoef^\Earthrad}\lVert
\Ypoints^\pointT\Amat\estsphcoef^\Earthrad - \dpoints_r \rVert^2=
\arg\min_{\estsphcoef^\Earthrad}\lVert
 \Ypoints^\pointT \Gmat \Gmat^\matT\Amat \estsphcoef^\Earthrad - \dpoints_r \rVert^2= 
\arg\min_{\srs}\lVert
\Gpoints^\pointT 
\srs - \dpoints_r \rVert^2,
\end{equation}
where we used the orthogonality $\Gmat\Gmat^\matT=\Imat$, the
definition Eq.~(\ref{definition Gpoints}), and identified the Slepian
expansion coefficients at satellite altitude through transformation of
the bandlimited vector~(\ref{ure}) into the $(L+1)^2$-dimensional vector
\begin{equation}\label{scalar full Slepian problem}
\srs=\Gmat^\matT\Amat\estsphcoef^\Earthrad.
\end{equation}
We invoke our regularization of only solving for the coefficients of
the $J$ best-concentrated Slepian functions at satellite altitude by
defining the $J\times \npoints$ dimensional matrix containing the
point evaluations of the $J$ best-concentrated Slepian functions on
the unit sphere
\begin{equation}\label{definition GpointsJ} 
\Gpoints_J=\GmatJ^\matT\Ypoints
,
\end{equation}
and by solving, instead of Eq.~(\ref{point data}),
\begin{equation}\label{scalar numerical Slepian problem}
\arg\min_{\srsj}\lVert
\Gpoints_J^\pointT \srsj -
\dpoints_r\rVert^2
,
\end{equation}
for the $J$-dimensional vector $\srsj$
containing the coefficients of the approximation at satellite
altitude in the bandlimited Slepian basis. When $J\leq
\npoints$ we have the solution
\begin{equation}\label{kukuk}
\srsj=(\Gpoints_J^{} \Gpoints_J^\pointT)^{-1} \Gpoints_J \dpoints_r
,
\end{equation}
which we then downward transform to the $(L+1)^2$ spherical-harmonic
coefficients $\estsphcoef^\Earthrad$ of the field on Earth's surface
$\Omega_\Earthrad$ as
\begin{equation}\label{solution 2 to problem 2}
\estsphcoef^\Earthrad =  \Amat^{-1}\GmatJ^{} \srsj= 
\Amat^{-1}\GmatJ 
(\Gpoints_J^{} \Gpoints_J^\pointT)^{-1} \Gpoints_J \dpoints_r
\qquad \text{(solution 2 to noisy problem $\Ptwo$)}. 
\end{equation}
The numerical conditioning of the matrix
$(\Gpoints_J^{}\Gpoints_J^\pointT)$ is determined by the truncation
parameter~$J$, and we require the inverse of the matrix
$\Amat$ defined in Eq.~(\ref{definition Amat}).

\begin{figure}
\begin{centering}
 \includegraphics[width=0.96\textwidth,
  trim = 10mm 50mm 6mm 18.5mm, clip]{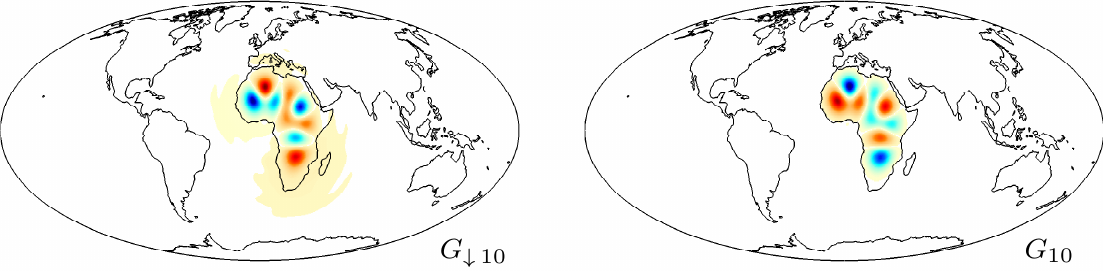}
 \caption{\label{figure Slepian down} Downward transformation of the
   10th best-concentrated scalar Slepian function for Africa and a
   maximum spherical-harmonic degree $L=30$. The right panel shows the
   concentrated scalar Slepian function
   $\Gfun_{10}=\Yfunvec^\funT\slepfuncoef_{10}$ for the radial component at
   an altitude of 500~km. The left panel shows the equivalent
   downward-transformed function $\Gfun_{\downarrow\,10}=\Yfunvec^ \funT
   \Amat^{-1}\slepfuncoef_{10}$ to describe a scalar potential on
   Earth's surface ($ \Earthrad=6371$~km). The concentration
   coefficient for the Slepian function~$\Gfun_{10}$ at altitude is
   $\lambda=0.99985$.  }
\end{centering}
\end{figure}

The resulting approximation $\estsignal(\Earthrad\rvec)$ of the
potential field $\signal(\Earthrad\rvec)$ at any point of interest
on~$\Omega_\Earthrad$ can be calculated as  
\begin{equation}\label{scalar field estimation}
\estsignal(\Earthrad \rvec)=\Yfunvec^\funT\estsphcoef^\Earthrad 
= \Gfunvec_{\downarrow J}^\funT(\Gpoints_J^{}\Gpoints_J^\pointT)^{-1} \Gpoints_J \dpoints_r
=  \Gfunvec_{\downarrow J}^\funT \srsj,
\end{equation}
where we have defined the vector of the $J$ best-concentrated (and its complement)
downward-transformed scalar Slepian functions as 
\begin{equation}\label{downward gay jay}
\Gfunvec_{\downarrow J}=\GmatJ^\matT\Amat^{-1}\Yfunvec,
\qquad
\Gfunvec_{\downarrow >J}=\GmatuJ^\matT\Amat^{-1}\Yfunvec,
\end{equation}
an example of which is plotted in Fig.~\ref{figure Slepian down}. We
reserve for later use the vectors of upward-transformed
Slepian functions,
\begin{equation}\label{definition Gfunvec upJ}
\Gfunvec_{\uparrow J}=\GmatJ^\matT\Amat\Yfunvec,\qquad
\Gfunvec_{\uparrow >J}=\GmatuJ^\matT\Amat\Yfunvec
.
\end{equation}
From Eqs~(\ref{downward gay jay})--(\ref{definition Gfunvec upJ})
and~(\ref{definition Gmat}) or~(\ref{definition Gfunvec})
we also obtain the equivalencies
\begin{equation}\label{up and down scalar slepians}
\Gfunvec_{\downarrow}^\funT\hsomm(\rvec)\hsom\Gfunvec_{\uparrow}^{}(\rvec')=
\Yfunvec^\funT\hsomm(\rvec)\hsom\Amat^{-1}\Gmat^{}\Gmat^\matT\hsom\Amat\Yfunvec(\rvec')=
\Yfunvec^\funT\hsomm(\rvec)\Yfunvec(\rvec')=
\Gfunvec^\funT\hsomm(\rvec)\hsom\Gfunvec(\rvec'),
\end{equation}
in the ``silent'' $J=(L+1)^2$ notation of Eq.~(\ref{definition
  Gfunvec}), but noting that Eq.~(\ref{up and down scalar slepians})
does have an equivalent in truncated form when $J\ne(L+1)^2$. Evidently, we also have
\begin{equation}\label{fixit1}
\Gfunvec_{\downarrow}^\funT\Gfunvec_{\uparrow}^{}=
\Gfunvec_{\downarrow J}^\funT\Gfunvec_{\uparrow J}^{}+
\Gfunvec_{\downarrow >J}^\funT\Gfunvec_{\uparrow >J^{}}.
\end{equation}

\subsection{Continuous Formulation and Statistical Considerations}
\label{section analytical formulation}

In this section we provide a formulation of the approach described in
Section~\ref{section scalar numerical} that considers the data in
their functional form instead of being given as point values.  In this
formalism we will then express the estimation variance, bias, and mean
squared error for the methods presented under some special cases. Our
results will generalize the scalar treatment of \citet{Simons+2006b}
in whose work we will point out a misprint that we correct here.

\subsubsection{Continuous Formulation}

The analytical counterpart to the pointwise data from Eq.~(\ref{scalar
  noisy data}) known (or desired) only within the target region
$\region$ is  
\begin{equation}\label{data only in region}
\datarad(\rvec)=\begin{cases}
\scalsignal(\satalt\rvec) + \noise(\rvec) &\text{if } \rvec \in \region\\
\text{unknown} &\text{if } \rvec \in \Omega \setminus \region,
\end{cases}
\end{equation}
where $\noise(\rvec)$ is the spatial noise function. The estimation
problem equivalent to Eq.~(\ref{point data}) can now be formulated as  
\begin{equation}\label{scalar Slepian signal up decomposition}
\arg\min_{\estsphcoef^\Earthrad}\intrnorm{
\Yfunvec^\funT\Amat\estsphcoef^\Earthrad-\datarad}=
\arg\min_{\estsphcoef^\Earthrad}\intrnorm{
\Yfunvec^\funT\Gmat\Gmat^\matT\Amat\estsphcoef^\Earthrad-\datarad}=
\arg\min_{\srs}\intrnorm{
\Gfunvec^\funT\srs-\datarad}, 
\end{equation}
where the vector of Slepian functions $\Gfunvec$ is defined in
Eq.~(\ref{definition Gfunvec}) and the estimated coefficients at satellite
altitude $\srs$ are in Eq.~(\ref{scalar full
  Slepian problem}). The problem is regularized by solving exclusively for
the $J$ best-concentrated Slepian coefficients that describe the data
in Eq.~(\ref{data only in region}), which transforms Eq.~(\ref{scalar
  Slepian signal up decomposition}) into the estimation problem
\begin{equation}\label{scalar functional estimation problem}
\arg\min_{\srsj}
\intrnorm{\Gfunvec_J^\funT\srsj-\datarad}
.
\end{equation}
Differentiating with respect to~$\srsj$ to
find the stationary points, and making use of Eq.~(\ref{scalar
  integration Slepian function}), the solution is given by
\begin{equation}
\srsj=
\left(\int_\region \Gfunvec_J^{}
\Gfunvec_J^\funT\dOmega\right)^{-1}\int_\region
\Gfunvec_J\datarad\dOmega
=\Lamat_J^{-1}\int_\region \Gfunvec_J\datarad\dOmega. 
\end{equation}

As with the estimation of the spherical-harmonic coefficients of the
potential field from the Slepian coefficients at altitude obtained
from pointwise data in Eq.~(\ref{solution 2 to problem 2}) we can
estimate the vector containing the $(L+1)^2$ spherical-harmonic
coefficients~$\estsphcoef^\Earthrad$ from the $J$-dimensional vector
of Slepian coefficients~$\srsj$ by first transforming it to the
$(L+1)^2$-dimensional vector of spherical-harmonic coefficients
$\GmatJ^{}\srsj$ and then downward transforming it using the matrix
$\Amat$ defined in Eq.~(\ref{definition Amat}). We thereby obtain the
spherical-harmonic coefficients $\estsphcoef^\Earthrad$ for the
estimation $\estsignal(\Earthrad\rvec)$ of the potential field on
Earth's surface $\Omega_\Earthrad$ as
\begin{equation} \label{analytical s2 to p2}
\estsphcoef^\Earthrad = \Amat^{-1} \GmatJ^{} \Lamat^{-1}_J \int_\region
\Gfunvec_J \datarad\dOmega \qquad (\text{analytical solution 2 to
  problem $\Ptwo$}). 
\end{equation} 
We can expand the coefficients $\estsphcoef^\Earthrad$ obtained
  from the data $\datarad$  by Eq.~(\ref{analytical s2 to p2}) to evaluate the
potential field anywhere on  Earth's surface as
\begin{equation}\label{scalar analytical potential estimation}
\estsignal(\Earthrad\rvec)=
\Yfunvec^\funT\estsphcoef^\Earthrad
=\Yfunvec^\funT\Amat^{-1}\GmatJ^{} \Lamat^{-1}_J \int_\region \Gfunvec_J \datarad\dOmega 
= \Gfunvec_{\downarrow J}^\funT \Lamat^{-1}_J  \int_\region \Gfunvec_J
\datarad\dOmega
,
\end{equation}
where the truncated vector of downward transformed Slepian functions
$\Gfunvec_{\downarrow J}$ is defined in Eq.~(\ref{downward gay jay}). 

\subsubsection{Effects of Bandlimiting the Scalar Estimates}\label{bl1}

The estimate given in Eq.~(\ref{scalar analytical potential
  estimation}) has a bandlimited representation of the unknown
potential at its heart, though the actual potential that we are
attempting to estimate will generally not be bandlimited, see
Eqs~(\ref{continued potential}) and~(\ref{bloL}), nor will the noise
be. To isolate the effects of the bandlimitation, we write the data as
the sum of a bandlimited part (which is expanded globally in Slepian
functions of the same bandwidth), its wideband complement, which
contains spherical harmonics with degree greater than~$L$ introduced
in Eq.~(\ref{Yfunvec all degrees}), and the noise
contribution. Eq.~(\ref{data only in region}) then becomes
\begin{equation}\label{scalar bias}
\datarad = \scalsignal(\satalt\rvec) + \noise =
\Gfunvec^\funT\int_\Omega \Gfunvec\hsom\scalsignal(\satalt\rvec) \dOmega +
\hat\Yfunvec_{>L}^ 
\funT\int_\Omega 
\hat\Yfunvec_{>L}\scalsignal(\satalt\rvec) \dOmega + \noise.
\end{equation}
To this we apply the integral transform of Eq.~(\ref{scalar analytical
  potential estimation}) using the $J$ best-concentrated Slepian functions
$\Gfunvec_J$, and we make use of the orthogonality Eq.~(\ref{slepor}),
Eqs~(\ref{definition GfunvecJ})--(\ref{scalar integration Slepian
  function}) and Eq.~(\ref{definition hatGfun J L}), to obtain the
expression
\begin{align}\label{yestonumber}
\int_\region \Gfunvec_J \datarad\dOmega&=
\int_\region  \Gfunvec_J^{}\Gfunvec^\funT \dOmega \int_\Omega 
\Gfunvec\hsom\scalsignal(\satalt\rvec)\dOmega 
+\int_\region  \Gfunvec_J^{}  \hat\Yfunvec_{>L}^\funT \dOmega 
\int_\Omega \hat\Yfunvec_{>L}\scalsignal(\satalt\rvec)\dOmega
+\int_\region \Gfunvec_J\hsom\noise\dOmega \\
&= \Lamat_J \int_\Omega \Gfunvec_J \scalsignal(\satalt\rvec) \dOmega + 
\GmatJ^\matT \DmatLL^\matT
\int_\Omega  \hat\Yfunvec_{>L} \scalsignal(\satalt\rvec) \dOmega +  \int_\region \Gfunvec_J\noise\dOmega
\\
&= \Lamat_J \int_\Omega \Gfunvec_J \scalsignal(\satalt\rvec) \dOmega + 
\GmatLJ^\matT
\int_\Omega  \hat\Yfunvec_{>L} \scalsignal(\satalt\rvec) \dOmega +  \int_\region \Gfunvec_J\noise\dOmega
\\\label{scalar concentration at alt expanded data}
&= \Lamat_J \int_\Omega \Gfunvec_J \scalsignal(\satalt\rvec) \dOmega + \int_\Omega 
 \GLJ \scalsignal(\satalt\rvec) \dOmega +  \int_\region \Gfunvec_J\noise\dOmega.
\end{align}
Finally we can insert the result~(\ref{scalar concentration at alt expanded
  data}) into Eq.~(\ref{scalar analytical potential estimation}) to
discover the contributions to the bandlimited estimate
$\estsignal(\Earthrad\rvec)$ from signal with energy in
the spherical-harmonic degree range~$l>L$ and the presence of noise:
\begin{equation}\label{scalar estimation full}
\estsignal(\Earthrad\rvec) = \Gfunvec_{\downarrow J}^\funT \int_\Omega
\Gfunvec_J \scalsignal(\satalt\rvec) \dOmega  
+  \Gfunvec_{\downarrow J}^\funT \Lamat^{-1}_J \left( \int_\Omega 
 \GLJ \scalsignal(\satalt\rvec) \dOmega
 + \int_\region \Gfunvec_J\hsom\noise\dOmega\right),
\end{equation}
an expression equivalent to eq.~(136) of
\citet{Simons+2006b}. Ultimately, Eq.~(\ref{scalar estimation full})
is derived from an estimate of the spherical-harmonic potential
coefficients, Eq.~(\ref{analytical s2 to p2}), that uses a truncated
(to~$J$) set of bandlimited (to~$L$) spatially concentrated (to~$R$)
Slepian functions. Keeping with the terminology introduced by
\citet{Simons+2006b}, the \textit{truncation bias} in the bandlimited
part of the estimate (the first right-hand-side term in
Eq.~\ref{scalar estimation full}) diminishes as~$J$ increases, but the
second, parenthetical, term grows, very unfavorably fast, with the
inverse-eigenvalue matrix~$\Lamat_J^{-1}$. This term contains the
\textit{broadband leakage}, which is captured from the non-bandlimited
part of the signal by the nonvanishing regional product integral in
the second term of Eq.~(\ref{yestonumber}), and the contribution due
to the noise in the region over which data are available.  Comparison
of the bandlimited estimate~(\ref{scalar estimation full}) with the
wideband original form~(\ref{bloL}) will furthermore identify a
\textit{broadband bias} that arises from the outright neglect of the
necessary basis functions, and is thus, essentially, unavoidable. The
broadband leakage can be controlled under some theoretical or
numerical schemes
\cite[e.g.][]{Hwang93,Trampert+96b,Albertella+2008}. Oftentimes,
however, those fail to be practically successful at the desired level
of accuracy of the solution~\cite[e.g.][]{Slobbe+2012}.


\subsubsection{Statistical Analysis for Scalar Bandlimited-White Processes}

The complete assessment of the statistical performance of the
estimators~(\ref{analytical s2 to p2})--(\ref{scalar analytical
  potential estimation}) is an ambitious objective. It is difficult to
go beyond Eq.~(\ref{scalar estimation full}) without making detailed
assumptions about the underlying statistics of both signal and noise,
not to mention the specifics of the region of data coverage and the
satellite altitude
\cite[e.g.][]{Kaula67a,Whaler+81,Xu92a,Xu92b,Xu98,Schachtschneider+2010,Schachtschneider+2012,Slobbe+2012}. However,
as shown by~\citet{Simons+2006b}, special cases are easy to come by
and learn from. We recall the standard definitions for the estimation
error, bias and variance,
\begin{align}\label{error}
\esterr&=\estsignal(\Earthrad\rvec) - \signal(\Earthrad\rvec),\\
\label{bias}
\bias &=  \big\langle \estsignal(\Earthrad\rvec) \big\rangle -
\signal(\Earthrad\rvec),\\
\variance &= 
\label{variance} \big\langle
\estsignal^2(\Earthrad\rvec)\big\rangle - \big\langle\estsignal(\Earthrad\rvec)
\big\rangle^2,
\end{align}
and, typically the quantity to be minimized, the mean squared error:
\begin{equation}\label{mse}
\langle \esterr^2 \rangle= \variance + \langle\bias^2\rangle.
\end{equation}
The angular brackets in Eq.~(\ref{mse}) refer to
averaging over a hypothetical ensemble of repeated observations,
treating \textit{both} signal and noise as stochastic processes
\cite[see][]{Simons+2006b}. 
We make the following four oversimplified assumptions by which to
obtain simple and insightful expressions for
$\variance,\bias$, and $\langle\esterr^2\rangle$:
\begin{enumerate}
\item\label{assume bandlimit} The signal $\signal(\Earthrad\rvec)$ is
  bandlimited, as are the Slepian functions $\Gfunvec$, with the same
  bandwidth $L$.
\item The signal is --- almost, given the incompatible
  stipulation~\ref{assume bandlimit} --- ``white'' on Earth's surface,
  with power $\sigpower$, in the sense
  $\langle\signal(\Earthrad\rvec)\signal(\Earthrad\rvec')\rangle=\sigpower\hsom\delta(\rvec,\rvec')$,
  and with $\delta(\rvec,\rvec')$ the scalar spherical delta function \cite[see][]{Simons+2006a}.
\label{assume white signal}
\item The noise is white at the observation level, with power
  $\noisepower$, as $\langle
  \noise(\rvec)\noise(\rvec')\rangle=\noisepower\delta(\rvec,
  \rvec')$.
\label{assume white noise}
\item The noise has zero mean and is uncorrelated with the signal,
  $\langle \noise(\rvec)\rangle=0=\langle
  \noise(\rvec)\signal(\rvec')\rangle$. \label{assume zero mean}
\end{enumerate}
To honor \ref{assume bandlimit} we insert the bandwidth-restricted
version of Eq.~(\ref{popo}) into Eq.~(\ref{scalar estimation full}),
observe the cancellation, via the whole-sphere orthogonality of
$\hat\Gfunvec_{>L}$ and $\Yfunvec$, of the first term inside of the
parentheses in Eq.~(\ref{scalar estimation full}), and then apply the
relation~(\ref{definition GfunvecJ}) and the
orthogonality~(\ref{definition Gmat}), to arrive at
\begin{align}\nonumber
\estsignal(\Earthrad\rvec) &= 
\Gfunvec_{\downarrow J}^\funT\left( \int_\Omega \Gfunvec_J\Yfunvec^\funT\hsomm\Amat \sphcoef^
\Earthrad \dOmega + \Lamat^{-1}_J \int_\region
\Gfunvec_J\hsom\noise\dOmega\right) = \Gfunvec_{\downarrow
  J}^\funT\left(\GmatJ^\matT\Amat\sphcoef^\Earthrad +
\Lamat^{-1}_J\int_\region \Gfunvec_J\hsom\noise\dOmega\right) \\ 
&= \Gfunvec_{\downarrow J}^\funT\left( \int_\Omega \Gfunvec_{\uparrow
  J} \signal(\Earthrad\rvec) \dOmega + \Lamat^{-1}_J \int_\region
\Gfunvec_J\hsom\noise\dOmega \right).\label{scalar special estsignal} 
\end{align}
The last equality follows from the bandlimited identification
$\signal(\Earthrad\rvec)=\Yfunvec^\funT\sphcoef^\Earthrad$ as from
Eq.~(\ref{non-bandlimited scalar}), global orthogonality of
the~$\Yfunvec$, and by substitution of Eq.~(\ref{definition Gfunvec
  upJ}). From Eqs~(\ref{up and down scalar slepians})--(\ref{fixit1})
we furthermore know that the unknown bandlimited
signal~$\signal(\Earthrad\rvec)$ can be represented using the up- and
downward transformed Slepian functions as
\begin{equation}\label{scalar up and down representation}
\signal(\Earthrad\rvec) = \Gfunvec_{\downarrow}^\funT\int_\Omega
\Gfunvec_{\uparrow} \signal(\Earthrad\rvec) \dOmega
=\Gfunvec_{\downarrow J}^\funT\int_\Omega
\Gfunvec_{\uparrow J} \signal(\Earthrad\rvec) \dOmega
+\Gfunvec_{\downarrow >J}^\funT\int_\Omega
\Gfunvec_{\uparrow >J} \signal(\Earthrad\rvec) \dOmega. 
\end{equation}
We can now calculate the bias $\bias$ from Eq.~(\ref{bias}) by
applying the averaging operation to Eq.~(\ref{scalar special
  estsignal}), using assumption~\ref{assume zero  
mean}, and then subtracting Eq.~(\ref{scalar up and down
  representation}), to give the result, which grows with diminishing
truncation~$J$, 
\begin{equation}\label{corr1}
\bias = - \Gfunvec_{\downarrow >J}^\funT \int_\Omega \Gfunvec_{\uparrow >J}\signal(\Earthrad\rvec)
\dOmega.
\end{equation} 
In order to calculate the variance $\variance$ we use Eq.~(\ref{scalar
  special estsignal}) to obtain the squared
\begin{align} 
\estsignal^2(\Earthrad\rvec)&= \Gfunvec_{\downarrow J}^\funT
\left(\int_\Omega \Gfunvec_{\uparrow J}\signal(\Earthrad\rvec)\dOmega + \Lamat^{-1}_J\int_\region \Gfunvec_J 
\hsom\noise\dOmega \right)
\left(\int_\Omega \signal(\Earthrad\rvec)\hsom\Gfunvec_{\uparrow
  J}^\funT\dOmega + \Lamat^{-1}_J\int_\region \noise
\hsom\Gfunvec_J^\funT
\dOmega\right)
\Gfunvec_{\downarrow J}\\\nonumber
& = \Gfunvec_{\downarrow J}^\funT\left( \int_\Omega
\int_\Omega\Gfunvec_{\uparrow  J}(\rvec)
\signal(\Earthrad\rvec)\signal(\Earthrad\rvec')\hsom\Gfunvec_{\uparrow  J}^{\funT}(\rvec')\dOmega'\dOmega \right.
\label{scalar altitude variance calculation}
 +\Lamat^{-1}_J\int_\region \int_\region \Gfunvec_{J}(\rvec)\hsom\noise(\rvec)\noise(\rvec')\hsom
 \Gfunvec_{J}^{\funT}(\rvec')\dOmega'\dOmega \hsom\Lamat^{-1}_J\\
&\left.\qquad+ \int_\Omega \int_\region \Gfunvec_{\uparrow J}(\rvec)
\signal(\Earthrad\rvec)\noise(\rvec')\hsom\Gfunvec_{J}^{\funT}(\rvec')\dOmega'\dOmega \hsom\Lamat^{-1}_J
+  \Lamat^{-1}_J\int_\region \int_\Omega
\Gfunvec_{ J}(\rvec)\hsom\noise(\rvec)\signal(\Earthrad\rvec')
\hsom\Gfunvec_{\uparrow J}^{\funT}(\rvec')\dOmega'\dOmega  \right)
\Gfunvec_{\downarrow J}.
\end{align}
We apply the averaging over the different realizations of the noise in
Eq.~(\ref{scalar altitude variance calculation}), and use
assumptions~\ref{assume white noise}--\ref{assume zero mean} and
Eq.~(\ref{scalar integration Slepian function}), from which we
subtract the square of the average of Eq.~(\ref{scalar special
  estsignal}) to obtain the variance in Eq.~(\ref{variance}), which
grows with~$J$, as
\begin{equation}\label{scalar special variance}
\variance=\noisepower\Gfunvec_{\downarrow J}^\funT \Lamat_J^{-1} \Gfunvec_{\downarrow J}^{}.
\end{equation}
The squared bias averaged over all realizations of the signal, using
assumption~\ref{assume white signal}, making the
substitution~(\ref{definition Gfunvec upJ}), and using the
whole-sphere orthogonality~(\ref{definition Gmat}) of the spherical
harmonics $\Yfunvec$, yields
\begin{equation}\label{corr2}
\langle\bias^2\rangle=
\sigpower\hsom 
\Gfunvec_{\downarrow  >J}^\funT
\big(\GmatuJ^\matT\Amat^2\GmatuJ^{}\big)\hsom\Gfunvec_{\downarrow >J}
,
\end{equation}
which leads, together with the variance in Eq.~(\ref{scalar special
  variance}), via Eq.~(\ref{mse}) to the mean squared estimation error
\begin{equation}\label{scalar mse equation}
\langle \esterr^2 \rangle=\noisepower\Gfunvec_{\downarrow J}^\funT  \Lamat_J^{-1}
\Gfunvec_{\downarrow J}^{} + \sigpower \hsom
\Gfunvec_{\downarrow >J}^\funT
\big(\GmatuJ^\matT\Amat^2\GmatuJ^{}\big)\hsom\Gfunvec_{\downarrow >J}^{}.
\end{equation}
With Eqs~(\ref{corr1}), (\ref{corr2}) and~(\ref{scalar mse equation})
we correct Eqs~(143)--(145) of \citet{Simons+2006b}. We can understand
their typo by writing Eq.~(\ref{corr2}) using Eq.~(\ref{downward gay
  jay}) as $\langle\bias^2\rangle=\sigpower \Yfunvec^\funT
\Amat^{-1}\GmatuJ^{}\GmatuJ^\matT\Amat^2\GmatuJ^{}\GmatuJ^\matT\Amat^{-1}\Yfunvec$
and recognizing that the terms $\GmatuJ^{}\GmatuJ^\matT$ are never
identities, and that the interior term
$\GmatuJ^\matT\Amat^2\GmatuJ^{}$ is an identity only when $\Amat$
itself is an identity, which is never the case in this chapter, but
would apply in the zero-altitude scalar case considered by
\citet{Simons+2006b}. Another way of stating it is that
\citet{Simons+2006b} mistakenly applied their identity~(93), which is
our~(\ref{up and down scalar slepians}), in the case of truncated
sums, for which it does not hold. The typos do not affect any of their
further analysis or conclusions, which were conducted at zero
altitude.

\section{Potential-Field Estimation from Vectorial Data Using Slepian
  Functions}\label{section vectorial data inversion} 

In this section we present a method to solve problem~$\Pfour$, the
estimation of the potential field on Earth's surface from noisy
(three-component) vectorial data at satellite altitude
\cite[e.g.][]{Arkani-Hamed2002}. The method is constructed in a
similar fashion to the scalar solutions to problem~$\Ptwo$ described
in Section~\ref{section radial data inversion}. We will use the
gradient-vector Slepian functions introduced in Section~\ref{section
  gradient vector Slepian functions} to fit the local data at
satellite altitude and then downward transform the gradient-vector
spherical harmonic coefficients thus obtained. As for the scalar case
we will first present the numerical method applicable to pointwise
data and then develop a functional formulation that will allow us to
analyze the effect of non-bandlimited signal and noise on the
estimation.

\subsection{Discrete Formulation and Truncated Solutions}\label{vector Slepian numerical} 

Given pointwise data values of the gradient of the potential that are
polluted by noise at the points
$\satalt\rvec_1,\ldots,\satalt\rvec_\npoints$,
\begin{equation}\label{vectorial noisy data}
\dpoints = \gradsigpoints + \noisepoints,
\end{equation}
where $\gradsigpoints$ is defined in Eq.~(\ref{full vectorial data}),
and $\noisepoints$ is a vector of noise values at the evaluation
points for the individual components, we seek to estimate the
spherical-harmonic coefficients $\sphcoef^\Earthrad=(
\ssphcoef_{00}^\Earthrad\,\cdots\,\ssphcoef_{LL}^\Earthrad)^T$ of the
scalar potential~$\signal$ on Earth's surface~$\Omega_\Earthrad$, as in the
statement~(\ref{Equation problem 4}) of problem~$\Pfour$. The solution
Eq.~(\ref{solution 1 to problem 4}) contains the matrix inverse
$(\Epoints\Epoints^\pointT)^{-1}$ which, like its counterpart
Eq.~(\ref{definition vector Kernel}), is intrinsically poorly
conditioned. To regularize the problem we transform the problem into
the gradient-vector Slepian basis for the relevant bandwidth and the
chosen target region~$\region$, and focus on estimating only the $J$
best-concentrated gradient-vector Slepian coefficients. We leave the
choice of the value~$J$ for later.

We define the $(L+1)^2\times 3\npoints$ dimensional matrix containing the
 $(L+1)^2$ gradient-vector Slepian functions
 $\Hfun_1,\ldots,\Hfun_{(L+1)^2}$ evaluated at the unit-sphere longitudes and
latitudes of the data,
\begin{equation}\label{definition Hpoints}
\Hpoints=\Hmat^\matT\Epoints,
\end{equation}
where the gradient-vector Slepian transformation matrix~$\Hmat$ is
defined in Eq.~(\ref{definition Hmat}) and the matrix $\Epoints$
containing the values of the gradient-vector spherical harmonics
evaluated at the data locations on the unit sphere is defined in
Eq.~(\ref{E point matrix}). Problem $\Pfour$ is rewritten from
Eq.~(\ref{Equation problem 4}) via the gradient-vector Slepian
transformation $\Hmat$ at altitude, to  
\begin{equation}\label{vector point data}
\arg\min_{\estsphcoef^\Earthrad} \lVert
\Epoints^\pointT\Bmat\estsphcoef^\Earthrad - \dpoints \rVert^2=
\arg\min_{\estsphcoef^\Earthrad} \lVert
\Epoints^\pointT \Hmat \Hmat^\matT  \Bmat\estsphcoef^\Earthrad-\dpoints \rVert^2=  
\arg\min_{\trs} \lVert
\Hpoints^\pointT \trs -\dpoints \rVert^2,
\end{equation} 
where we used the orthogonality $\Hmat\Hmat^\matT=\Imat$, the
definition Eq.~(\ref{definition Hpoints}) and introduced the 
gradient-vector Slepian coefficients at satellite altitude 
\begin{equation}\label{definition upvecSlepian}
\trs =  \Hmat^\matT  \Bmat\estsphcoef^\Earthrad.
\end{equation}

\begin{figure}
\begin{centering}
 \includegraphics[width=0.98\textwidth,
 trim = 12mm 0mm 4mm 9mm, clip]{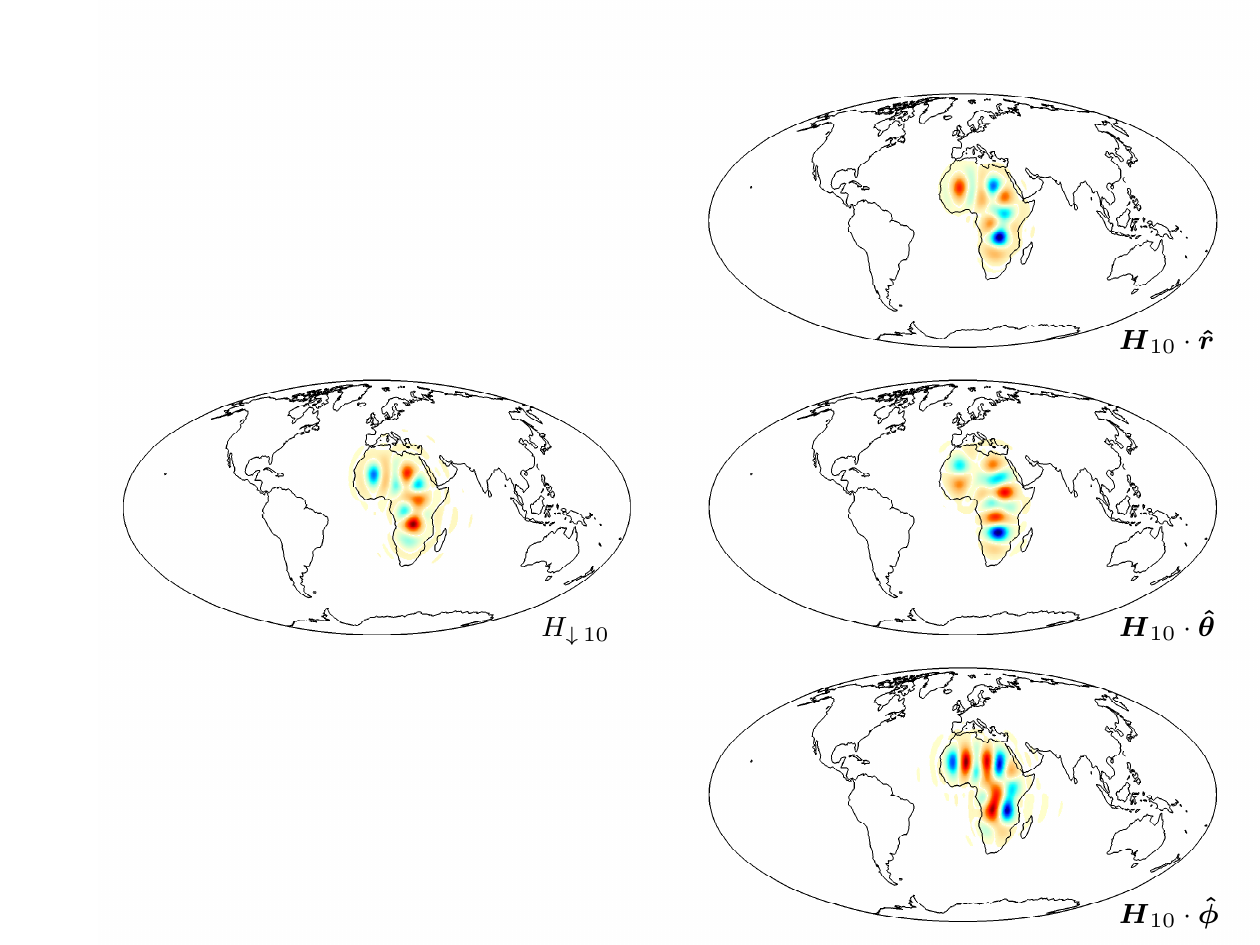}
 \caption{\label{figure GVS down} Downward transformation of the 10th
   best-concentrated gradient-vector Slepian function for Africa and a
   maximum spherical-harmonic degree $L=30$. The right panels show the
   concentrated gradient-vector Slepian function $\Hfun_{10} =
   \Efunvec^\funT\vslepfuncoef_{10}$.  Top right panel shows the
   radial component $\Hfun_{10}\cdot\rvec$, middle right panel the
   tangential (colatitudinal) component $\Hfun_{10} \cdot \thvec$, and the
   lower right panel the tangential (longitudinal) component
   $\Hfun_{10}\cdot\phvec$. The left panel shows the downward transformed
   scalar potential $\Hpfun_{\downarrow\, 10} = \Yfunvec^\funT \Bmat^{-1}
   \vslepfuncoef_{10}$ on Earth's surface ($\Earthrad = 6371$~km) that
   corresponds to the field $\Hfun_{10}$ at satellite altitude
   $500$~km. The concentration coefficient for the gradient-vector
   Slepian function~$\Hfun_{10}$ at satellite altitude is $\lambda=0.93$.}
\end{centering}
\end{figure}

As for the scalar case we apply regularization by only estimating the
coefficients for the $J$ best-concentrated gradient-vector Slepian
functions. We define the $J\times 3\npoints$ dimensional matrix containing
the point evaluations of those, 
\begin{equation}\label{definition HpointsJ}
\Hpoints_J=\HmatJ^\matT\Epoints
\end{equation}
and then solve 
\begin{equation}\label{vector numerical least squares problem}
\arg\min_{\trsj} \lVert
\Hpoints_J^\pointT \trsj -\dpoints
\rVert^2
\end{equation}
for the $J$-dimensional vector $\trsj$ of gradient-vector Slepian
coefficients at satellite altitude.  For $J\leq 3\npoints$ the
minimizer
\begin{equation}\label{vector numerical least squares solution}
\trsj = (\Hpoints_J^{} \Hpoints_J^\pointT) ^{-1}\Hpoints_J \dpoints
\end{equation}
is subsequently downward transformed to the $(L+1)^2$ spherical-harmonic
coefficients $\estsphcoef^\Earthrad$ of the field on Earth's surface
$\Omega_\Earthrad$ as
\begin{equation}\label{solution 2 problem 4}
\estsphcoef^\Earthrad = \Bmat^{-1}\HmatJ\trsj= \Bmat^{-1}\HmatJ (\Hpoints_J 
\Hpoints_J^\pointT)^{-1}  
\Hpoints_J^{} \dpoints  \qquad \text{(solution 2 to noisy problem $\Pfour$)},
\end{equation}
using matrix $\Bmat$ defined in Eq.~(\ref{definition B}). The
conditioning of the matrix $(\Hpoints_J \Hpoints_J^\pointT)$ is
determined by the truncation level~$J$. The local approximation
$\estsignal(\Earthrad\rvec)$ of the potential field
$\signal(\Earthrad\rvec)$ can now be calculated by 
\begin{equation}\label{vectorial field estimation}
\estsignal(\Earthrad\rvec)=\Yfunvec^\funT\estsphcoef^{\Earthrad}=
\Hpfunvec^\funT_{\downarrow J}
 (\Hpoints_J 
\Hpoints_J^\pointT)^{-1}  
\Hpoints_J \dpoints= \Hpfunvec^\funT_{\downarrow J}\trsj,
\end{equation}
where we have defined the vector of the $J$ best-concentrated gradient
vector Slepian functions (and its complement) that are downward
transformed (hence, expanded in scalar spherical harmonics) as 
\begin{equation}\label{definition Hpfunvec down}
\Hpfunvec_{\downarrow J}=\HmatJ^\matT\Bmat^{-1}\Yfunvec,\qquad
\Hpfunvec_{\downarrow >J}=\Hmat_{>J}^\matT\Bmat^{-1}\Yfunvec.
\end{equation} 
Fig.~\ref{figure GVS down} shows an example. Similarly, we will be
needing the upward-transformed pair of vectors 
\begin{equation}\label{definition Hpfunvec up}
\Hpfunvec_{\uparrow J} = \HmatJ^\matT\Bmat\Yfunvec,\qquad
\Hpfunvec_{\uparrow >J} = \Hmat_{>J}^\matT\Bmat\Yfunvec,
\end{equation}
and the relation derived from them when $J=(L+1)^2$ and
Eq.~(\ref{definition Hmat}) or Eq.~(\ref{definition Hfunvec}),  the
equivalent of Eq.~(\ref{up and down scalar slepians}), namely
\begin{equation}\label{up and down vector slepians}
\Hpfunvec_\downarrow^\funT(\rvec)\hsom\Hpfunvec_\uparrow^{}(\rvec')=
\Yfunvec^\funT\hsomm(\rvec)\hsom\Bmat^{-1}\Hmat\Hmat^\matT\hsom\Bmat\hsom\Yfunvec(\rvec') 
=\Yfunvec^\funT\hsomm(\rvec)\Yfunvec(\rvec')=
\Hpfunvec^\funT\hsomm(\rvec)\hsom\Hpfunvec^{}(\rvec').
\end{equation}
Once again we stress that we cannot derive such an equality after any
truncation of the Slepian function set. We do have
\begin{equation}\label{fixit2}
\Hpfunvec_\downarrow^\funT(\rvec)\hsom\Hpfunvec_\uparrow^{}(\rvec')=
\Hpfunvec_{\downarrow J}^\funT(\rvec)\hsom\Hpfunvec_{\uparrow J}^{}(\rvec')+
\Hpfunvec_{\downarrow >J}^\funT(\rvec)\hsom\Hpfunvec_{\uparrow >J}^{}(\rvec').
\end{equation}

\subsection{Continuous Formulation and Statistical Considerations}
\label{section vector analytical formulation}

In this section we reformulate the method described in
Section~\ref{vector Slepian numerical} such that instead of estimating
the potential field from pointwise data, we estimate the field from
functional data that is only available in the target region
$\region$. This will then enable us to analyze the effect of a
non-bandlimited signal and general noise on the estimation of the
potential field on Earth's surface~$\Omega_\Earthrad$.

\subsubsection{Continuous Formulation}

The data that are the functional equivalent of the point
values~(\ref{vectorial noisy data}) in the target region~$R$ are now
expressed as
\begin{equation}\label{vector data in region}
\datavec(\rvec)=\begin{cases}
\vecsignal(\satalt\rvec) + \vecnoise(\rvec) &\text{if } \rvec \in \region\\
\text{unknown} &\text{if } \rvec \in \Omega\setminus\region,
\end{cases}
\end{equation}
where $\vecnoise(\rvec)$ is a vector valued function of space
describing the noise at satellite altitude~$\satalt$.  The problem
equivalent to Eq.~(\ref{vector point data}),
\begin{equation}\label{vector Slepian signal up decomposition}
\arg\min_{\estsphcoef^\Earthrad} \intrnorm{
\Efunvec^\funT\Bmat\estsphcoef^\Earthrad - \datavec}=
\arg\min_{\estsphcoef^\Earthrad} \intrnorm{
\Efunvec^\funT \Hmat \Hmat^\matT  \Bmat\estsphcoef^\Earthrad-\datavec}=  
\arg\min_{\trs} \intrnorm{
\Hfunvec^\funT \trs -\datavec},
\end{equation} 
where the vector of gradient-vector Slepian functions $\Hfunvec$
is defined in Eq.~(\ref{definition Hfunvec})  and the estimated vector of
coefficients for the gradient-vector Slepian functions at satellite
altitude~$\trs$ is defined in Eq.~(\ref{definition upvecSlepian}). 

As for the numerical formulation we apply regularization by solving
only for the coefficients of the $J$ best-concentrated gradient-vector
Slepian functions at altitude to fit the data $\datavec$ given in
Eq.~(\ref{vector data in region}). We thence turn Eq.~(\ref{vector
  Slepian signal up decomposition}) into the estimation problem
\begin{equation}\label{vector analytical estimation problem}
\arg\min_{\trsj}\intrnorm{\Hfunvec_J^\funT\trsj-\datavec}=
\arg\min_{\trsj} \int_\region 
\left(\Hfunvec_J^\funT \trsj - \datavec\right)
\cdot
\left(\Hfunvec_J^\funT \trsj - \datavec\right)\dOmega
,
\end{equation}
which is solved by 
\begin{equation}
\trsj = \left(\int_\region \Hfunvec_J^{}
\cdot\Hfunvec_J^\funT\dOmega\right) ^{-1}
\int_\region \Hfunvec_J \cdot \datavec \dOmega
=\Sigmat_J^{-1}\int_\region\Hfunvec_J \cdot \datavec \dOmega,
\end{equation}
where we have used Eq.~(\ref{definition SigmatJ}). As for the
pointwise data case shown in Eq.~(\ref{solution 2 problem 4}) we
obtain and estimate $\estsphcoef^\Earthrad$ for the spherical-harmonic
coefficients of the potential field on Earth's
surface~$\Omega_\Earthrad$ as
\begin{equation}\label{analytical solution 2 to problem 4} 
\estsphcoef^\Earthrad= \Bmat^{-1}\HmatJ^{}\Sigmat_J^{-1}\int_\region
\Hfunvec_J \cdot \datavec \dOmega \qquad \text{(analytical solution 2
  to problem $\Pfour$)}.
\end{equation}
We can transform the coefficients $\estsphcoef^\Earthrad$ obtained
from the data $\datavec$ by Eq.~(\ref{analytical solution 2 to problem
  4}) into a local estimate of the potential field at the Earth's
surface as
\begin{equation}\label{vector analytical potential estimation}
\estsignal(\Earthrad\rvec) = \Yfunvec^\funT\estsphcoef^\Earthrad =
\Yfunvec^\funT\Bmat^{-1}\HmatJ^{}\Sigmat_J^{-1}\int_\region
\Hfunvec_J \cdot \datavec \dOmega = \Hpfunvec_{\downarrow J}^\funT
\Sigmat^{-1} \int_\region \Hfunvec_J \cdot \datavec \dOmega,
\end{equation}
where the vector containing the downward transformed gradient-vector
Slepian functions $\Hpfunvec_{\downarrow J}$ was defined in
Eq.~(\ref{definition Hpfunvec down}). 

\subsubsection{Effects of Bandlimiting the Vector Estimates}\label{bl2}

The estimate~(\ref{vector analytical potential estimation}) is
bandlimited but neither the data nor the noise usually would be. To
study the leakage and bias that arise from this discrepancy in the
representation, we separate the data explicitly into a bandlimited and
a broadband signal part, and the noise, much like we did for the
scalar case in Section~\ref{bl1}, as
 \begin{equation}
\datavec = \vecsignal(\satalt\rvec) + \vecnoise =
\Hfunvec^\funT\int_\Omega\Hfunvec\cdot \vecsignal(\satalt\rvec)\dOmega
+ \hat\Efunvec_{>L}^\funT\int_\Omega \hat\Efunvec_{>L}\cdot
\vecsignal(\satalt\rvec) \dOmega + \vecnoise. 
\end{equation}
To work towards Eq.~(\ref{vector analytical potential estimation}) we
multiply the data with the vector $\Hfunvec_J$ containing the $J$
best-concentrated gradient-vector Slepian functions and integrate over
the region. We make use of the orthogonality Eq.~(\ref{foxot}), and
Eqs~(\ref{definition HfunvecJ})--(\ref{definition SigmatJ}), and the
relations Eq~(\ref{definition hat Kmat L+1 >L})--(\ref{definition hatHfun J L}), to arrive at
\begin{align}\nonumber
\int_\region \Hfunvec_J \cdot \datavec\dOmega & =
\int_\region\Hfunvec_J \cdot \Hfunvec^\funT\dOmega \int_\Omega
\Hfunvec\cdot\vecsignal(\satalt\rvec) \dOmega\\ 
&\qquad+\int_\region
\Hfunvec_J^{} \cdot \hat\Efunvec_{>L}^\funT\dOmega \int_\Omega
\hat\Efunvec_{>L}\cdot \vecsignal(\satalt\rvec) \dOmega + \int_\region
\Hfunvec_J\cdot\vecnoise\dOmega \\ 
&=\Sigmat_J\int_\Omega
\Hfunvec_J\cdot \vecsignal(\satalt\rvec)\dOmega + \HmatJ^\matT
\KmatLL^\matT \int_\Omega \hat\Efunvec_{>L}\cdot
\vecsignal(\satalt\rvec) \dOmega + \int_\region
\Hfunvec_J\cdot\vecnoise\dOmega, \\ 
&=\Sigmat_J\int_\Omega
\Hfunvec_J\cdot \vecsignal(\satalt\rvec)\dOmega +
\HmatLJ^\matT
 \int_\Omega \hat\Efunvec_{>L}\cdot
\vecsignal(\satalt\rvec) \dOmega + \int_\region
\Hfunvec_J\cdot\vecnoise\dOmega,\\
&=\Sigmat_J\int_\Omega
\Hfunvec_J\cdot \vecsignal(\satalt\rvec)\dOmega + \int_\Omega
\HLJ\cdot \vecsignal(\satalt\rvec)\dOmega + \int_\region
\Hfunvec_J\cdot\vecnoise\dOmega,
\label{vectorial bias integration R}
\end{align}
Substituting Eq.~(\ref{vectorial bias integration R}) into the
expression for our estimate Eq.~(\ref{vector analytical potential
  estimation}) exposes its bandlimited and broadband constituent terms 
\begin{equation}
\estsignal(\Earthrad\rvec) = \Hpfunvec_{\downarrow J}^\funT\int_\Omega
\Hfunvec_J\cdot \vecsignal(\satalt\rvec) \dOmega
+\Hpfunvec_{\downarrow J}\Sigmat_J^{-1}\left( \int_\Omega \HLJ\cdot
\vecsignal(\satalt\rvec)\dOmega + \int_\region
\Hfunvec_J\cdot\vecnoise\dOmega\right).\label{vector estimation full}
\end{equation}
The convenience of our notation is apparent from the comparison of
this equation with Eq.~(\ref{scalar estimation full}), which is
functionally very similar. Here, as there, the estimation error of the
bandlimited part of the signal (the first term in Eq.~\ref{vector
  estimation full}) becomes smaller with less truncation (larger~$J$),
but the bias from the non-bandlimited part of the signal and the noise
(second term) grows, amplified by the concentration factor
$\Sigmat_J^{-1}$ which becomes less well conditioned with growing~$J$,
as Slepian functions with ever smaller eigenvalues are being included
into the estimate.

\subsubsection{Statistical Analysis for Vectorial Bandlimited-White Processes}

Even more so than for the scalar case described in
Section~\ref{section analytical formulation}, the calculation of the
variance, bias, and mean squared error of the
estimates~(\ref{analytical solution 2 to problem 4})--(\ref{vector
  analytical potential estimation}), in the general sense of
Eq.~(\ref{vector estimation full}), would be very involved without
imparting much insight. Instead, as for the scalar case, we narrow our
scope to vectorial data~$\datavec$ that satisfy some special
properties.  Because the field $\estsignal(\Earthrad\rvec)$ that we
estimate from these data is still a scalar function we can retain the
definitions of variance, bias, and mean squared error given in
Eqs~(\ref{error})--(\ref{mse}). We update the list of assumptions as
follows:
\begin{enumerate}
\item The signal $\signal(\Earthrad\rvec)$ is bandlimited with the
  same bandlimit $L$ as the Slepian functions $\Hfunvec$.\label{vector
    assume bandlimited}
\item The signal is white on the surface
  $\langle\signal(\Earthrad\rvec)\signal(\Earthrad\rvec')\rangle=\sigpower\hsom\delta(\rvec, 
\rvec')$.\label{vector assume white signal} 
\item The noise is white at the observation level, $\langle
  \vecnoise(\rvec)\vecnoise(\rvec')\rangle=\noisepower\di(\rvec,
  \rvec')$, with $\di(\rvec,\rvec')$ the vectorial delta function
  \cite[see][]{Plattner+2013}.
\label{vector assume white noise}
\item The noise has zero mean and none of its components are
  correlated with the signal,
  $\langle\vecnoise(\rvec)\rangle=\zerovec=\langle
  \vecnoise(\rvec)\signal(\rvec')\rangle.$
\label{vector assume zero mean noise}
\end{enumerate}

Following assumption \ref{vector assume bandlimited} we insert the
bandlimited portion of Eq.~(\ref{popa}) into Eq.~(\ref{vector
  estimation full}), supply the form of Eq.~(\ref{definition
  HfunvecJ}), observe the cancellation of the whole-sphere inner
product between $\HLJ$ and $\Efunvec$ inside the parentheses in
eq.~(\ref{vector estimation full}), and then use the
relations~(\ref{definition HfunvecJ}) and~(\ref{definition Hmat}) to write
\begin{align}\nonumber
\estsignal(\Earthrad\rvec) & = \Hpfunvec_{\downarrow J}^\funT\left(
\int_\Omega \Hfunvec_J\cdot \Efunvec^\funT
\Bmat\sphcoef^\Earthrad\dOmega + \Sigmat_J^{-1}
\int_\region\Hfunvec_J\cdot\vecnoise \dOmega \right)=
\Hpfunvec_{\downarrow J}^\funT\left(  \HmatJ^\matT
\Bmat\sphcoef^\Earthrad + \Sigmat_J^{-1}
\int_\region\Hfunvec_J\cdot\vecnoise \dOmega \right)\nonumber\\ 
&= \Hpfunvec_{\downarrow J}^\funT\left( \int_\Omega
\Hpfunvec_{\uparrow J} \signal(\Earthrad\rvec) \dOmega +
\Sigmat_J^{-1} \int_\region\Hfunvec_J\cdot\vecnoise \dOmega 
\right),\label{vector special estsignal}
\end{align}
the last equality following from Eq.~(\ref{non-bandlimited scalar}),
global orthogonality of the~$\Efunvec$,  and Eq.~(\ref{definition
  Hpfunvec up}). From Eqs~(\ref{up and down vector slepians})~and~(\ref{fixit2}) 
  we learn
that the unknown bandlimited true signal $\signal(\Earthrad\rvec)$
can be represented by 
\begin{equation}\label{vector up and down representation}
\signal(\Earthrad\rvec)=\Hpfunvec_{\downarrow}^\funT\int_\Omega
\Hpfunvec_{\uparrow} \signal(\Earthrad\rvec) \dOmega 
=\Hpfunvec_{\downarrow J}^\funT\int_\Omega
\Hpfunvec_{\uparrow J} \signal(\Earthrad\rvec) \dOmega
+\Hpfunvec_{\downarrow >J}^\funT\int_\Omega
\Hpfunvec_{\uparrow >J} \signal(\Earthrad\rvec) \dOmega. 
\end{equation}
The bias of Eq.~(\ref{bias}) derives from averaging Eq.~(\ref{vector
  special estsignal}), using assumption~\ref{vector assume zero mean
  noise}, and then subtracting Eq.~(\ref{vector up and down
  representation}) to yield a term that grows as $J$ gets lowered, 
\begin{equation}\label{vector bias equation}
\bias = -\Hpfunvec_{\downarrow >J}^\funT \int_\Omega\Hpfunvec_{\uparrow >J} \signal(\Earthrad\rvec)\dOmega.
\end{equation}
The variance $\variance$ requires the square of Eq.~(\ref{vector
  special estsignal}), that is,
\begin{align}\nonumber
\estsignal^2(\Earthrad\rvec)&=\Hpfunvec_{\downarrow J}^\funT\left(
\int_\Omega\Hpfunvec_{\uparrow J} \signal(\Earthrad\rvec)\dOmega +
\Sigmat_J^{-1}\int_\region\Hfunvec_J\cdot\vecnoise\dOmega  \right)
\left( \int_\Omega \signal(\Earthrad\rvec) \Hpfunvec_{\uparrow
  J}^\funT \dOmega + \Sigmat_J^{-1}\int_\region\vecnoise \cdot
\Hfunvec_J^\funT \dOmega  \right)\Hpfunvec_{\downarrow J}\\ 
&= \Hpfunvec_{\downarrow J}^\funT\left(
\int_\Omega\int_\Omega  \Hpfunvec_{\uparrow
  J}(\rvec)\signal(\Earthrad\rvec)\signal(\Earthrad\rvec')
\Hpfunvec_{\uparrow J}^\funT(\rvec')\dOmega' \dOmega  + \Sigmat_J^{-1}\int_\region\int_\region [\Hfunvec_J(\rvec)
  \cdot \vecnoise(\rvec) ] [\vecnoise(\rvec') \cdot
  \Hfunvec_J^\funT(\rvec')] \dOmega'\dOmega \hsom  \Sigmat_J^{-1}\right.\label{vector squared estimator}\\ 
&\nonumber
\left.\qquad+ \int_\Omega\int_\region \Hpfunvec_{\uparrow
  J}(\rvec)\signal(\Earthrad\rvec)[\vecnoise(\rvec') \cdot
  \Hfunvec_J^\funT(\rvec')] \dOmega'\dOmega  \hsom\Sigmat_J^{-1}
 +  \Sigmat_J^{-1} \int_\region\int_\Omega
    [\Hfunvec_J(\rvec) \cdot \vecnoise(\rvec)
    ]\signal(\Earthrad\rvec')  \Hpfunvec_{\uparrow
      J}^\funT(\rvec')\dOmega'\dOmega\right) 
\Hpfunvec_{\downarrow J}.
\end{align}
After averaging Eq.~(\ref{vector squared estimator}) under the
assumptions~\ref{vector assume white noise}--\ref{vector assume zero
  mean noise}, using Eq.~(\ref{definition SigmatJ}), and subtracting
the square of the average of Eq.~(\ref{vector special estsignal}), we
get  the estimation variance of Eq.~(\ref{variance}), which grows
with~$J$, in the form
\begin{equation}\label{vector variance}
\variance= \noisepower\Hpfunvec_{\downarrow
  J}^\funT\Sigmat_J^{-1}\Hpfunvec_{\downarrow J}
.
\end{equation}
The average squared bias under the assumption~\ref{vector assume
  white signal}, with Eq.~(\ref{definition Hpfunvec up}) and the global
orthogonality of the spherical harmonics~$\Yfunvec$, is written as
\begin{equation}
\langle\bias^2\rangle= \sigpower \Hpfunvec_{\downarrow > J}^\funT
\big(\Hmat_{> J}^\matT \Bmat^2\Hmat_{> J}\big)\Hpfunvec_{\downarrow > J}
,
\end{equation}
which, together with the variance in Eq.~(\ref{vector variance}),
leads to the mean squared error defined in Eq.~(\ref{mse}), in the form
\begin{equation}\label{vector mean squared error special case}
\langle\esterr^2\rangle = \noisepower\Hpfunvec_{\downarrow
  J}^\funT\Sigmat_J^{-1}\Hpfunvec_{\downarrow J} + \sigpower
\Hpfunvec_{\downarrow > J}^\funT 
\big(\Hmat_{> J}^\matT\Bmat^2\Hmat_{> J}\big)
\Hpfunvec_{\downarrow > J}. 
\end{equation}

\section{Numerical Examples}\label{section examples}

In this section we illustrate the use of Eqs~(\ref{solution 2 to
  problem 2})--(\ref{scalar field estimation}) to solve the noisy
scalar problem~$\Ptwo$, and Eqs~(\ref{solution 2 problem
  4})--(\ref{vectorial field estimation}) for the noisy vectorial
problem~$\Pfour$. In both cases our aim is to estimate the scalar
potential field on Earth's surface from noisy scalar and vectorial
data, synthetically generated at a representative altitude. Throughout
the section we assume the Earth to be a sphere of radius
$\Earthrad=6371$~km and the satellite to fly in a spherical orbit at
$(\satalt-\Earthrad)=500$~km above Earth's surface.  We implemented
the numerical algorithms in Matlab, and wherever the solution of a
linear system of equations was required, such as in Eq.~(\ref{kukuk})
or Eq.~(\ref{vector numerical least squares solution}), we used the
operator \texttt{mldivide}, e.g. $(\Gpoints_J^{} \Gpoints_J^\pointT)
\backslash (\Gpoints_J \dpoints_r)$ and $(\Hpoints_J^{}
\Hpoints_J^\pointT)\backslash(\Hpoints_J \dpoints)$.
 
The ``true'' potential field
$\signal(\Earthrad\rvec)=\Yfunvec^\funT\sphcoef^\Earthrad$ in our
numerical experiments is bandlimited to degree $L=72$ and its
isotropic signal power is constant within the bandlimit by satisfying
$ \frac{1}{2l+1}\sum_{m=-l}^l(\ssphcoef_{lm}^\Earthrad)^2=1 $ for
$1\leq l \leq L$. We ensured that the signal had zero mean over the
entire Earth's surface by setting
$\ssphcoef_{00}^\Earthrad=0$. Figs~\ref{example radial data
  inversion}~and~\ref{example vectorial data inversion} show the
potential-field signal in their upper-left panels. 

The bandlimited scalar quantity at satellite altitude
$\scalsignal(\satalt \rvec)$ is defined by the bandlimited version of
Eq.~(\ref{popo}), and likewise, the vectorial
quantity~$\vecsignal(\satalt\rvec)$ by the bandlimited restriction of
Eq.~(\ref{bliblu}). In each of the experiments in this section we
sampled the fields at altitude at the same set of 2217 points which
were uniformly distributed (equal surface area) over the target region
$\region$, Africa, of solid-angle area~$a=\int_R d\Omega$. From these
points we created vectors with the data~$\dpoints_r$ or
$\dpoints$ as in Eqs~(\ref{scalar noisy data})
and~(\ref{vectorial noisy data}).

The noise for the scalar problem was generated at every location of
the data points by independent sampling from a zero-mean Gaussian
distribution with a variance equal to 2.5\% of the numerical signal
power at satellite altitude $\satalt$ given by
$(1/\npoints)\|\gradsigpoints_r\|^2=(1/\npoints)\sum_{i=1}^k
[\scalsignal(\satalt \rvec_i)]^2 $.  For the vectorial problem we
generated the noise for each of the three signal components at
satellite altitude, $\scalsignal(\satalt \rvec)$,
$\scalsignalt(\satalt \rvec)$, and $\scalsignalp(\satalt \rvec)$,
independently from zero-mean Gaussian distributions with identical
variances equal to 2.5\% of the numerical power of the signal in each
of the components separately.  

At each fixed Slepian-basis truncation level~$J$, the scalar estimates
in Eq.~(\ref{solution 2 to problem 2}) are derived from the
solutions~(\ref{kukuk}) which minimize the quadratic
misfit~(\ref{scalar numerical Slepian problem}) that is our
regularized proxy for the noisy problem~(\ref{point data}). Similarly,
the vectorial estimates Eq.~(\ref{solution 2 problem 4}) derive from
the solutions~(\ref{vector numerical least squares solution}) to the
misfit~(\ref{vector numerical least squares problem}) which is our
regularized version of the noisy problem~(\ref{vector point data}). As
we have seen in the theoretical treatment of the problem, the
truncation regularization biases the estimates (see Eqs~\ref{corr1}
and~\ref{vector bias equation}) by an amount that grows when
lowering~$J$ (more truncation), but the estimation variances (see
Eqs~\ref{scalar special variance} and~\ref{vector variance}) are
positively affected by lowering $J$ (which leads to smaller
variance). In all this, our ultimate objective is to control the
trade-off between bias and variance and make our estimates of the
potential field at the surface of the Earth as \textit{efficient} as
possible \cite[][]{Cox+74,Davison2003}. We thus need to evaluate the
quality of the estimates made using different truncation levels~$J$ in
terms of their mean squared errors (see Eqs~\ref{scalar mse equation}
and~\ref{vector mean squared error special case}).


For each experiment we will compute as a measure of efficiency the
mean squared error between the estimated potential-field and the
(bandlimited) truth, at the Earth's surface, averaged over the area of
interest, as follows
\begin{equation}\label{mse calc}
\mse =\frac{1}{a} \int_\region
\big[\signal(\Earthrad\rvec)-\estsignal(\Earthrad\rvec)\big]^2
\dOmega =\frac{1}{a}
\big(\sphcoef^\Earthrad-\estsphcoef^\Earthrad\big)^\matT \Dmat \hsom
\big(\sphcoef^\Earthrad-\estsphcoef^\Earthrad\big).
\end{equation}
With the truth $\signal(\Earthrad\rvec) =
\Yfunvec^\funT\sphcoef^{\Earthrad}$, and the estimates in the common
form $\estsignal(\Earthrad\rvec) =
\Yfunvec^\funT\estsphcoef^{\Earthrad}$ as given by either
Eqs~(\ref{scalar field estimation}) and~(\ref{vectorial field
  estimation}), the truncation-level $J$-dependent Eq.~(\ref{mse
  calc}) can be calculated directly with the aid of the localization
kernel Eq.~(\ref{definition scalar kernel}), as shown.  We will
express the regional mean squared error relative to the mean squared
signal strength over the same area, which is given by
\begin{equation}\label{mss calc}
\mss =\frac{1}{a} \int_\region
\signal^2(\Earthrad\rvec)
\dOmega =\frac{1}{a}
\big(\sphcoef^\Earthrad)^\matT \Dmat \hsom
\big(\sphcoef^\Earthrad\big).
\end{equation}
We will call the relative measure 
\begin{equation}\label{solution relative mse}
\varphi(J)=\frac{\mse}{\mss},
\end{equation} 
and plot it in function of the Slepian-function truncation
level~$J$. Finally, we will also quote the relative quadratic measure
of data misfit, Eq.~(\ref{scalar numerical Slepian problem}), between
the given data $\dpoints_r$ and the simulated 
data, $\Ypoints^\pointT\Amat\estsphcoef^\Earthrad$,
\begin{equation}\label{data relative mse}
\psi(J)
=\frac{\lVert\Ypoints^\pointT\Amat\estsphcoef^\Earthrad - \dpoints_r \rVert^2}
{\lVert\dpoints_r \rVert^2}
,
\end{equation}
where we recall that the prediction~$\estsphcoef^\Earthrad$ is given
by Eq.~(\ref{solution 2 to problem 2}) and thereby remains a function of
the truncation level~$J$. In the vectorial case, the equivalent metric
is the relative mean squared data misfit, Eq.~(\ref{vector numerical least squares problem}), between
the three vectorial components of the given data $\dpoints$ and the
three vectorial components of the simulated data,
$\Epoints^\pointT\Bmat\estsphcoef^\Earthrad$, 
\begin{equation}\label{vector data relative mse}
\psi(J)
=\frac{\lVert\Epoints^\pointT\Bmat\estsphcoef^\Earthrad - \dpoints \rVert^2}{\lVert\dpoints \rVert^2}.
\end{equation}

\subsection{Estimating the Potential Field at the Surface from
  Radial-Component Data at Satellite Altitude}
\label{section example radial}

\begin{figure}
\begin{centering}
 \includegraphics[width=\textwidth,trim = 10mm 0mm 0mm 0mm,clip]
{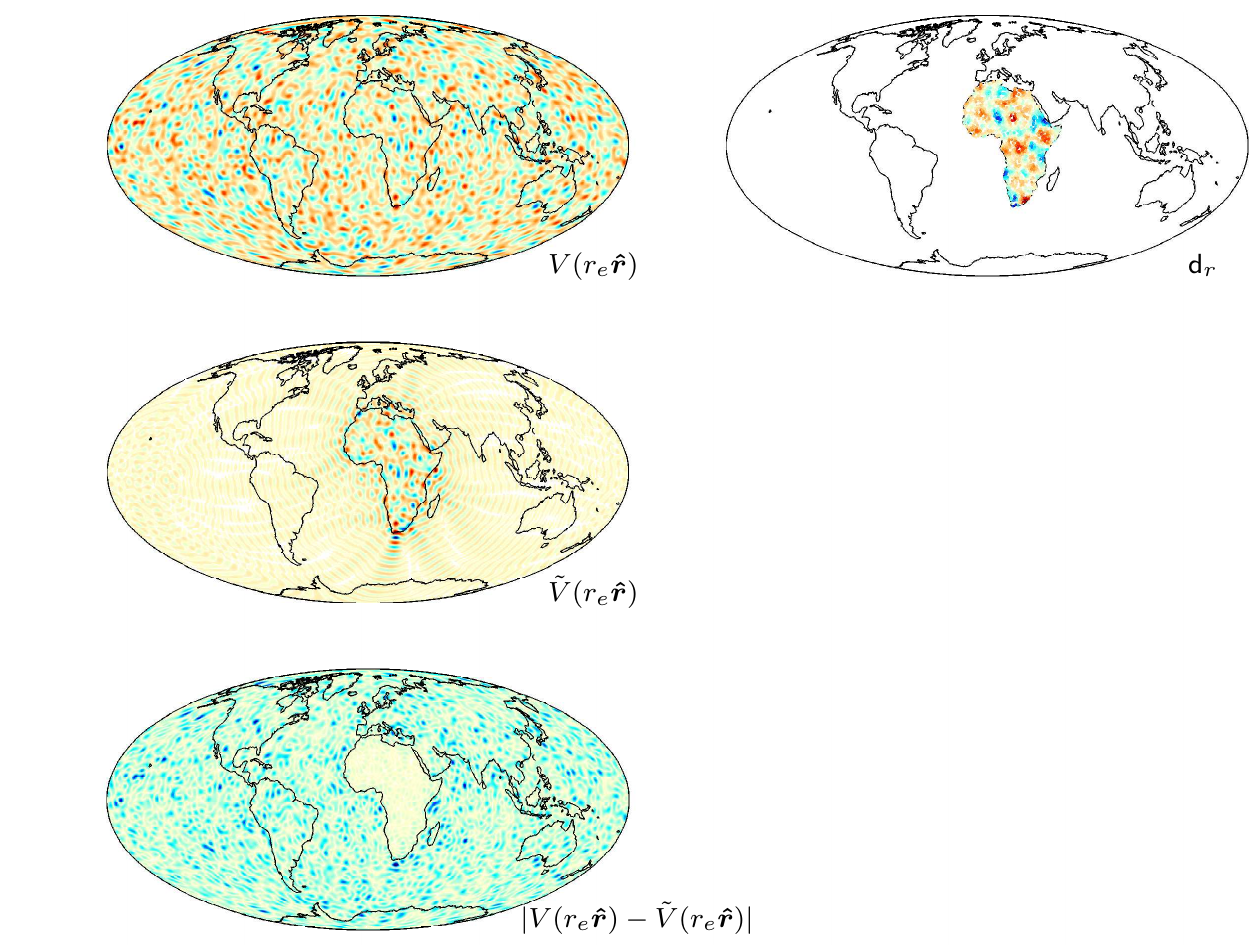}
 \caption{\label{example radial data inversion} Example of the
   estimation of a potential field on Earth's surface from noisy
   radial-derivative data at satellite altitude
   $\satalt=\Earthrad+500$~km, using Slepian functions bandlimited to
   $L=72$ and spatially concentrated to the target region Africa. The
   upper left panel shows the true potential field
   $\signal(\Earthrad\rvec)$ on Earth's surface. The upper right panel
   shows the \numpoints{} noisy data $\dpoints_r$ at satellite
   altitude. The middle left panel shows the estimated potential field
   $\estsignal(\Earthrad\rvec)$ calculated from the data using
   Eq.~(\ref{scalar field estimation}), with Slepian-function
   truncation level $J=\optimJo$. The lower left panel shows the
   absolute value of the difference $\lvert\signal(\Earthrad\rvec) -
   \estsignal(\Earthrad\rvec)\rvert$ between the true and the
   estimated potential fields.}
\end{centering}
\end{figure}


Fig.~\ref{example radial data inversion} shows the results from a
suite of experiments with noisy scalar data. For generality we omitted
a color bar and legend. We used the same linear color scale,
normalized to the maximum absolute $\signal(\Earthrad\rvec)$ value,
for all three panels on the left side. Blue is positive, red is
negative and all points with absolute value smaller than 1\% of the
maximum are left white.  The data, shown on the right, are also
color-coded in the same colormap, but the colors are scaled with
respect to the scale of the panels in the left column to account for
the reduced data values at satellite altitude.

The true potential field, $\signal(\Earthrad\rvec)$, is displayed in
the upper left panel of Fig.~\ref{example radial data inversion} and
one realization of the the noisy radial-derivative data at altitude,
$\dpoints_r$, are shown in the upper right panel. In the middle left
panel we plot the estimate $\estsignal(\Earthrad\rvec)$, at Earth's
surface $\Omega_\Earthrad$, from Eq.~(\ref{scalar field estimation}),
with $J=\optimJo$. In the bottom left panel we show the absolute value
of the difference between the truth and the estimate. The relative
mean squared error, following Eq.~(\ref{data relative mse}),
is~\rmseo. The Slepian-function truncation level $J=\optimJo$ was
chosen based on the numerical experiment shown in Fig.~\ref{many
  radial data inversion}. For this value of~$J$ the estimated
potential field $\estsignal(\Earthrad\rvec)$ approximates the true
potential field $\signal(\Earthrad\rvec)$ very well within Africa, and
it has almost no energy outside the region of interest.


\begin{figure}
\begin{centering}
\includegraphics[width=0.8\textwidth]
{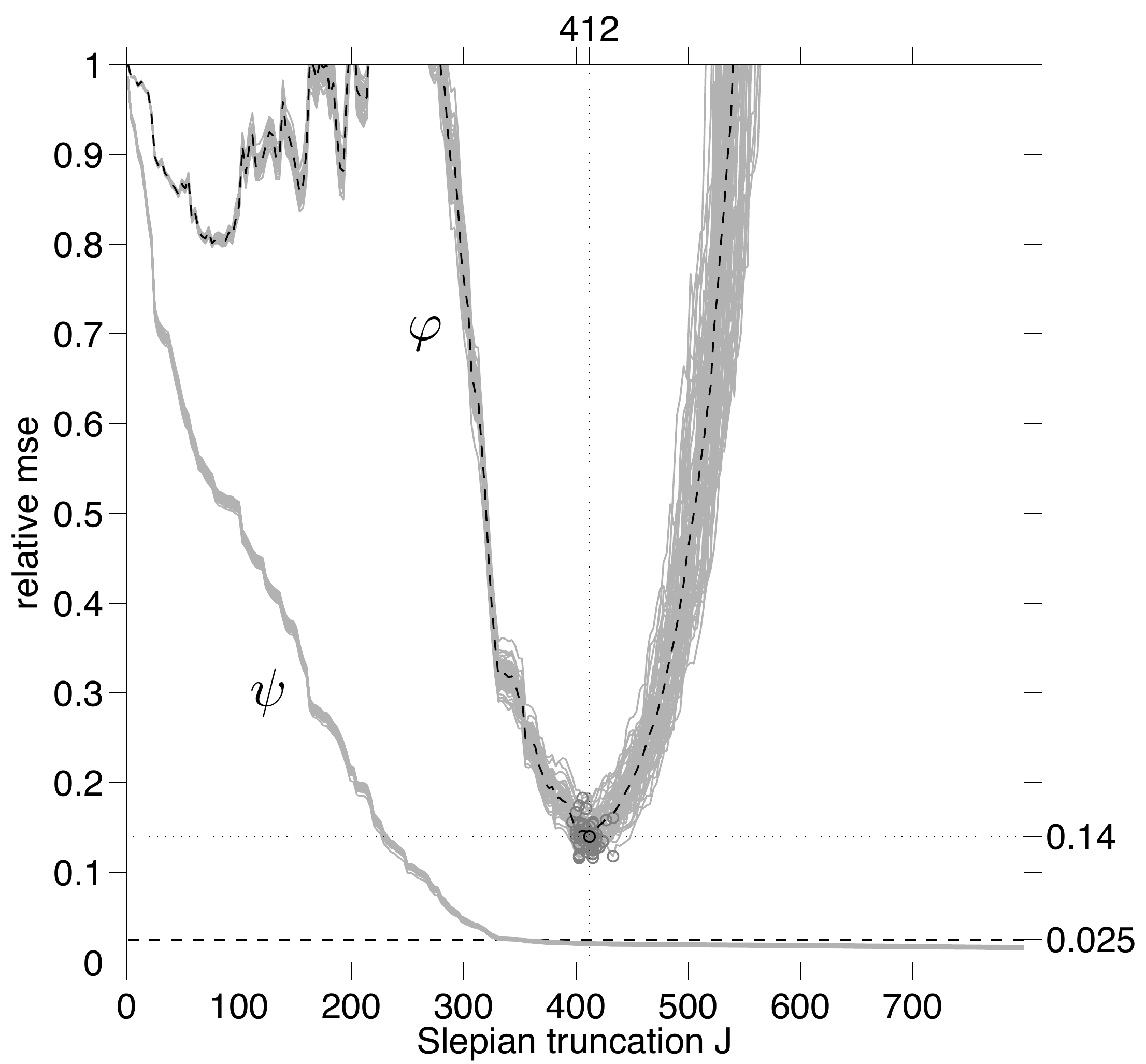}
 \caption{\label{many radial data inversion} Relative regional mean
   squared model errors $\varphi(J)$, from Eq.~(\ref{solution relative
     mse}), and relative mean squared data misfit $\psi(J)$, from
   Eq.~(\ref{data relative mse}), for potential field estimation from
   radial-derivative data as described in Eq.~(\ref{scalar field
     estimation}).  The true signal is the one shown in
   Fig.~\ref{example radial data inversion}.  Each of the 64
   realizations of noise leads to a gray $\varphi(J)$-curve and a gray
   $\psi(J)$. The optimal truncation points are indicated by gray
   circles, the average optimal truncation point by a black circle and
   the average $\varphi(J)$ behavior is the black dashed line. The
   dashed horizontal line is the relative energy of the noise.}
\end{centering}
\end{figure}

In Fig.~\ref{many radial data inversion}, each of the 64 gray lines
labeled~$\varphi$ is a curve of $\varphi(J)$, the regional relative
mean squared model error calculated as in Eq.~(\ref{solution relative
  mse}). The same true signal values $\gradsigpoints_r$ were used, but
every experiment used data~$\dpoints_r$, as given by Eq.~(\ref{scalar
  noisy data}), that were contaminated by a different realization of
the noise field $\noisepoints_r$, as described at the beginning of
this section. Every curve starts at $\varphi(0)=1$, as without any
basis functions, only the zero model is obtained. The relative \mse{}
decreases dramatically after about $J=250$ and the estimation improves
as more Slepian functions are involved. As we have explained earlier
for the theoretical behavior in Eq.~(\ref{scalar mse equation}), the
squared bias term~$\bias^2$ diminishes in value with
increasing~$J$. Less truncation (larger~$J$) reduces the estimation
bias, but this decrease is in competition with the
variance~$\variance$ term, which increases with $J$. The influence of
data noise is felt more and more with the inclusion of additional
basis functions.

The turning points of minimum relative mean squared estimation error
for each of the experiments are indicated by a gray circle. At the
corresponding value~$J$, the optimal Slepian truncation level for each
specific data set is reached. The average of all of the $\varphi(J)$
curves shown is represented by a black dashed line.  All individual
turning points are clustered around the average ideal truncation
point, which is the $J=\optimJo$ indicated by the black circle.  The
relative regional mean squared model errors~$\varphi$ do not improve
immediately after $J=1$, unlike the data errors~$\psi$. There is a
local minimum, followed by a rise, and a precipitous decline after
$J=250$ or thereabouts. We explain this behavior theoretically by our
minimizing the misfit of the upward-transformed potential field at the
altitude of the data (see Eq.~\ref{scalar functional estimation
  problem}) instead of the misfit on the surface, which is measured
by~$\varphi$. To obtain the potential field on the surface we need to
downward-transform the radial-field estimate at altitude, obtained by
truncation, as shown by Eq.~(\ref{analytical s2 to p2}). The downward
transformation operator $\Amat$ defined in Eq.~(\ref{definition Amat})
is poorly conditioned for high maximum degrees~$L$ and large relative
satellite altitudes $\satalt/\Earthrad$.  The interaction between all
of the terms altogether displays a complex behavior that, however, has
a clear global minimum which leads to a working algorithm and an
objective decision as to the optimal Slepian function truncation
level.

Because the noise level is relatively small compared to the signal
strength, and because we use the same \numpoints{} data locations, the
$\psi$-lines with the data fits are close together. The relative mean
squared data misfit curves $\psi(J)$ in Fig.~\ref{many radial data
  inversion} are decreasing fast until their values reach the relative
energy of the noise, 2.5\%, indicated by the dashed horizontal black
line. At this point the relative mean squared data misfit decreases
much slower, or almost not at all. We recall that the noise is
generated in the spatial domain, and is therefore not
bandlimited. Hence, the noise has appreciable energy at the degrees
larger than~72 which cannot be fit by the $L=72$ bandlimited Slepian
functions.

\subsection{Estimating the Potential Field at the Surface from
 Gradient-Vector Data at Satellite Altitude}
\label{section example gradient}

\begin{figure}
\center
 \includegraphics[width=\textwidth,trim = 10mm 0mm 0mm 0mm, clip]
                 {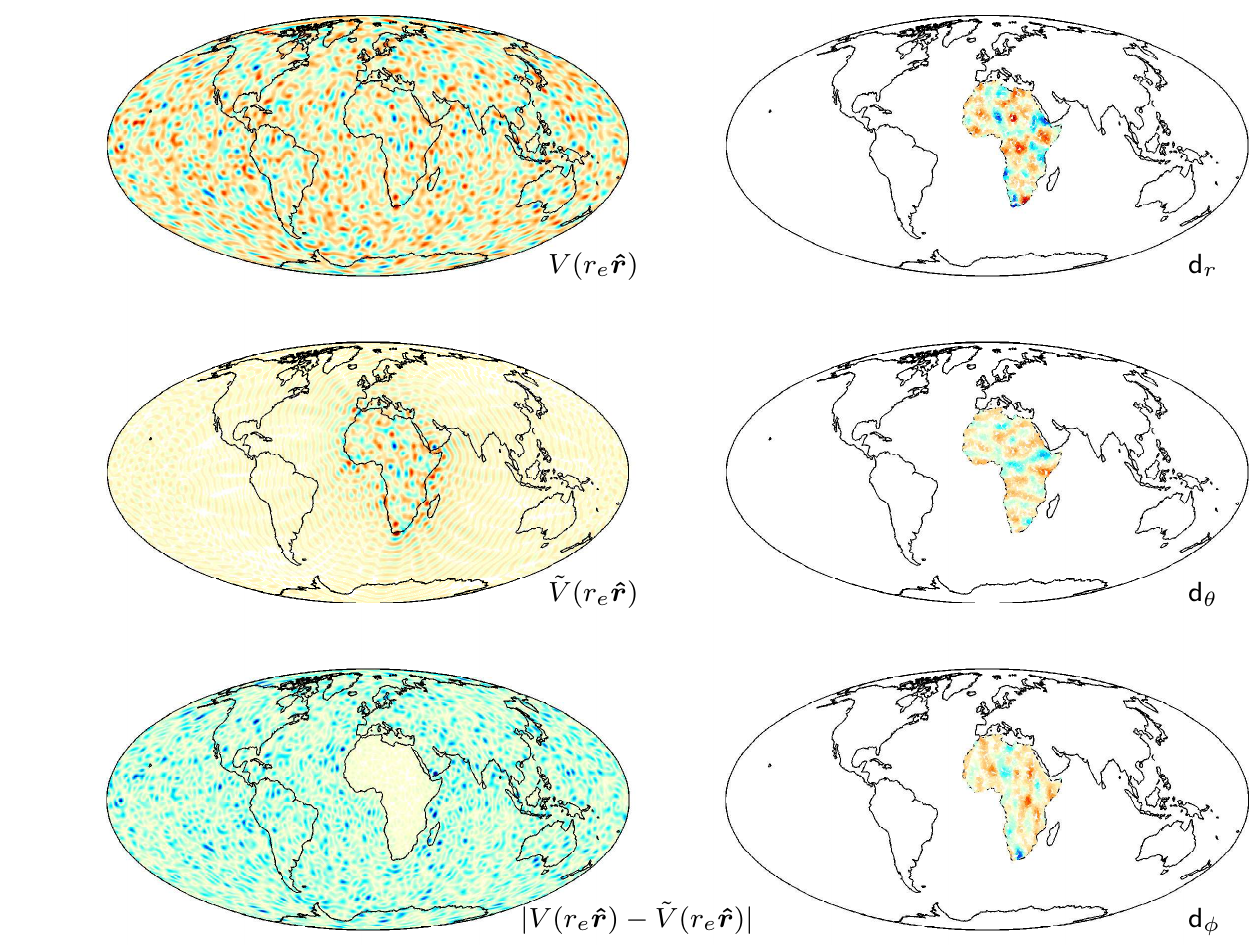}
 \caption{\label{example vectorial data inversion} Example of a
   potential field estimation on Earth's surface from noisy gradient
   data at altitude $\satalt=\Earthrad+500$ km for Slepian functions
   with maximum degree $L=72$ and target region Africa. The upper left
   panel shows the true potential field $\signal(\Earthrad\rvec)$ on
   Earth's surface. The three right panels show the noisy data
   $\dpoints$ at satellite altitude given by \numpoints{} data
   values. The top right panel depicts the radial component
   $\dpoints_r$, the middle right panel the tangential colatitudinal
   component $\dpoints_\theta$, and the lower right panel the
   tangential longitudinal component $\dpoints_\phi$. The middle left
   panel shows the estimated potential field
   $\estsignal(\Earthrad\rvec)$ calculated from the data with Slepian
   truncation $J=\optimJt$. The lower left panel shows the absolute
   difference $\vert \signal(\Earthrad\rvec) -
   \estsignal(\Earthrad\rvec) \rvert$ between the true and the
   estimated potential fields.}
\end{figure}

\begin{figure}
\begin{centering}
 \includegraphics[width=0.8\textwidth]
{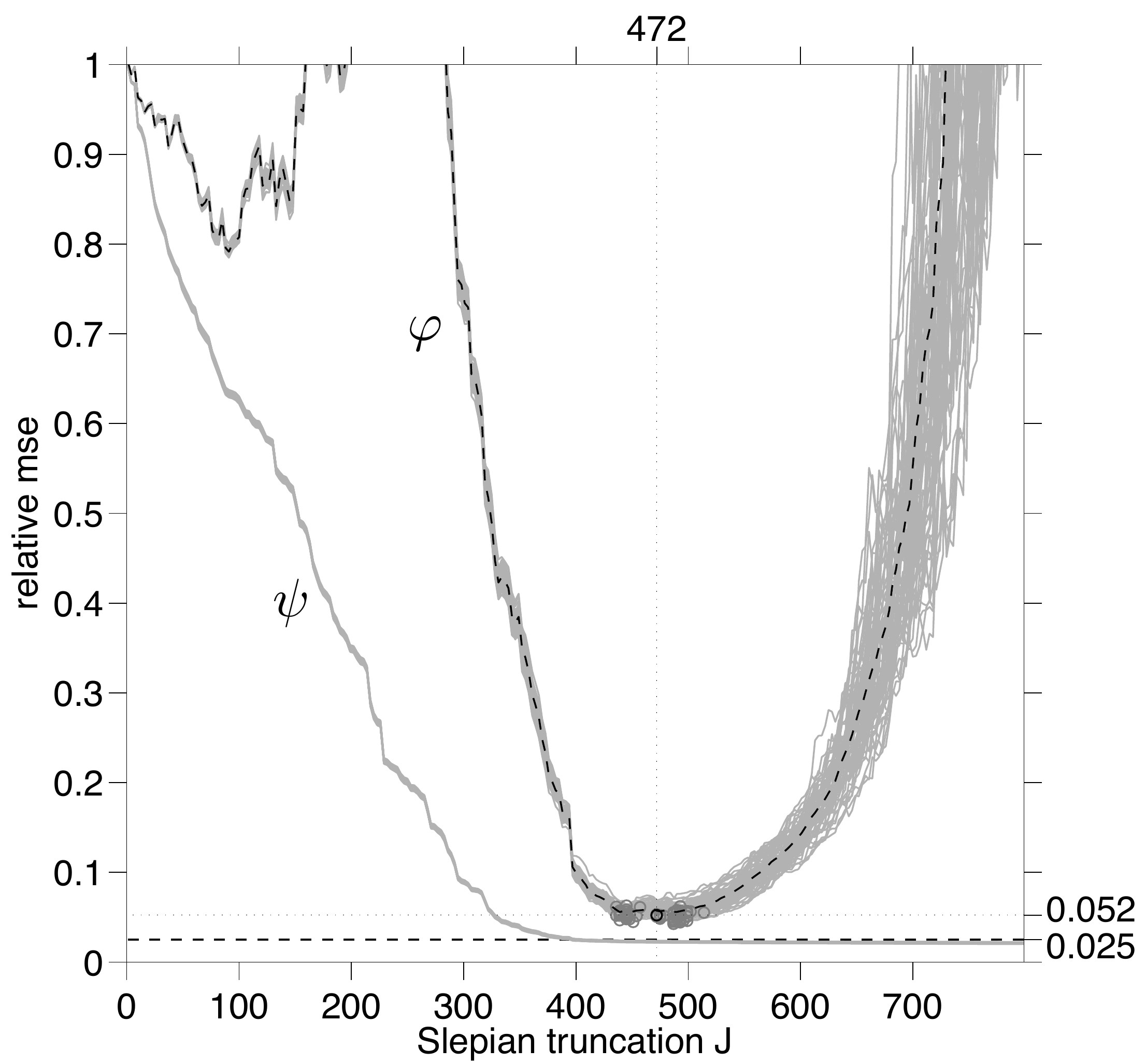}
 \caption{\label{many vectorial data inversion} Relative regional mean
   squared model error~$\varphi(J)$, from Eq.~(\ref{solution relative
     mse}), and relative mean squared data misfit~$\psi(J)$, from
   Eq.~(\ref{data relative mse}), for potential-field estimation from
   vectorial data described in Eq.~(\ref{vectorial field estimation}).
   The true signal is the same as for Fig.~\ref{example vectorial data
     inversion}.  Each of the 64 realizations of noise leads to a gray
   $\varphi(J)$-line and a gray~$\psi(J)$. The optimal truncation
   points are indicated by the gray circles, the average optimal
   truncation point by the black circle and the average $\varphi(J)$
   line by the black dashed line. The dashed horizontal line is the
   relative energy of the noise.}
\end{centering}
\end{figure}

Fig.~\ref{example vectorial data inversion} shows the results from an
experiment with noisy vectorial data. Our plot color conventions are
unchanged from those in Section~\ref{section example radial}, except
now the three panels on the right are scaled to the maximum absolute
vectorial data value at satellite altitude.  The true potential field
$\signal(\Earthrad\rvec)$ is found in the upper left panel of
Fig.~\ref{example vectorial data inversion}, and the noisy data at
altitude $\dpoints$ are shown on the right. The top right panel shows
the radial component~$\dpoints_r$, the middle right panel the
tangential colatitudinal component~$\dpoints_\theta$, and the lower
right panel the tangential longitudinal component~$\dpoints_\phi$.

We use Eq.~(\ref{vectorial field estimation}) to calculate an estimate
$\estsignal(\Earthrad\rvec)$ for the potential field on Earth's
surface, choosing the Slepian truncation $J=\optimJt$ based on the
numerical experiments shown in Fig.~\ref{many vectorial data
  inversion}. The estimated scalar potential field on Earth's surface
$\estsignal(\Earthrad\rvec)$ is shown in the middle left panel of
Fig.~\ref{example vectorial data inversion}. The lower left panel of
Fig.~\ref{many vectorial data inversion} shows the absolute difference
$\lvert \signal(\Earthrad\rvec) - \estsignal(\Earthrad\rvec)\rvert$
between the true and the estimated signal. The estimated field
$\estsignal(\Earthrad\rvec)$ approximates the true signal
$\signal(\Earthrad\rvec)$ well within Africa and is close to zero
outside of that target region. The relative regional mean squared model
error calculated using Eq.~(\ref{solution relative mse}) is~$\rmset$.

In Fig.~\ref{many vectorial data inversion} we plot the relative
regional mean squared model errors $\varphi(J)$ defined in
Eq.~(\ref{solution relative mse}) as a function of the truncation
level~$J$, for each of the 64 experiments. Each data set $\dpoints$ is
generated from the same true vector field $\sigpoints'$ using
Eq.~(\ref{vectorial noisy data}), but differs by the realization of
the noise $\noisepoints$, as discussed at the top of this
section. Each experiment starts at $\varphi(0)=1$ and descends from about
$J=250$ into a deep valley with increasing number of Slepian
functions. The theoretical relation in Eq.~(\ref{vector mean squared
  error special case}) explains how the decreasing bias and increasing
variance trade off as a function of the increasing number $J$ of
Slepian functions. The turning points are indicated by gray circles;
they all cluster around the same truncation value. The average
relative regional mean squared model error is shown by a dashed black
line, and the average optimal Slepian truncation level $J=\optimJt$ by a black
circle.  As in the scalar case the curves $\varphi(J)$ go through a
local minimum before reaching the global optimum truncation
level. Indeed, since we minimized Eq.~(\ref{vector analytical
  estimation problem}) at altitude, in order to obtain the estimate
$\estsignal(\Earthrad\rvec)$ at the Earth's surface we need to apply
the downward transformation operator $\Bmat$ defined in
Eq.~(\ref{definition B}). At high maximum degrees~$L$ and high
relative satellite altitudes $\satalt/\Earthrad$ this operator is
poorly conditioned. The interaction between the various competing
effects produces a complex but reproducible error behavior.

The 64 curves for the relative mean squared data misfit in
Fig.~\ref{many vectorial data inversion} are close together because
the signal-to-noise level is high, and because we reuse the same
\numpoints{} data locations. 
As for the scalar case, the relative mean squared data misfit $\psi(J)$ 
decreases fast until it reaches the relative energy of the noise, $2.5\%$,
indicated by the dashed horizontal black line.

\section{Conclusions}

We presented two methods to estimate a potential field from gradient
data at satellite altitude that are concentrated over a certain
region. At the heart of both methods lies the use of spatiospectrally
concentrated spherical basis functions. The first method only
considered the radial component of the data and used scalar Slepian
functions. The second method considered all three vectorial components
of the data and used gradient-vector Slepian functions, a special case
of vector Slepian functions. From the theoretical analysis of both
methods, and through extensive experimentation, we show how the mean
squared reconstruction error depends on the number of Slepian or
gradient-vector Slepian functions used for the estimation. The more
Slepian functions involved, the smaller the bias, but the larger the
variance in the presence of noise.


\acknowledgments 

A.P. thanks the Ulrich Schmucker Memorial Trust and the Swiss National
Science Foundation, the National Science Foundation, and Princeton
University for funding, and the Smart family of Cape Town for their
hospitality while writing this manuscript. This research was sponsored
by the U.S. National Science Foundation under grants EAR-1150145 and
EAR-1245788 to F.J.S.

\bibliography{/Users/alainplattner/Desktop/Princeton/OwnPapers/bibliography/bib,/Users/alainplattner/Desktop/Princeton/OwnPapers/bibliography/bib_Handbook}

\begin{thebibliography}{63}
\expandafter\ifx\csname natexlab\endcsname\relax\def\natexlab#1{#1}\fi

\bibitem[Albertella et~al.(1999)Albertella, Sans{\`o}, \&
  Sneeuw]{Albertella+99}
Albertella, A., Sans{\`o}, F., \& Sneeuw, N., 1999.
\newblock Band-limited functions on a bounded spherical domain: the {S}lepian
  problem on the sphere, {\it J.~Geodesy\/}, {\bf 73}, 436--447.

\bibitem[Albertella et~al.(2008)Albertella, Savcenko, Bosch, \&
  Rummel]{Albertella+2008}
Albertella, A., Savcenko, R., Bosch, W., \& Rummel, R., 2008.
\newblock {Dynamic Ocean Topography -- {T}he Geodetic Approach}, Tech. Rep.~27,
  Institut f\"{u}r Astronomische und Physikalische Geod\"{a}sie,
  Forschungseinrichtung Satellitengeod\"{a}sie, M\"{u}nchen.

\bibitem[Arkani-Hamed(2001)]{Arkani-Hamed2001}
Arkani-Hamed, J., 2001.
\newblock A 50-degree spherical harmonic model of the magnetic field of {M}ars,
  {\it J.~Geophys.~Res.\/}, {\bf 106}(E10), 23197--23208, doi:
  10.1029/2000JE001365.

\bibitem[Arkani-Hamed(2002)]{Arkani-Hamed2002}
Arkani-Hamed, J., 2002.
\newblock An improved 50-degree spherical harmonic model of the magnetic field
  of {M}ars derived from both high-altitude and low-altitude data, {\it
  J.~Geophys.~Res.\/}, {\bf 107}(E10), 5083, doi: 10.1029/2001JE001835.

\bibitem[Arkani-Hamed(2004)]{Arkani-Hamed2004}
Arkani-Hamed, J., 2004.
\newblock A coherent model of the crustal magnetic field of {M}ars, {\it
  J.~Geophys.~Res.\/}, {\bf 109}, E09005, doi: 10.1029/2004JE002265.

\bibitem[Arkani-Hamed \& Strangway(1986)]{Arkani-Hamed+86}
Arkani-Hamed, J. \& Strangway, D.~W., 1986.
\newblock Band-limited global scalar magnetic anomaly map of the {E}arth
  derived from {M}agsat data, {\it J.~Geophys.~Res.\/}, {\bf 91}(B8),
  8193--8203.

\bibitem[Backus et~al.(1996)Backus, Parker, \& Constable]{Backus+96}
Backus, G.~E., Parker, R.~L., \& Constable, C.~G., 1996.
\newblock {\it Foundations of geomagnetism\/}, Cambridge Univ.~Press,
  Cambridge, UK.

\bibitem[Beggan et~al.(2013)Beggan, Saarim\"aki, Whaler, \&
  Simons]{Beggan+2013}
Beggan, C.~D., Saarim\"aki, J., Whaler, K.~A., \& Simons, F.~J., 2013.
\newblock Spectral and spatial decomposition of lithospheric magnetic field
  models using spherical {S}lepian functions, {\it Geophys.~J.~Int.\/}, {\bf
  193}(1), 136--148, 10.1093/gji/ggs122.

\bibitem[Blakely(1995)]{Blakely95}
Blakely, R.~J., 1995.
\newblock {\it Potential Theory {i}n Gravity {a}nd Magnetic Applications\/},
  Cambridge Univ.~Press, New York.

\bibitem[B{\"o}lling \& Grafarend(2005)]{Boelling+2005}
B{\"o}lling, K. \& Grafarend, E.~W., 2005.
\newblock Ellipsoidal spectral properties of the {E}arth’s gravitational
  potential and its first and second derivatives, {\it J.~Geodesy\/}, {\bf
  79}(6--7), 300--330, doi: 10.1007/s00190--005--0465--y.

\bibitem[Chambodut et~al.(2005)Chambodut, Panet, Mandea, Diament, Holschneider,
  \& Jamet]{Chambodut+2005}
Chambodut, A., Panet, I., Mandea, M., Diament, M., Holschneider, M., \& Jamet,
  O., 2005.
\newblock Wavelet frames: an alternative to spherical harmonic representation
  of potential fields, {\it Geophys.~J.~Int.\/}, {\bf 163}(3), 875--899.

\bibitem[Cox \& Hinkley(1974)]{Cox+74}
Cox, D.~R. \& Hinkley, D.~V., 1974.
\newblock {\it Theoretical Statistics\/}, Chapman and Hall, London, UK.

\bibitem[Dahlen \& Tromp(1998)]{Dahlen+98}
Dahlen, F.~A. \& Tromp, J., 1998.
\newblock {\it Theoretical Global Seismology\/}, Princeton Univ.~Press,
  Princeton, N.~J.

\bibitem[Davison(2003)]{Davison2003}
Davison, A.~C., 2003.
\newblock {\it Statistical Models\/}, Cambridge Univ.~Press, Cambridge, UK.

\bibitem[de~Santis(1991)]{DeSantis91}
de~Santis, A., 1991.
\newblock Translated originspherical cap harmonic analysis, {\it
  Geophys.~J.~Int.\/}, {\bf 106}, 253--263.

\bibitem[Eshagh(2009)]{Eshagh2009b}
Eshagh, M., 2009.
\newblock Comparison of two approaches for considering laterally varying
  density in topographic effect on satellite gravity gradiometric data, {\it
  Acta Geophysica\/}, pp. 10.2478/s11600--009--0057--y.

\bibitem[Fengler et~al.(2007)Fengler, Freeden, Kohlhaas, Michel, \&
  Peters]{Fengler+2006a}
Fengler, M.~J., Freeden, W., Kohlhaas, A., Michel, V., \& Peters, T., 2007.
\newblock Wavelet modeling of regional and temporal variations of the earth's
  gravitational potential observed by {GRACE}, {\it J.~Geodesy\/}, {\bf 81}(1),
  5--15, doi: 10.1007/s00190--006--0040--1.

\bibitem[Freeden \& Michel(2004)]{Freeden+2004b}
Freeden, W. \& Michel, V., 2004.
\newblock {\it Multiscale Potential Theory\/}, Birkh\"auser, Boston, Mass.

\bibitem[Freeden \& Schreiner(2009)]{Freeden+2009}
Freeden, W. \& Schreiner, M., 2009.
\newblock {\it Spherical Functions of Mathematical Geosciences: {A} Scalar,
  Vectorial, and Tensorial Setup\/}, Springer, Berlin.

\bibitem[Gubbins et~al.(2011)Gubbins, Ivers, Masterton, \&
  Winch]{Gubbins+2011b}
Gubbins, D., Ivers, D., Masterton, S.~M., \& Winch, D.~E., 2011.
\newblock Analysis of lithospheric magnetization in vector spherical harmonics,
  {\it Geophys.~J.~Int.\/}, {\bf 187}, 99--117, doi:
  10.1111/j.1365--246X.2011.05153.x.

\bibitem[Haines(1985)]{Haines85a}
Haines, G.~V., 1985.
\newblock Spherical cap harmonic analysis, {\it J.~Geophys.~Res.\/}, {\bf
  90}(B3), 2583--2591.

\bibitem[Harig \& Simons(2012)]{Harig+2012}
Harig, C. \& Simons, F.~J., 2012.
\newblock Mapping {G}reenland's mass loss in space and time, {\it
  Proc.~Natl.~Acad.~Sc.\/}, {\bf 109}(49), 19934--19937, doi:
  10.1073/pnas.1206785109.

\bibitem[Hwang(1993)]{Hwang93}
Hwang, C., 1993.
\newblock Spectral analysis using orthonormal functions with a case study on
  sea surface topography, {\it Geophys.~J.~Int.\/}, {\bf 115}, 1148--1160.

\bibitem[Hwang \& Chen(1997)]{Hwang+97}
Hwang, C. \& Chen, S.-K., 1997.
\newblock Fully normalized spherical cap harmonics: {A}pplication to the
  analysis of sea-level data from {TOPEX}/{POSEIDON} and {ERS}-1, {\it
  Geophys.~J.~Int.\/}, {\bf 129}, 450--460.

\bibitem[Jahn \& Bokor(2012)]{Jahn+2012}
Jahn, K. \& Bokor, N., 2012.
\newblock Vector slepian basis functions with optimal energy concentration in
  high numerical aperture focusing, {\it Optics Comm.\/}, {\bf 285},
  2028--2038, doi: 10.1016/j.optcom.2011.11.107.

\bibitem[Kaula(1967)]{Kaula67a}
Kaula, W.~M., 1967.
\newblock Theory of statistical analysis of data distributed over a sphere,
  {\it Rev.~Geophys.\/}, {\bf 5}(1), 83--107.

\bibitem[Kennedy \& Sadeghi(2013)]{Kennedy+2013}
Kennedy, R.~A. \& Sadeghi, P., 2013.
\newblock {\it Hilbert Space Methods in Signal Processing\/}, Cambridge
  Univ.~Press, Cambridge, UK.

\bibitem[Korte \& Holme(2003)]{Korte+2003}
Korte, M. \& Holme, R., 2003.
\newblock Regularization of spherical cap harmonics, {\it Geophys.~J.~Int.\/},
  {\bf 153}, 253--262, doi: 10.1046/j.1365--246X.2003.01898.x.

\bibitem[Langel \& Hinze(1998)]{Langel+98}
Langel, R.~A. \& Hinze, W.~J., 1998.
\newblock {\it The Magnetic Field of the Earth's Lithosphere: The Satellite
  Perspective\/}, Cambridge Univ.~Press, Cambridge, UK.

\bibitem[Lewis \& Simons(2012)]{Lewis+2012}
Lewis, K.~W. \& Simons, F.~J., 2012.
\newblock Local spectral variability and the origin of the {M}artian crustal
  magnetic field, {\it Geophys.~Res.~Lett.\/}, {\bf 39}, L18201, doi:
  10.1029/2012GL052708.

\bibitem[Lowes \& Winch(2012)]{Lowes+2012}
Lowes, F.~J. \& Winch, D.~E., 2012.
\newblock Orthogonality of harmonic potentials and fields in spheroidal and
  ellipsoidal coordinates: {a}pplication to geomagnetism and geodesy, {\it
  Geophys.~J.~Int.\/}, {\bf 191}(2), 491--507, doi:
  10.1111/j.1365--246X.2012.05590.x.

\bibitem[Lowes et~al.(1995)Lowes, de~Santis, \& Duka]{Lowes+95}
Lowes, F.~J., de~Santis, A., \& Duka, B., 1995.
\newblock A discussion of the uniqueness of a {L}aplacian potential when given
  only partial field information on a sphere, {\it Geophys.~J.~Int.\/}, {\bf
  121}(2), 579--584.

\bibitem[Mallat(2008)]{Mallat2008}
Mallat, S., 2008.
\newblock {\it A Wavelet Tour {o}f Signal Processing, {T}he Sparse Way\/},
  Academic Press, San Diego, Calif., 3rd edn.

\bibitem[Maus(2010)]{Maus2010}
Maus, S., 2010.
\newblock An ellipsoidal harmonic representation of {E}arth’s lithospheric
  magnetic field to degree and order 720, {\it Geochem.~Geophys.~Geosys.\/},
  {\bf 11}(6), Q06015, doi: 10.1029/2010GC003026.

\bibitem[Maus et~al.(2006{\natexlab{a}})Maus, L\"uhr, \& Purucker]{Maus+2006b}
Maus, S., L\"uhr, H., \& Purucker, M., 2006{\natexlab{a}}.
\newblock Simulation of the high-degree lithospheric field recovery for the
  \textit{Swarm} constellation of satellites, {\it Earth Planets Space\/}, {\bf
  58}, 397--407.

\bibitem[Maus et~al.(2006{\natexlab{b}})Maus, Rother, Hemant, Stolle, L{\"u}hr,
  Kuvshinov, \& Olsen]{Maus+2006c}
Maus, S., Rother, M., Hemant, K., Stolle, C., L{\"u}hr, H., Kuvshinov, A., \&
  Olsen, N., 2006{\natexlab{b}}.
\newblock Earth's lithospheric magnetic field determined to spherical harmonic
  degree 90 from {CHAMP} satellite measurements, {\it Geophys.~J.~Int.\/}, {\bf
  164}, 319--330, doi: 10.1111/j.1365--246X.2005.02833.x.

\bibitem[Maus et~al.(2006{\natexlab{c}})Maus, Rother, Stolle, Mai, Choi,
  L{\"u}hr, Cooke, \& Roth]{Maus+2006a}
Maus, S., Rother, M., Stolle, C., Mai, W., Choi, S., L{\"u}hr, H., Cooke, D.,
  \& Roth, C., 2006{\natexlab{c}}.
\newblock Third generation of the {P}otsdam {M}agnetic {M}odel of the {E}arth
  ({POMME}), {\it Geochem.~Geophys.~Geosys.\/}, {\bf 7}, Q07008, doi:
  10.1029/2006GC001269.

\bibitem[Mayer \& Maier(2006)]{Mayer+2006}
Mayer, C. \& Maier, T., 2006.
\newblock Separating inner and outer {E}arth's magnetic field from {CHAMP}
  satellite measurements by means of vector scaling functions and wavelets,
  {\it Geophys.~J.~Int.\/}, {\bf 167}, 1188--1203, doi:
  10.1111/j.1365--246X.2006.03199.x.

\bibitem[Moritz(2010)]{Moritz2010}
Moritz, H., 2010.
\newblock Classical physical geodesy, in {\em Handbook of Geomathematics\/},
  chap.~6, pp. 130--158, doi: 10.1007/978--3--642--01546--5\_6, eds Freeden,
  W., Nashed, M.~Z., \& Sonar, T., Springer, Heidelberg, Germany.

\bibitem[Nutz(2002)]{Nutz2002}
Nutz, H., 2002.
\newblock {\it A Unified Setup of Gravitational Field Observables\/}, Ph.D.
  thesis, Univ.~Kaiserslautern, Germany.

\bibitem[O'Brien \& Parker(1994)]{OBrien+94}
O'Brien, M.~S. \& Parker, R.~L., 1994.
\newblock Regularized geomagnetic field modelling using monopoles, {\it
  Geophys.~J.~Int.\/}, {\bf 118}(3), 566--578, doi:
  10.1111/j.1365--246X.1994.tb03985.x.

\bibitem[Olsen et~al.(2009)Olsen, Mandea, Sabaka, \&
  T{\o}ffner-Clausen]{Olsen+2009}
Olsen, N., Mandea, M., Sabaka, T.~J., \& T{\o}ffner-Clausen, L., 2009.
\newblock {CHAOS-2}---a geomagnetic field model derived from one decade of
  continuous satellite data, {\it Geophys.~J.~Int.\/}, {\bf 179}, 1477--1487,
  doi: 10.1111/j.1365--246X.2009.04386.x.

\bibitem[Olsen et~al.(2010)Olsen, Hulot, \& Sabaka]{Olsen+2010}
Olsen, N., Hulot, G., \& Sabaka, T.~J., 2010.
\newblock Sources of the geomagnetic field and the modern data that enable
  their investigation, in {\em Handbook of Geomathematics\/}, chap.~5, pp.
  105--124, doi: 10.1007/978--3--642--01546--5\_5, eds Freeden, W., Nashed,
  M.~Z., \& Sonar, T., Springer, Heidelberg, Germany.

\bibitem[Plattner \& Simons(2013)]{Plattner+2013}
Plattner, A. \& Simons, F.~J., 2013.
\newblock Spatiospectral concentration of vector fields on a sphere, {\it
  Appl.~Comput.~Harmon.~Anal.\/}, p. doi:10.1016/j.acha.2012.12.001.

\bibitem[Rowlands et~al.(2005)Rowlands, Luthcke, Klosko, Lemoine, Chinn,
  McCarthy, Cox, \& Anderson]{Rowlands+2005}
Rowlands, D.~D., Luthcke, S.~B., Klosko, S.~M., Lemoine, F. G.~R., Chinn,
  D.~S., McCarthy, J.~J., Cox, C.~M., \& Anderson, O.~B., 2005.
\newblock Resolving mass flux at high spatial and temporal resolution using
  {GRACE} intersatellite measurements, {\it Geophys.~Res.~Lett.\/}, {\bf 32},
  L04310, doi: 10.1029/2004GL021908.

\bibitem[Rummel \& van Gelderen(1995)]{Rummel+95}
Rummel, R. \& van Gelderen, M., 1995.
\newblock Meissl scheme --- {s}pectral characteristics of physical geodesy,
  {\it Manuscr.~Geod.\/}, {\bf 20}(5), 379--385.

\bibitem[Sabaka et~al.(2010)Sabaka, Hulot, \& Olsen]{Sabaka+2010}
Sabaka, T.~J., Hulot, G., \& Olsen, N., 2010.
\newblock Mathematical properties relevant to geomagnetic field modeling, in
  {\em Handbook of Geomathematics\/}, chap.~17, pp. 503--538, doi:
  10.1007/978--3--642--01546--5\_17, eds Freeden, W., Nashed, M.~Z., \& Sonar,
  T., Springer, Heidelberg, Germany.

\bibitem[Schachtschneider et~al.(2010)Schachtschneider, Holschneider, \&
  Mandea]{Schachtschneider+2010}
Schachtschneider, R., Holschneider, M., \& Mandea, M., 2010.
\newblock Error distribution in regional inversion of potential field data,
  {\it Geophys.~J.~Int.\/}, {\bf 181}, 1428--1440, doi:
  10.1111/j.1365--246X.2010.04598.x.

\bibitem[Schachtschneider et~al.(2012)Schachtschneider, Holschneider, \&
  Mandea]{Schachtschneider+2012}
Schachtschneider, R., Holschneider, M., \& Mandea, M., 2012.
\newblock Error distribution in regional modelling of the geomagnetic field,
  {\it Geophys.~J.~Int.\/}, {\bf 191}, 1015--1024, doi:
  10.1111/j.1365--246X.2012.05675.x.

\bibitem[Simons \& Dahlen(2006)]{Simons+2006b}
Simons, F.~J. \& Dahlen, F.~A., 2006.
\newblock Spherical {S}lepian functions and the polar gap in geodesy, {\it
  Geophys.~J.~Int.\/}, {\bf 166}, 1039--1061, doi:
  10.1111/j.1365--246X.2006.03065.x.

\bibitem[Simons et~al.(2006)Simons, Dahlen, \& Wieczorek]{Simons+2006a}
Simons, F.~J., Dahlen, F.~A., \& Wieczorek, M.~A., 2006.
\newblock Spatiospectral concentration on a sphere, {\it SIAM Rev.\/}, {\bf
  48}(3), 504--536, doi: 10.1137/S0036144504445765.

\bibitem[Simons et~al.(2009)Simons, Hawthorne, \& Beggan]{Simons+2009b}
Simons, F.~J., Hawthorne, J.~C., \& Beggan, C.~D., 2009.
\newblock Efficient analysis and representation of geophysical processes using
  localized spherical basis functions, in {\em Wavelets~{XIII}\/}, vol. 7446,
  pp. 74460G, doi: 10.1117/12.825730, SPIE.

\bibitem[Slepian(1964)]{Slepian64}
Slepian, D., 1964.
\newblock Prolate spheroidal wave functions, {F}ourier analysis and uncertainty
  --- {IV}: {E}xtensions to many dimensions; generalized prolate spheroidal
  functions, {\it Bell Syst.~Tech.~J.\/}, {\bf 43}(6), 3009--3057.

\bibitem[Slepian(1983)]{Slepian83}
Slepian, D., 1983.
\newblock Some comments on {F}ourier analysis, uncertainty and modeling, {\it
  SIAM Rev.\/}, {\bf 25}(3), 379--393.

\bibitem[Slepian \& Pollak(1961)]{Slepian+61}
Slepian, D. \& Pollak, H.~O., 1961.
\newblock Prolate spheroidal wave functions, {F}ourier analysis and uncertainty
  --- {I}, {\it Bell Syst.~Tech.~J.\/}, {\bf 40}(1), 43--63.

\bibitem[Slobbe et~al.(2012)Slobbe, Simons, \& Klees]{Slobbe+2012}
Slobbe, D.~C., Simons, F.~J., \& Klees, R., 2012.
\newblock The spherical {S}lepian basis as a means to obtain spectral
  consistency between mean sea level and the geoid, {\it J.~Geodesy\/}, {\bf
  86}(8), 609--628, doi: 10.1007/s00190--012--0543--x.

\bibitem[Th{\'e}bault et~al.(2006)Th{\'e}bault, Schott, \&
  Mandea]{Thebault+2006}
Th{\'e}bault, E., Schott, J.~J., \& Mandea, M., 2006.
\newblock Revised spherical cap harmonic analysis ({R-SCHA}): Validation and
  properties, {\it J.~Geophys.~Res.\/}, {\bf 111}(B1), B01102, doi:
  10.1029/2005JB003836.

\bibitem[Trampert \& Snieder(1996)]{Trampert+96b}
Trampert, J. \& Snieder, R., 1996.
\newblock Model estimations biased by truncated expansions: {P}ossible
  artifacts in seismic tomography, {\it Science\/}, {\bf 271}(5253),
  1257--1260, doi: 10.1126/science.271.5253.1257.

\bibitem[Whaler \& Gubbins(1981)]{Whaler+81}
Whaler, K.~A. \& Gubbins, D., 1981.
\newblock Spherical harmonic analysis of the geomagnetic field: an example of a
  linear inverse problem, {\it Geophys.~J.~Int.\/}, {\bf 65}(3), 645--693, doi:
  10.1111/j.1365--246X.1981.tb04877.x.

\bibitem[Wieczorek \& Simons(2007)]{Wieczorek+2007}
Wieczorek, M.~A. \& Simons, F.~J., 2007.
\newblock Minimum-variance spectral analysis on the sphere, {\it J.~Fourier
  Anal.~Appl.\/}, {\bf 13}(6), 665--692, doi: 10.1007/s00041--006--6904--1.

\bibitem[Xu(1992{\natexlab{a}})]{Xu92a}
Xu, P., 1992{\natexlab{a}}.
\newblock Determination of surface gravity anomalies using gradiometric
  observables, {\it Geophys.~J.~Int.\/}, {\bf 110}, 321--332.

\bibitem[Xu(1992{\natexlab{b}})]{Xu92b}
Xu, P., 1992{\natexlab{b}}.
\newblock The value of minimum norm estimation of geopotential fields, {\it
  Geophys.~J.~Int.\/}, {\bf 111}, 170--178.

\bibitem[Xu(1998)]{Xu98}
Xu, P., 1998.
\newblock Truncated {SVD} methods for discrete linear ill-posed problems, {\it
  Geophys.~J.~Int.\/}, {\bf 135}(2), 505--514, doi:
  10.1046/j.1365--246X.1998.00652.x.

\end{thebibliography}
\bibliographystyle{gji}

\newpage

\section*{Table of Symbols}

\begin{center}
\begin{longtable}{c|l|c}
symbol&description&Eq.\\\hline
\endfirsthead

symbol&description&Eq.\\\hline
\endhead

\multicolumn{3}{r}{{\textit{Continued on next page}}} \\ 
\endfoot

\hline
\endlastfoot

\hline
$L$&spherical-harmonic bandwidth&\\
$\region$& target region of data availability and for Slepian function concentration&\\
$\signal(\Earthrad\rvec)$& three-dimensional potential-field function at Earth's surface $\Earthrad$&(\ref{scalar decomposition at radius randrada})\\
$\signal(\satalt\rvec)$& three-dimensional potential-field function at satellite altitude $\satalt$&(\ref{continued potential})\\
$\ssphcoef_{lm}^\Earthrad$& expansion coefficients of $\signal(\Earthrad\rvec)$ in the basis of spherical harmonics  $\Yfun_{lm}$ &(\ref{definition uEarthrad})\\
$\ssphcoef_{lm}^\satalt$& expansion coefficients of $\signal(\satalt\rvec)$ in the basis of spherical harmonics  $\Yfun_{lm}$ &(\ref{bloL})\\
$\partial_r\signal(\satalt\rvec)$ & radial derivative of the potential field at satellite altitude $\satalt$&(\ref{radial derivative})\\
$\vecsignal(\satalt\rvec)$& three-dimensional gradient of the potential field at satellite altitude $\satalt$&(\ref{potential gradient})\\
$\svsphcoef^\satalt_{lm}$& expansion coefficients of $\vecsignal(\satalt\rvec)$ in the basis of gradient-vector harmonics $\Efun_{lm}$& (\ref{vector decomposition at radius randrada pt2})\\
$\sphcoef^\randrada$&  vector containing the coefficients $\ssphcoef_{lm}^\randrada$ at radius $\randrada$ &(\ref{definition uEarthrad})\\
$\vsphcoef^\randrada$& vector containing the coefficients $\svsphcoef^\randrada_{lm}$ at radius $\randrada$ &(\ref{vector decomposition at radius randrada pt2})\\
$\Amat$& $(L+1)^2\times(L+1)^2$ diagonal matrix transforming the $\sphcoef^\Earthrad$ to the $\Yfun_{lm}$ coefficients of $\partial_r\signal(\satalt\rvec)$&(\ref{definition Amat})\\
$\Bmat$& $(L+1)^2\times(L+1)^2$ diagonal matrix transforming the  $\sphcoef^\Earthrad$  to the $\Efun_{lm}$ coefficients of $\vecsignal(\satalt\rvec)$&(\ref{definition B})\\
\hline
$\Yfun_{lm}$& scalar spherical-harmonic function for degree $l$ and order $m$\hspace{2.6cm} &(\ref{Y definition})\\
$\Yfunvec$& vector of all $(L+1)^2$ scalar spherical-harmonic functions to degree $L$& (\ref{Yfunvec definition})\\
 $\Ypoints$& $(L+1)^2\times\npoints$ matrix of $\Yfun_{lm}$ with bandwidth $L$ evaluated at $\rvec_1,\ldots,\rvec_\npoints$ &(\ref{Y point matrix})\\
 $\hat\Yfunvec$& vector of all scalar spherical-harmonic functions to degree $\infty$ &(\ref{Yfunvec all degrees})\\
$\hat\Yfunvec_{>L}$& vector of all scalar spherical-harmonic functions for degrees $L<l\le \infty$& (\ref{Yfunvec all degrees})\\ \hline
$\Efun_{lm}$& gradient-vector spherical-harmonic function for degree $l$ and order $m$ &(\ref{E definition})\\
$\Efunvec$& $(L+1)^2$ vector of all $\Efun_{lm}$ up to degree $L$&(\ref{Efunvec definition})\\
$\Epoints$&$(L+1)^2\times3\npoints$ matrix of all of the $\Efun_{lm}$ evaluated at $\rvec_1,\ldots,\rvec_\npoints$ &(\ref{E point matrix})\\
$\hat\Efunvec$&vector of all gradient-vector spherical harmonics up to degree $\infty$&(\ref{Efunvec all degrees})\\
$\hat\Efunvec_{>L}$&vector of all gradient-vector spherical harmonics for degrees $L<l\le \infty$&(\ref{Efunvec all degrees})\\
\hline
$\Gfun_\alpha$& $\alpha$th best spatially concentrated (within $\region$) bandlimited (to $L$) scalar spherical Slepian function  &(\ref{definition Gfun})\\ 
$\hat\Gfun_\alpha$&$\alpha$th best spectrally concentrated (within $L$) spacelimited (to $R$) scalar spherical Slepian function &(\ref{scalar spatial truncation})\\
$\slepfuncoef_\alpha$&$(L+1)^2\times 1$ vector containing the $\Yfun_{lm}$ coefficients of one of the $\Gfun_\alpha$&(\ref{definition Gfun})\\
$\hat\slepfuncoef_\alpha$&infinite-dimensional vector containing the $\Yfun_{lm}$ coefficients of one of the $\hat\Gfun_\alpha$&(\ref{definition hat slepfuncoef})\\ 
$\Gfunvec$& $(L+1)^2\times 1$ vector containing all of the $\Gfun_\alpha$ ordered with decreasing concentration ratio $\lambda_\alpha$ &(\ref{definition Gfunvec})\\
$\Gfunvec_J$& $J\times 1$  vector of functions containing the $\Gfun_1,\ldots,\Gfun_J$ &(\ref{definition GfunvecJ})\\
$\Gfunvec_{\downarrow J}$&$J\times 1$ vector of localized downward-transformed scalar Slepian functions &(\ref{downward gay jay})\\
$\Gfunvec_{\uparrow J}$&$J\times 1$ vector of localized upward-transformed scalar Slepian functions &(\ref{definition Gfunvec upJ})\\
$\Gfunvec_{\downarrow >J}$&$\left[(L+1)^2-J+1\right]\times 1$ vector complementing $\Gfunvec_{\downarrow J}$  &(\ref{downward gay jay})\\
$\Gfunvec_{\uparrow >J}$&$\left[(L+1)^2-J+1\right]\times 1$ vector complementing $\Gfunvec_{\uparrow J}$  &(\ref{definition Gfunvec upJ})\\
$\Gpoints$&$(L+1)^2\times\npoints$ matrix of all of the $\Gfun_{\alpha}$ evaluated at
$\rvec_1,\ldots,\rvec_\npoints$&(\ref{definition Gpoints})\\ 
$\Gpoints_J$&$J\times\npoints$ matrix of $\Gfun_{1},\ldots,\Gfun_J$ evaluated at $\rvec_1,\ldots,\rvec_\npoints$&(\ref{definition GpointsJ})\\
$\Gmat$&$(L+1)^2\times(L+1)^2$ matrix containing the $\Yfun_{lm}$ coefficients for all of the $\Gfun_\alpha$&(\ref{definition Gmat})\\
$\GmatJ$&$(L+1)^2\times J$ matrix containing the $\Yfun_{lm}$ coefficients for the $\Gfun_1,\ldots,\Gfun_J$&(\ref{definition GmatJ})\\
$\lambda_\alpha$& energy concentration ratio of one of the $\Gfun_\alpha$ &(\ref{scalar concentration problem})\\
$\Lamat$&$(L+1)^2\times(L+1)^2$ diagonal  matrix containing all of the  $\lambda_\alpha$&(\ref{scalar eigenvalue problem})\\
$\Lamat_J$&$J\times J$ diagonal matrix containing the $J$ largest $\lambda_1,\ldots,\lambda_J$&(\ref{scalar integration Slepian function})\\
$\Dmat$&$(L+1)^2\times(L+1)^2$ localization matrix diagonalized by
$\Gmat$ & (\ref{definition scalar kernel})\\
$\DmatL$&$\infty\times(L+1)^2$  matrix extending $\Dmat$ down to contain the inner products of $\hat\Yfunvec$ and $\Yfunvec$ &(\ref{definition DmatL})\\
$\DmatLL$&$\infty\times(L+1)^2$  matrix containing the lowermost portion of $\DmatL$ for degrees $l>L$ &(\ref{definition hatDmat L+1 L})\\
$\GLa$&scalar function made from the degrees  $l>L$ of $\hat\Gfun_{\alpha}$&(\ref{definition Gfun >L})\\
$\hat\slepfuncoef_{>L,\alpha}$& infinite-dimensional vector containing the $l>L$ entries of $\hat\slepfuncoef_\alpha$&(\ref{definition single g >L})\\
$\GLJ$&$J\times 1$ vector of functions containing the first $J$ of the $\GLa$&(\ref{definition hatGfun J L})\\
$\GmatLJ$&$\infty\times J$  matrix containing the $\Yfun_{lm}$ coefficients for $l>L$ of the $\hat\Gfun_{\alpha}$ &(\ref{definition hatGmat J L})\\\hline\newpage
$\Hfun_\alpha$&$\alpha$th best-concentrated gradient-vector Slepian function for bandwidth $L$ and region $\region$&(\ref{vslepor})\\
$\vslepfuncoef_\alpha$&$(L + 1)^2\times 1$ vector containing the
$\Efun_{lm}$ coefficients of one of the $\Hfun_\alpha$&(\ref{definition hvec})\\ 
$\Hfunvec$&$(L+1)^2\times 1$ vector containing all of the
$\Hfun_\alpha$ ordered with decreasing concentration ratio $\sigma_\alpha$&(\ref{definition Hfunvec})\\
$\Hfunvec_J$& $J\times 1$  vector of functions containing the $\Hfun_1,\ldots,\Hfun_J$&(\ref{definition HfunvecJ})\\
$\Hpfunvec_{\downarrow J}$&$J\times 1$ vector of scalar-valued downward-transformed gradient-vector Slepian functions &(\ref{definition Hpfunvec down})\\
$\Hpfunvec_{\uparrow J}$&$J\times 1$ vector of scalar-valued upward-transformed gradient-vector Slepian functions  &(\ref{definition Hpfunvec up})\\
$\Hpfunvec_{\downarrow >J}$&$\left[(L+1)^2-J+1\right]\times 1$ vector complementing $\Hpfunvec_{\downarrow J}$  &(\ref{definition Hpfunvec down})\\
$\Hpfunvec_{\uparrow >J}$&$\left[(L+1)^2-J+1\right]\times 1$ vector complementing $\Hpfunvec_{\uparrow J}$  &(\ref{definition Hpfunvec up})\\
$\Hpoints$&$(L+1)^2\times3\npoints$ matrix of all of the $\Hfun_{\alpha}$ evaluated at
$\rvec_1,\ldots,\rvec_\npoints$&(\ref{definition
  Hpoints})\\
$\Hpoints_J$&$J\times3\npoints$ matrix of $\Hfun_{1},\ldots,\Hfun_J$ evaluated at
$\rvec_1,\ldots,\rvec_\npoints$&(\ref{definition HpointsJ})\\ 
$\Hmat$&$(L+1)^2\times(L+1)^2$ matrix containing the
$\Efun_{lm}$ coefficients for all of the
$\Hfun_\alpha$&(\ref{definition Hmat})\\
$\HmatJ$&$(L+1)^2\times J$ matrix containing the $\Efun_{lm}$ coefficients for the $\Hfun_1,\ldots,\Hfun_J$&(\ref{definition HmatJ})\\
$\sigma_\alpha$&energy concentration ratio of $\Hfun_\alpha$ over $\region$
&(\ref{vector Slepian optimization problem})\\
$\Sigmat$&$(L+1)^2\times(L+1)^2$ diagonal matrix containing all of the $\sigma_\alpha$&
(\ref{definition gradient vector Slepian function coefficients})\\
$\Sigmat_J$&$J\times J$ diagonal matrix containing the $J$ largest $\sigma_1,\ldots,\sigma_J$&(\ref{definition SigmatJ})\\
$\Kmat$&$(L+1)^2\times(L+1)^2$ localization matrix diagonalized by
$\Hmat$ &(\ref{definition vector Kernel})\\
$\KmatLL$&$\infty\times(L+1)^2$ matrix containing the inner products of $\hat\Efunvec_{>L}$ with $\Efunvec$&(\ref{definition hat Kmat L+1 >L})\\
$\HLJ$&$J\times 1$ vector of functions containing the $\hat\Efunvec_{>L}$ components of the spacelimited $\Hfunvec$   &(\ref{definition hatHfun J L})\\
\hline
$\datarad(\rvec)$&scalar data function at satellite altitude $\satalt$&(\ref{data only in region})\\
$\datavec(\rvec)$&gradient data function at satellite altitude $\satalt$&(\ref{vector data in region})\\
$\dpoints_r$&$\npoints\times 1$ vector of measured radial data values at satellite altitude $\satalt$&(\ref{scalar noisy data})\\
$\dpoints$&$3\npoints\times 1$ vector of measured gradient data values at satellite altitude $\satalt$&(\ref{vectorial noisy data})\\
$\noise(\rvec)$&scalar noise function at satellite altitude $\satalt$&(\ref{data only in region})\\
$\vecnoise(\rvec)$&vectorial noise function at satellite altitude $\satalt$&(\ref{vector data in region})\\
$\noisepoints_r$&$\npoints\times 1$ vector of radial-derivative noise at satellite altitude $\satalt$&(\ref{scalar noisy data})\\
$\noisepoints$&$3\npoints\times 1$ vector of vectorial noise at satellite altitude $\satalt$&(\ref{vectorial noisy data})\\
$\sigpoints$& $\npoints\times 1$  vector containing the potential-field signal points $\signal(\satalt\rvec),\ldots,\signal(\satalt\rvec)$ &(\ref{definition sigpoints})\\ 
$\gradsigpoints_r$&$\npoints\times 1$  vector containing the radial-derivative signal points  $\partial_r\signal(\satalt\rvec),\ldots,\partial_r\signal(\satalt\rvec)$&(\ref{radial data})\\ 
$\gradsigpoints$&$3\npoints\times 1$  vector containing the full gradient signal points  $\gradsigpoints_r, \gradsigpoints_\theta$, and $\gradsigpoints_\phi$ at satellite altitude&(\ref{full vectorial data})\\ 
\hline
$\estsignal(\Earthrad\rvec)$&potential field at the Earth's surface estimated from the radial-derivative data at altitude&(\ref{scalar field estimation})\\
                           &potential field at the Earth's surface estimated from the full gradient data at altitude&(\ref{vectorial field estimation})\\
$\srsj$&$J\times 1$ vector of $\Gfun_\alpha$ coefficients of $\signal(\satalt\rvec)$  estimated from the scalar data $\dpoints_r$&(\ref{kukuk})\\
$\trsj$&$J\times 1$ vector of $\Hfun_\alpha$ coefficients of $\vecsignal(\satalt\rvec)$  estimated from the vector data $\dpoints$&(\ref{vector numerical least squares solution})\\
$\estsphcoef^\Earthrad$&$(L+1)^2\times 1$ vector of $\Yfun_{lm}$ coefficients of the estimate $\estsignal(\Earthrad\rvec)$  derived from the $\srs$ &(\ref{solution 2 to problem 2})\\
&$(L+1)^2\times 1$ vector of $\Yfun_{lm}$ coefficients of the estimate $\estsignal(\Earthrad\rvec)$ derived from  the $\trsj$ &(\ref{solution 2 problem 4})\\
\hline
$\variance$&variance of the estimate  $\estsignal(\Earthrad\rvec)$ from the scalar data $\datarad(\rvec)$ in truncated Slepian estimation &(\ref{scalar special variance})\\
           &variance of the estimate  $\estsignal(\Earthrad\rvec)$ from the vector data $\datavec(\rvec)$ in truncated Slepian estimation &(\ref{vector variance})\\
$\bias$&bias of the estimate $\estsignal(\Earthrad\rvec)$ from the scalar data $\datarad(\rvec)$ in truncated Slepian estimation &(\ref{corr1})\\
       &bias of the estimate $\estsignal(\Earthrad\rvec)$ from the vector data $\datavec(\rvec)$ in truncated Slepian estimation &(\ref{vector bias equation})\\
$\langle\esterr^2\rangle$&mean squared error of the estimate  $\estsignal(\Earthrad\rvec)$ from the scalar data $\datarad(\rvec)$ &(\ref{scalar mse equation})\\
                         &mean squared error of the estimate  $\estsignal(\Earthrad\rvec)$ from the vector data $\datavec(\rvec)$  &(\ref{vector mean squared error special case})\\
$\varphi(j)$&relative regional mean squared model error between $\estsignal(\Earthrad\rvec)$ and  $\signal(\Earthrad\rvec)$
&(\ref{solution relative mse})\\
$\psi(J)$&relative mean squared data misfit between $\dpoints_r$ and
$\Ypoints^\pointT\Amat\estsphcoef^\Earthrad$  for the scalar case &(\ref{data relative mse})\\
&relative mean squared data misfit between $\dpoints$ and $\Epoints^\pointT\Bmat\sphcoef^\Earthrad$ or 
 for the vector case&(\ref{vector data relative mse})\\
\hline
\end{longtable}
\end{center}

\end{document}